\documentclass[conference]{IEEEtran}
\IEEEoverridecommandlockouts
\usepackage{cite}
\usepackage{amsmath,amssymb,amsfonts}
\usepackage{algorithmic}
\usepackage{graphicx}
\usepackage{textcomp}
\usepackage{xcolor}
\usepackage{subfiles} 
\usepackage{hyperref}

\def\BibTeX{{\rm B\kern-.05em{\sc i\kern-.025em b}\kern-.08em
    T\kern-.1667em\lower.7ex\hbox{E}\kern-.125emX}}

\newcommand{\ra}[1]{\textcolor{black}{ #1}}
\newcommand{\rabiah}[1]{\textcolor{black}{ #1}}

\newcommand{\getsr}{\ensuremath{\leftarrow_\$}}
\newcommand{\tor}{\ensuremath{\rightarrow_\$}}

\newcommand{\Zq}{\mathbb{Z}_q}
\newcommand{\NN}{\mathbb{N}}
\newcommand{\bits}[1]{\{0,1\}^{#1}}
\newcommand{\secpar}{\lambda}
\mathchardef\mhyphen="2D

\newtheorem{theorem}{\textbf{Theorem}}
\newtheorem{definition}{\textbf{Definition}}
\newtheorem{remark}{\textbf{Remark}}

\newcommand{\AuC}{\mathrm{AuC}}
\newcommand{\UE}{\mathrm{UE}}
\newcommand{\gNB}{\mathrm{gNB}}

\newcommand{\numParties}{\ensuremath{n_P}}
\newcommand{\numSessions}{\ensuremath{n_S}}

\newcommand{\Alg}[1]{\ensuremath{\mathsf{#1}}}
\newcommand{\SanSig}{\Alg{SanSig}}
\newcommand{\AuthEnc}{\Alg{AE}}

\newcommand{\KeyGen}{\Alg{KGen}}
\newcommand{\Sign}{\Alg{Sign}}
\newcommand{\Sanit}{\Alg{Sanit}}
\newcommand{\Verify}{\Alg{Verify}}
\newcommand{\proof}{\Alg{Proof}}

\newcommand{\Judge}{\Alg{Judge}}

\newcommand{\Enc}{\Alg{Enc}}
\newcommand{\Dec}{\Alg{Dec}}
\newcommand{\KDF}{\Alg{KDF}}
\newcommand{\DDH}{\Alg{DDH}}
\newcommand{\EUFCMA}{\Alg{EUFCMA}}
\newcommand{\KGenacc}{\Alg{KGen}_{\ddot{C}}}
\newcommand{\Genacc}{\Alg{Gen}_{\ddot{C}}}
\newcommand{\Updateacc}{\Alg{Update}_{\ddot{C}}}
\newcommand{\Nwitcreate}{\Alg{NonWitCreate}}
\newcommand{\Nwitupdate}{\Alg{NonWitUpdate}}
\newcommand{\Verifyacc}{\Alg{Verify}_{\ddot{C}}}
\newcommand{\UniHand}{\ensuremath{\mathsf{UniHand}}}

\newcommand{\challenger}{\mathcal{C}}
\newcommand{\adversary}{\mathcal{A}}
\newcommand{\chal}{\ensuremath{\mathcal{C}}}

\newcommand{\adv}{\ensuremath{\mathcal{A}}}
\newcommand{\bdv}{\ensuremath{\mathcal{B}}}
\newcommand{\Unlink}{\Alg{Unlink}}
\newcommand{\Prot}{\Pi}

\newcommand{\msgsent}{msg_s}
\newcommand{\msgrecd}{msg_r}
\newcommand{\sid}{sid}

\newcommand{\id}{\mathit{id}}
\newcommand{\inprogress}{\mathtt{in\mhyphen progress}}
\newcommand{\acceptflag}{\mathtt{accepted}}
\newcommand{\rejectflag}{\mathtt{rejected}}
\newcommand{\testsess}{\pi_i^s}
\newcommand{\partsess}{\pi_j^t}

\newcommand{\session}{\pi}
\newcommand{\cleanpredicate}{\mathbf{clean}}

\newcommand{\status}{\ensuremath{\alpha}}

\newcommand{\MA}{\ensuremath{\mathsf{MA}}}
\newcommand{\KIND}{\ensuremath{\mathsf{KIND}}}
\newcommand{\Conf}{\ensuremath{\mathsf{Conf}}}

\newcommand{\pk}{\ensuremath{\mathit{pk}}}
\newcommand{\sk}{\ensuremath{\mathit{sk}}}
\newcommand{\adm}{\ensuremath{\mathit{ADM}}}
\newcommand{\modd}{\ensuremath{\mathit{MOD}}}
\newcommand{\cert}{\ensuremath{\mathbb{C}}}
\newcommand{\pid}{\ensuremath{\mathit{pid}}}
\newcommand{\gid}{\ensuremath{\mathit{gid}}}
\newcommand{\role}{\ensuremath{\mathit{\rho}}}
\newcommand{\tid}{\ensuremath{\mathit{T_{ID}}}}

\newcommand{\skacc}{\ensuremath{\mathit{sk}_{\ddot{C}}}}
\newcommand{\GID}{\mathrm{GID}}

\newcommand{\EID}{\mathrm{EID}}

\newcommand{\UID}{\mathrm{UID}}
\newcommand{\RUID}{\mathrm{RUID}}

\newcommand{\RL}{\mathit{RL}}

\newcommand{\Send}{\ensuremath{\Alg{\textbf{Send}}}}
\newcommand{\Reveal}{\ensuremath{\Alg{\textbf{Reveal}}}}
\newcommand{\Corrupt}{\ensuremath{\Alg{\textbf{Corrupt}}}}
\newcommand{\CorruptLTK}{\ensuremath{\Alg{\textbf{CorruptLTK}}}}
\newcommand{\CorruptASK}{\ensuremath{\Alg{\textbf{CorruptASK}}}}
\newcommand{\StateReveal}{\Alg{\textbf{StateReveal}}}
\newcommand{\SendTest}{\Alg{\textbf{SendTest}}}
\newcommand{\Test}{\Alg{\textbf{Test}}}
\newcommand{\Create}{\Alg{\textbf{Create}}}

\newcommand{\Adv}[2]{\ensuremath{\mathbf{Adv}^{#1}_{#2}}}
\newcommand{\Exp}[2]{\ensuremath{\mathbf{Exp}^{#1}_{#2}}}

\newcommand{\ClientAction}[1]{ 
	\node[right] at (\InitX, \Y) {#1};
}
\newcommand{\ServerAction}[1]{
	\node[left] at (\RespX, \Y) {#1};
}

\newcommand{\AdversaryAction}[1]{
	\node at ($1/2*(\InitX, \Y)+1/2*(\RespX, \Y)$) {#1};
}
\newcommand{\ClientToServer}[3][->]{
	\NextLine[0.5]
	\draw[#1] (\ArrowLeft,\Y) -- node[above] {#2} node[below] {#3} (\ArrowRight,\Y) ;
	\NextLine[0.5]
}
\newcommand{\ServerToClient}[3][->]{
	\NextLine[0.5]
	\draw[#1] (\ArrowRight,\Y) -- node[above] {#2} node[below] {#3} (\ArrowLeft,\Y) ;
	\NextLine[0.5]
}
\newcommand{\ClientToAdversary}[3][->]{
	\NextLine[0.5]
	\draw[#1] (\ArrowLeft,\Y) -- node[above] {#2} node[below] {#3} (\ArrowCenter,\Y) ;
	\NextLine[0.5]
}
\newcommand{\ServerToAdversary}[3][->]{
	\NextLine[0.5]
	\draw[#1] (\ArrowRight,\Y) -- node[above] {#2} node[below] {#3} (\ArrowCenter,\Y) ;
	\NextLine[0.5]
}

\newcommand{\AdversaryToClient}[3][->]{
	\NextLine[0.5]
	\draw[#1] (\ArrowCenter,\Y) -- node[above] {#2} node[below] {#3} (\ArrowLeft,\Y) ;
	\NextLine[0.5]
}
\newcommand{\AdversaryToServer}[3][->]{
	\NextLine[0.5]
	\draw[#1] (\ArrowCenter,\Y) -- node[above] {#2} node[below] {#3} (\ArrowRight,\Y) ;
	\NextLine[0.5]
}

\newcommand{\NextLine}[1][1.0]{
	\pgfmathparse{\Y+#1}
	\edef\Y{\pgfmathresult}
}
\newcommand{\linkgame}[2]{\hyperref[#1]{G#2}}

\newcounter{Bdversary}

\newcommand\orcidicon[1]{\href{https://orcid.org/#1}{\mbox{\scalerel*{
\begin{tikzpicture}[yscale=-1,transform shape]
\pic{orcidlogo};
\end{tikzpicture}
}{|}}}}

\newcommand*{\addFileDependency}[1]{% argument=file name and extension
  \typeout{(#1)}% latexmk will find this if $recorder=0 (however, in that case, it will ignore #1 if it is a .aux or .pdf file etc and it exists! if it doesn't exist, it will appear in the list of dependents regardless)

  \IfFileExists{#1}{}{\typeout{No file #1.}}% latexmk will find this message if #1 doesn't exist (yet)
}

\usepackage{comment}
\usepackage{adjustbox}
\usepackage{tikz}
\usetikzlibrary{svg.path}
\usetikzlibrary{calc}
\usepackage{color}
\usepackage{soul,xcolor}
\usepackage{makecell}
\usepackage{enumitem}
\usepackage{mathtools}
\usepackage{newtxmath}
\usepackage{comment}
\usepackage{multirow}
\usepackage{url}
\usepackage{wrapfig}
\usepackage{diagbox}
\usepackage{floatrow}
\usepackage{subcaption}
\usepackage{authblk}

\newif\iffullversion
\newif\ifsubmissionversion
\fullversionfalse
\submissionversiontrue

\usepackage{fancyhdr} 

\begin{document}

\title{UniHand: Privacy-preserving Universal Handover for Small-Cell Networks in 5G-enabled Mobile Communication with KCI Resilience\\
}

\author[1,2]{Rabiah Alnashwan}
\author[1]{Prosanta Gope}
\author[1]{Benjamin Dowling}
\affil[1]{Department of Computer Science, The University of Sheffield, Sheffield, UK}
\affil[ ]{\textit {\{ralnashwan1,p.gope, b.dowling, \}@sheffield.ac.uk}}
\affil[2]{Department of Computer Science, Imam Mohammad Ibn Saud Islamic University,  Riyadh, Saudi Arabia}
\affil[ ]{\textit {ralnashwan@imamu.edu.sa}}

%\and
%\and
%\and


\maketitle
\thispagestyle{fancy}
\renewcommand{\headrulewidth}{0pt}

\begin{abstract}
Introducing Small Cell Networks (SCN) has significantly improved wireless link quality, spectrum efficiency and network capacity, which has been viewed as one of the key technologies in the fifth-generation (5G) mobile network. However, this technology increases the frequency of handover (HO) procedures caused by the dense deployment of cells in the network with reduced cell coverage, bringing new security and privacy issues. The current 5G-AKA and HO protocols are vulnerable to security weaknesses, such as the lack of forward secrecy and identity confusion attacks. The high HO frequency of HOs might magnify these security and privacy concerns in the 5G mobile network. This work addresses these issues by proposing 
a secure privacy-preserving universal HO scheme ($\UniHand$) for SCNs in 5G mobile communication. $\UniHand$ can achieve mutual authentication, strong anonymity, perfect forward secrecy, key-escrow-free and key compromise impersonation (KCI) resilience. To the best of our knowledge, this is the \textit{first} scheme to achieve secure, privacy-preserving universal HO with \textit{KCI} resilience for roaming users in 5G environment. We demonstrate that our proposed scheme is resilient against all the essential security threats by performing a comprehensive formal security analysis and conducting relevant experiments to show the cost-effectiveness of the proposed scheme.

\end{abstract}


\section{Introduction}
The generations of cellular network standards have evolved continuously, and each generation offers new improvements over its successor to cope with the market needs. The newly deployed cellular network (the fifth generation) provides its users with a better realisation of continuous access to networks worldwide. Increased network capacity guarantees faster speed, higher broadband, and lower latency. One promising way to achieve these network improvements is through increasing network density, i.e. increasing the number of base stations (cells) in the network, known as a Small-Cell Network (SCN). However, SCNs reduce the serving spectrum of each cell, causing the handover process to be triggered more frequently than in the previous mobile cellular generations (like GSM and 3G), aggravating handover privacy and security issues in 5G. For example, \cite{cremers_component-based_2019, braeken_symmetric_2020, basin_formal_2018, borgaonkar_new_2018} have analysed the 5G-AKA and HO protocols and identified security, privacy and efficiency weaknesses, such as identity confusion attacks, confidentiality attacks on sequence numbers (breaking untraceability) and confusion attacks. It follows that current solutions for 5G-AKA and HO security are insufficient. To overcome most of these weaknesses in 5G-AKA and HO, the following security and privacy requirements must be adequately addressed: mutual authentication (MA), Strong anonymity (SA) (user anonymity and unlinkability) and perfect forward secrecy (PFS). As important as these security features, key-escrow freeness (KEF) and Key Compromise Impersonation (KCI) resilience should also be supported in 5G networks. Both security features are essential to improve the security of the 5G network and avoid catastrophic security degradation in case of compromising a single private key in the network. KEF ensures that secret keys   are jointly computed between users and communicating partners and are not fully controlled by any third party. KCI vulnerabilities allow impersonation attacks that trigger if the private key of a network participant/entity (such as $\UE$, $\gNB$ or $\AuC$) is revealed, which enables the adversary to impersonate others to the compromised participant \cite{chalkias_two_2009}. The consequence of this attack will be catastrophic, especially if the adversary compromises a base station/$\gNB$ in a 5G network, which enables the adversary to impersonate other users to the compromised $\gNB$, or perform a MITM attack with all connected users in that cell. This consequence may be aggravated by the significant increase of small cells in the 5G environment, as it linearly increases the probability of compromising $\gNB$. Therefore, providing resilience against KCI attacks in 5G networks is essential. This attack represents a subtle, often underestimated threat and is difficult to counter. Unfortunately, existing and conventional 5G AKA and HO protocols cannot deal with this type of attack. 
%On the other hand, many real-world protocols, such as TLS \cite{rescorla_rfc_2018} and signal\cite{cohn-gordon_formal_2020}, have identified the impact of KCI attacks. Hence, they provided resilience against KCI attacks to overcome the severe security impact that KCI may cause. 
%Compared to the lack PFS in a protocol, KCI can potentially have more severe consequences. Besides having information from past sessions, as in a PFS attack, in KCI, an attacker could get more information from sessions that may never have been generated before by masquerading as a different legitimate party to the compromised party. Therefore, It is undoubtedly essential to provide resilience against KCI attacks in 5G networks. 
\ra{Therefore, in this article, we introduce $\UniHand$ scheme, an Authentication and Key Agreement (AKA) protocol and handover protocol that achieves all the desirable security requirements (such as PFS, SA, MA); it also supports KEF and provides resistance against KCI attacks.
To realise these feature $\UniHand$ utilises sanitisable signatures \cite{ateniese_sanitizable_2005} and universal accumulators \cite{li_universal_2007}. Sanitisable signatures provide $\UniHand$ with a modifiable signature, where two parties (i.e. signer and sanitiser) can generate a valid signature of a certificate. Hence,  the certificate initiator issues certificates with their signatures for network users. Users can modify a \textit{part} of their certificate and generate a valid signature of the modified certificate. By utilising $\SanSig$, $\UniHand$ preserves the following signature properties: immutability, privacy, accountability, transparency and unforgeability\cite{ateniese_sanitizable_2005, brzuska_security_2009}. On the other hand, allowing device revocation within $\UniHand$ is required to manage the legitimate users in the network. $\UniHand$ uses dynamic accumulators to provide efficient membership revocation while maintaining user privacy. However, the accumulator's efficiency depends on the frequency of updating the accumulator, which is itself correlated to the number of joining users. Usually, in a typical setting of mobile networks, the number of joining users exceeds the number of revoked users. Therefore, non-membership witnesses are significantly more efficient, as updating the revoked membership occurs less frequently than updating the membership accumulator. Thus, we utilise an accumulator that supports \textit{dynamic} and \textit{non-membership} witnesses, proposed by \cite{li_universal_2007}, to achieve privacy and increase the efficiency of $\UniHand$. }

%Accumulator has been used in $\UniHand$ to effectively manage user revocation in the network, particularly negative accumulators where valid users are equipped with non-membership to the revocation list (RL). The efficiency of accumulators relies on the frequency of updating the accumulator every time a user is added. Since the number of joining users is relatively higher than the number of revoked, it is more efficient to use the accumulator for the revoked users and generate non-membership to valid users. 

To the best of our knowledge, this is the \emph{first work} that provides KCI resilience in the authenticated key agreement and handover protocols in 5G settings.

\subsection{Related Works}
\label{related works}
Substantial research has been done on 5G authentication and handover protocols to analyse and identify security and privacy weaknesses in the conventional 5G protocols, i.e. 5G-AKA and 5G-HO. Peltonen, Sasse and Basin \cite{peltonen_comprehensive_2021} provide a comprehensive formal security analysis for the conventional 5G handover scheme. Basin et al. \cite{basin_formal_2018}, and Cremers and Dehnel-Wild \cite{cremers_component-based_2019} also provide a comprehensive formal security analysis for the standard 5G AKA protocol. These studies highlighted under-specified security requirements and identified security and privacy weaknesses in the current version of 5G-AKA and HO protocols \cite{3rd_generation_partnership_project_3gpp_security_2020}, such as traceability attacks from active adversaries, identity confusion attacks, lack of perfect forward secrecy and confidentiality attacks on sequence numbers.
Therefore, several related authentication schemes such as ReHand \cite{fan_rehand_2020}, RUSH \cite{zhang_robust_2021}, LSHA \cite{yan_lightweight_2021} CPPHA \cite{cao_cppha_2021} and Braeken \cite{braeken_symmetric_2020} have been proposed in the literature to improve the security and privacy of AKA and HO protocols. These schemes can be divided into two categories based on the underlying cryptography: symmetric-based and asymmetric-based schemes. Braeken \cite{braeken_symmetric_2020} proposed an efficient symmetric-based AKA protocol that improves the 5G-AKA and overcomes identity replay attacks discovered in 5G-AKA. Nevertheless, the proposed protocol suffers from an unlinkability attack, which only supports in-session unlinkability in case of successful authentication. However, if authentication fails, user identities can be linked due to the reused Globally Unique Temporary User Equipment Identity (GUTI) \cite{3rd_generation_partnership_project_3gpp_security_2020}. Additionally, Fan et al. \cite{fan_rehand_2020}, proposed ReHand, a symmetric-based protocol that supports a secure region-based handover scheme and fast revocation management for SCNs. This protocol provides a seamless handover for roaming inside a region only, where users roaming to a different region cannot perform the fast HO; instead, they must execute another initial authentication.
Yan and Ma \cite{yan_lightweight_2021} propose LSHA, a symmetric-based handover and authentication protocol exploiting neighbouring base stations in the 5G network. This protocol depends on two keys to secure HOs: a $\gNB$ secret key and a session key between $\gNB$s. However, LSHA relies on 5G-AKA, which is susceptible to several attacks, undermining the security of LSHA. \\
Some proposed schemes rely on asymmetric cryptography, such as RUSH \cite{zhang_robust_2021} and CPPHA \cite{cao_cppha_2021}. Zhang et al. \cite{zhang_robust_2021} propose RUSH, a universal handover authentication protocol that preserves user anonymity, achieving perfect forward secrecy and key-escrow freeness. The RUSH protocol provides universal handover in 5G HetNets through blockchains and leverages chameleon hashes (CH) to achieve user anonymity. However, their analysis omits the use of blockchain, and thus security and performance issues due to blockchain use may have been overlooked. Additionally, RUSH does not address CH's linkability issue. Cao et al.  \cite{cao_cppha_2021} propose CPPHA, an asymmetric-based scheme which utilises software-defined networks (SDN) in 5G. CPPHA relies on an authentication handover module (AHM) that resides in the SDN controller, which is responsible for monitoring and predicting users' future paths. The AHM pre-distributes (before the actual HO) user-related information to the predicted $\gNB$s, including the user's ID and first session key. Thus, it only supports weak anonymity against eavesdroppers, where all prospective $\gNB$s maintain user IDs. Finally, CPPHA is susceptible to sequence number desynchronisation attacks.

\subsection{Motivation and Contributions}
Authenticated key agreement is the initial and most crucial step for validating joining users before providing any services to them. Similarly, the handover procedure re-authenticates roaming users and ensures the continuity of network services. So guaranteeing a high level of security and privacy of AKA and HO is vital for a 5G network. Nevertheless, the existing 5G-AKA and HO protocols are susceptible to security and privacy issues, yet several existing works have attempted to address some of these issues. However, to the best of our knowledge, the existing works fail to address \textit{all} security and privacy issues in the 5G AKA and HO protocols, such as traceability attacks, identity confusion attacks, lack of PFS and confidentiality attacks on sequence numbers.
Most previous solutions use the existing 5G-AKA to build their HO protocol, inheriting all 5G-AKA weaknesses. In addition, no existing protocols have provably achieved KCI resilience and KEF. This creates a critical security vulnerability in the 5G network, especially when considering an active attacker who can impersonate an honest party to the compromised party. Both properties are essential to achieve higher security and fairness in key generation and reduce the probability of protocol failure due to a single key compromise.
Therefore, we propose $\UniHand$, a Universal Handover scheme achieving seamless user mobility and all required security and privacy properties (explained in Section \ref{Design Goals}), including KCI resilience and KEF. $\UniHand$ is the first to achieve secure, privacy-preserving universal HO with KCI resilience for roaming users in 5G without relying on additional infrastructural support (such as blockchain). Our contributions are as follows:

\begin{itemize}
    \item The \emph{first} standalone solution to achieve KCI resilience for roaming users in 5G SCNs, providing privacy-preserving and secure authentication and universal HO protocols;
    \item An effective user membership revocation scheme for efficiently managing a large number of users in 5G using dynamic universal accumulators \cite{li_universal_2007};
    \item A rigorous formal security analysis of our proposed scheme;
    \item A comparative analysis of $\UniHand$ with previously existing literature, demonstrating $\UniHand$ achieves all required security properties.
\end{itemize}

The rest of this paper is organised as follows: Section \ref{sec: Preliminaries} introduces sanitisable signatures and accumulators. Section \ref{sec: sys and adv} introduces the system architecture, adversary model and targeted security properties. Next, we present the secure Universal Handover scheme $\UniHand$, with a detailed description of all phases, including registration, initial authentication and universal  HO. Section \ref{Formal Security Analysis} conducts a formal security analysis of $\UniHand$. Section \ref{Discussion} compares the achieved security properties of $\UniHand$ with previous literature and evaluates $\UniHand$'s computational and communication cost, concluding with Section \ref{sec:conclusion}.

\section{Preliminaries} 
\label{sec: Preliminaries}
Here we introduce sanitisable signatures and accumulators, two cryptographic primitives underlying $\UniHand$. \rabiah{While both primitives (sanitisable signatures and accumulators) have more complex functionality and properties compared to conventional cryptographic primitives, they can be constructed using conventional cryptographic primitives.}

\subsection{Sanitisable signatures} 
Sanitisable signature schemes ($\SanSig$), introduced by Ateniese et al.\ra{\cite{ateniese_sanitizable_2005}} provide similar features to standard digital signatures while allowing the signing authority to delegate signing to another party, the so-called sanitiser. The sanitiser can modify specific blocks of the signed message and generate another valid signature over the modified (sanitised) message. The sanitiser can sanitise a signature without the original signer's participation, providing signature flexibility between the signer and sanitiser. \ra{In particular, let a message $m$ consisting of $n$ blocks $m=(m_1,.....,m_n)$, where $n \in \mathbb{Z}$ and $m_i \in {0,1}^*$. The signer allows the sanitiser to modify specific $j$ blocks of the message, i.e., $(m_j,...m_{j+i})$ that the signer defined as admissible blocks of the message. These admissible blocks are the only blocks that a sanitiser can modify in the message. In order to achieve that, the signer has to specify sections of a signed message that the sanitiser can modify using deterministic functions $\modd$ and $\adm$. The signer uses $\adm$ to specify the modifiable blocks of a message, and the sanitiser uses $\modd$ to specify the modified blocks of the messages, i.e., $m^{*}=\modd (m)$ and $\adm (m^{*},m)\to \{0,1\}$. We require $\SanSig$ to provide existential unforgeability under chosen message attack ($\EUFCMA$) security.
\rabiah{$\SanSig$ can be constructed using chameleon hashes associated with a standard digital signature as in \cite{ateniese_sanitizable_2005}, a digital signature and a group signature \cite{brzuska_unlinkability_2010}, or even from two standard digital signatures \cite{brzuska_efficient_2014, brzuska_santizable_2009, brzuska_non-interactive_2013}.} In general, $\SanSig$ consists of a tuple of algorithms $\SanSig=\{\KeyGen,\Sign, \Sanit,\Verify,\proof, \Judge\}$. We omit $\proof$ and $\Judge$ since they are not required in $\UniHand$.}
\begin{itemize}
    \item \ra{$\boldsymbol{\KeyGen(1^{n}) \to (\pk,\sk)}$ is a pair of key generation algorithms for the signer and the sanitiser respectively: $(\pk_{sig},\sk_{sig}) \getsr \KeyGen_{sig}(1^{n})$, 
    $(\pk_{san},\sk_{san}) \getsr \KeyGen_{san}(1^{n})$.}
    
    \item \ra{$\boldsymbol{\Sign(m,\sk_{sig},\pk_{san}, \adm) \tor (\sigma)}$:  $\Sign$ algorithm takes as input four parameters: a message $m \in \bits{*}$, a signer private key $\sk_{sig}$, sanitiser public key $\pk_{san}$ and the admissible modifiable message blocks ($\adm$).}
    
    \item \ra{$\boldsymbol{\Sanit(m,\modd, \sigma,  \pk_{sig}, \sk_{san} ) \tor (m^{*},\sigma^{*}) }$: $\Sanit$ algorithm takes as input five parameters: original message $m$, a modification of the original message $\modd$, a signature $\sigma$, signer public key $\pk_{sig}$ and sanitiser private key $\sk_{san}$. }
    
    \item \ra{$\boldsymbol{\Verify (m, \sigma, \pk_{sig}, \pk_{san} ) \tor b}$: $\Verify$ algorithm takes as input five parameters: a message $m$, a signature $\sigma$ and the public keys of the signer $\pk_{sig}$ and sanitiser $\pk_{san}$. Next it outputs a bit $b \in \bits{}$, where $b=1$ if $\sigma$ verifies message $m$ under $\pk_{san}$ and $\pk_{sig}$, and $b=0$ otherwise.}
 
\end{itemize}

\subsection{Accumulators}
\ra{Cryptographic accumulators was first introduced by Benaloh and de Mare \cite{benaloh_one-way_1994} in 1994. The proposed one-way accumulator was constructed using a hash function with quasi-commutativeness and one-way property. The generic concept of accumulators is to accumulate a number of elements from a finite set $X= {x_1,.... x_n}$ into one accumulated value $acc$ of constant size. Since the accumulator depends on a quasi-commutativeness property,  the order of the accumulated elements is not important. Cryptographic accumulators can be categorised based on the construction (symmetric and asymmetric), characteristics (i.e. static and dynamic) and security assumptions (i.e. RSA, Strong Diffie-Hellman and Bilinear Diffie-Hellman). Hence, accumulators can be constructed distinctly depending on the application and the required security features. \rabiah{In $\UniHand$, a universal accumulator that supports negative and positive membership witnesses can be built from RSA and relying on the strong RSA assumption \cite{baldimtsi_accumulators_2017, li_universal_2007}.} In general, accumulator $acc$ consists of a tuple of algorithms $acc=\{\KGenacc,\Genacc, \Updateacc,\Nwitcreate,\Verifyacc$,  $\Nwitupdate\}$.}
\begin{itemize}
    \item \ra{$\boldsymbol{\KGenacc(1^{n}) \tor (\skacc) }$:
    $\KGenacc$ generates a secret key $\skacc$. }

    \item \ra{$\boldsymbol{\Genacc(\skacc,X) \tor (\ddot{C})}$:
    $\Genacc$ algorithm takes as input an accumulator secret key $\skacc$, and all values to be accumulated $X= \{x_i,...x_n\}$ (where $X \gets \phi$ when initialised), returning an accumulator $\ddot{C}$.}
     
    \item  \ra{$\boldsymbol{\Updateacc(\skacc,\ddot{C},x^{*} ) \tor (\ddot{C}^{*})}$: $\Updateacc$ takes as input an accumulator secret key $\skacc$, an accumulator $\ddot{C}$ and all new values to be accumulated $x^{*}$, returning the updated accumulator $\ddot{C}^{*}$.}   
    
    \item \ra{$\boldsymbol{\Nwitcreate(\skacc,\ddot{C},X, x^{*})\tor (\omega_{x})} $: $\Nwitcreate$ takes as input an accumulator secret key $\skacc$, an accumulator $\ddot{C}$, previously accumulated values $X$ and value $x^{*}$ (where $x^{*} \notin X$) returning a non-membership witness $\omega_{x}$ for $x^{*}$.}

    \item \ra{$\boldsymbol{\Verifyacc(\ddot{C},\omega_{x},x ) \tor b}$: $\Verifyacc$ takes as input an accumulator $\ddot{C}$, a non-membership witness $\omega_{x}$ and queried value $x$, returning a bit $b \in \bits{}$, where $b=1$ if  $x \notin X$, and $b=0$ otherwise.}

    \item \ra{$\boldsymbol{\Nwitupdate(\ddot{C},\ddot{C}^{*},x^{*},x, \omega_{x}) \tor (c^{*}_{x})}$: $\Nwitupdate$ takes as input an accumulator $\ddot{C}$, an updated accumulator $\ddot{C}^{*}$, a (new) accumulated value $x^{*}$, an non-accumulated value $x$ and the non-membership witness $\omega_{x}$, returning a new non-membership witness $\omega^{*}_{x}$. }
\end{itemize}

\section{System Architecture, Design Goals and Adversary Model}
\label{sec: sys and adv}
In this section, we first describe the $\UniHand$ system architecture, then describe the adversary model and design security and privacy goals.

\subsection{System Architecture}
\begin{figure}
    \includegraphics [scale=0.33]{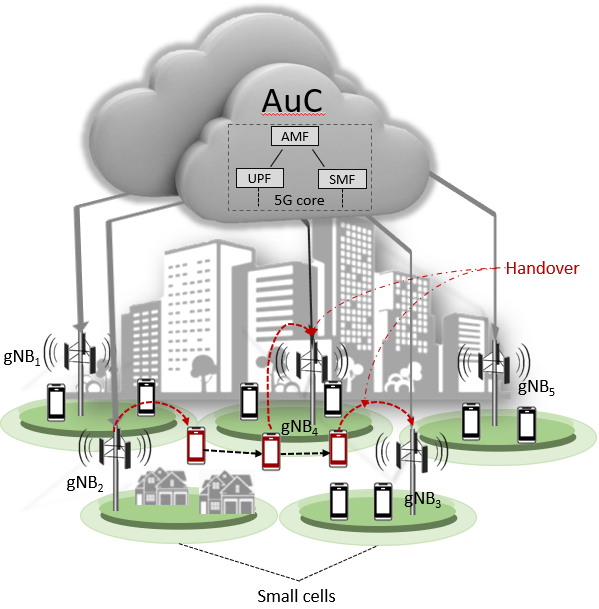}
    \caption{System Architecture}
    \label{fig:Sys Architecture}

\end{figure}

This section introduces the System Architecture of our proposed $\UniHand$ scheme. There are three major components of the system architecture: Authentication Center ($\AuC$), 5G radio base station (Next Generation NodeB $\gNB$) and User Equipment  ($\UE$), as shown in Figure \ref{fig:Sys Architecture}. $\AuC$ encapsulates the 5G core network entities, i.e. access and mobility management function, user plane function, session management function, and the authentication server function. $\AuC$ is responsible for configuring all parties in the network, including the $\gNB$ and $\UE$ and generating certificates for authentication purposes. Meanwhile, $\gNB$ connects $\UE$ to the $\AuC$. $\UniHand$ consists of two main protocols: An initial authentication and a universal handover. During the initial authentication, the $\AuC$ authenticates $\UE$s, generates certificates for $\UE$s for future handovers, and generates session keys for future communication. In the universal handover, the $\gNB$ authenticates the $\UE$s by validating their certificates and generating session keys. 

\subsection{Adversary Model}
Our security model is intended to capture security notions recommended by the 3GPP group \cite{3rd_generation_partnership_project_3gpp_security_2020} and \cite{peltonen_comprehensive_2021, basin_formal_2018, cremers_component-based_2019} for 5G authentication and handover protocols, described in Section \ref{Design Goals}. During the execution of $\UniHand$, all communication channels are public, i.e. the adversary can control the public channels fully. Our analysis combines three types of adversaries: Type 1 adversary $\adv_1$ controls the network and can intercept, insert, modify and delete any message. Type 2 adversary $\adv_2$ tries to break the linkability and key-indistinguishability proprieties of communicating parties. Type 3 adversary $\adv_3$ captures KCI attacks and is capable of compromising at most one of the following secret keys: 
\begin{enumerate}
    \item Long-term keys (LTK) shared between $\UE$ and $\AuC$.
    \item Asymmetric secret keys (ASK), signing and sanitising keys of the protocol participants, i.e. $\UE$, $\gNB$ and $\AuC$
    \item Session keys established between protocol participants.
\end{enumerate} 

An adversary may also try to launch several other attacks on $\UniHand$, including impersonation, replay, man-in-the-middle attacks, etc. However, the proposed $\UniHand$ scheme will be able to resist these attacks through mutual authentication, user anonymity, unlinkability, PFS, and KCI resilience. 

\subsection{Design Goals}
\label{Design Goals}
The newly deployed 5G  mobile network supports increasingly dense connections while providing optimised network efficiency. However, several requirements must be fulfilled to ensure security and privacy in 5G mobile communication. In this regard,  \cite{basin_formal_2018} and \cite{peltonen_comprehensive_2021} have reviewed the 5G-AKA and handover protocols and identified the following security: privacy requirements for authentication and handover protocols that need to be adequately addressed. Our security analysis (in Section \ref{Formal Security Analysis}) and the discussion (in Section \ref{Discussion}) ensure that $\UniHand$ can successfully achieve all security goals. 

\begin{enumerate}
\item \emph{\textbf{Mutual authentication:}}
Ensuring the authenticity of every party in the network is essential to reduce the risk of many security threats, such as MITM and impersonation attacks. More details about the MA experiment can be found in [\ref{ma-sec}].

\item \emph{\textbf{Strong anonymity:}}
Maintaining users' privacy is one of the main goals to be achieved in the 5G network. Most authentication and handover protocols address privacy by anonymising user identities (using pseudo-identities). However, providing user anonymity alone is insufficient, as the adversary may break privacy by linking users' protocol executions. In this regard, we introduce strong anonymity properties, capturing user anonymity and unlinkability. Additional details about this security feature can be found in \ref{UNlink-sec}. 

\item \emph{\textbf{Perfect forward secrecy:}}
PFS is essential in 5G to ensure the security of the previously-computed session keys after long-term secrets are compromised. This security feature is especially essential in 5G networks due to the high frequency of handovers caused by the increased deployment of cells, which generates many session keys with each handover. Thus, 5G protocols should ensure that any compromise does not affect the security of past session keys.

\item \emph{\textbf{Key-escrow free:}} 
There has been a lot of debate among researchers about the key escrow problem of using a trustworthy escrow to maintain a copy of all users’ secret keys. Even though providing escrow is essential in some domains, it brings issues regarding the trustworthiness of a specific third party controlling the escrow. In today’s information security, it isn't easy to fully trust one entity, as it might take advantage of this trust and reveal some of the encrypted information in the network. Another case is if this entity is compromised, it will cause a single point of failure to the entire network. Moreover, the idea of key escrow contradicts the end-to-end encryption and the strong security requirement of the 5G network. Therefore it is not recommended to give the authority to a single party to generate all secret keys or maintain any users’ keys in the network. Nevertheless, providing a key escrow-free property provide higher security, and it is essential to protect users in case of a compromised third party.
\item \emph{\textbf{Key compromise impersonation resilience:}} Although providing a PFS protects previous sessions' information, it doesn't prevent an attacker from leaking new information by masquerading as a legitimate party to the compromised one, forming a KCI attack. Therefore providing a KCI resilience scheme is essential in 5G, primarily to ensure the security and authenticity of communicating partners after a long-term secret is compromised. 

\item\emph{\textbf{Effective revocation management:}}
It is essential to provide effective revocation management to accommodate the extended capacity of connections in the 5G network. Therefore, our proposed protocols use a revocation list (RL) to manage the revoked users from the network. In general, the number of joined users supersedes the revoked ones. For this reason, our RL utilises a dynamic accumulator that supports non-membership witnesses for better efficiency.
\end{enumerate}

\section{The Proposed Uni-Hand Scheme}
\label{Proposed scheme}

In this section, we introduce $\UniHand$, which consists of a \emph{System Initialisation phase}, an \emph{initial authentication} protocol and \emph{universal HO} protocol. The first phase in  $\UniHand$ is \textit{System Initialisation}, responsible for registering $\gNB$s and new users into the network, generating the initialisation parameters for $\UE$ and $\gNB$. During this phase, the $\AuC$ generates certificates for all $\gNB$s in the network. Additionally, AuC generates a long-term secret key and shares it with the user along with pseudo-identities ($\pid$ and $\tid$) to preserve users' anonymity. Next, $\AuC$ creates an accumulator to manage user revocation. 
In the \textit{Initial Authentication} protocol, the $\AuC$ generates certificates for joining users, which are created using $\SanSig$ algorithm using their $\pid$ and users' sanitising public keys. Finally, the \textit{Universal HO} protocol provides roaming users with network access while maintaining network and users' security and privacy via assuring security features mentioned in Section \ref{Design Goals}.

\subsection{System Initialisation}
In this phase, we assume that all communication channels are secure. During this phase, all required parameters for $\UniHand$ are produced, such as the registration of $\UE$s and $\gNB$s in the network and initiating the accumulator. Hence this phase can be divided into three main parts:

\begin{enumerate}

\item \textbf{$\boldsymbol{\gNB}$ Registration}: Every $\gNB$ in the network has to do registration and authentication with the $\AuC$. For that, first, $\gNB$ has to generate asymmetric sanitising key pair $(\pk_{san}^{\gNB},\sk_{san}^{\gNB}) \getsr \SanSig.\KeyGen_{san}(1^{n})$. Next, the $\gNB$ will send a registration message to $\AuC$, which includes his/her sanitising public key $(\pk_{san}^{\gNB})$. Upon receiving the registration message, the $\AuC$ will authenticate the $\gNB$ and generate a certificate ($\cert_{G}, \sigma_{G}$) via $\SanSig$ algorithm, $\sigma_{G} \getsr \SanSig.\Sign(\cert_{G}, \sk_{sig}^{\AuC}, \pk_{san}^{\gNB}, \adm(\cert_{mod}^{G}) )$. To expedite this process, AuC may execute this step offline.

\item \textbf{$\boldsymbol {\UE}$ Registration}:
Similarly, every user in the network needs to register into the network to utilise the provided network services. A $\UE$ begins this phase by generating asymmetric sanitising key pair $(\pk_{san}^{\UE}, \sk_{san}^{\UE}) \getsr \SanSig.\KeyGen_{san}(1^{n})$. Next, the $\UE$ sends a registration request to the $\AuC$, which includes all $\UE$ credentials, including their sanitising public key $\pk_{san}^{\UE}$ via a secure channel. Upon receiving the registration message, the $\AuC$ generates two independent pseudo identities $\pid $ and $ \tid$ (i.e. there is no direct relationship between the aliases), in addition to a symmetric long-term key \ra{($k_i$)} for the intended user and shares them with the $\UE$, where $\pid_{i},\tid_{i} \getsr \bits{n}$.

\item\textbf{Accumulator Initialisation}:
To initiate the revocation list (RL) using an accumulator ($\ddot{C}$), first the $\AuC$ has to generate a secret key for the accumulator via $ \KGenacc(1^{n}) \tor (\skacc)$, then generate the accumulator using $ \Genacc(\skacc,X) \tor \RL$, where $X$ is initially empty.

\end{enumerate}

\iffullversion
Finally, $\UE, \gNB \& \AuC$ need to record the essential parameters and secret keys as shown in Table \ref{tab:Key Management}

\begin{table}[h!]
\centering
\caption{Key Management}
\label{tab:Key Management}
\begin{tabular}{l|l|l}
\hline
\textbf{UE} & \textbf{gNB} & \textbf{AuC} \\ \hline
 $k_i $  &  $\sk_{san}^{\gNB}$   & $k_i, \pid, \tid$     \\ \hline
  $\sk_{san}^{UE_i}$ & $\GID$  & $ \sk_{sig}^{\AuC}$    \\ \hline
 $\pid, \tid$ &    & $\skacc, \RL$   \\ \hline
\end{tabular}
\end{table}
\fi
\subsection{Initial authentication}
Each registered user needs to execute this protocol to join the network securely. During this protocol, each communicating party mutually authenticates their partner, and then the $\AuC$ generates a new certificate for the intended $\UE$, as illustrated in Figure \ref{fig:init-auth-alt}:

\begin{figure}[htp!]
	\centering
	\begin{adjustbox}{max width=0.8\textwidth, max height=1\textheight}
		\fbox{
	\begin{tikzpicture}[yscale=-0.33,>=latex]
    \tikzstyle{every node}=[font=\large]
	\edef\InitX{0}
	\edef\ArrowLeft{1}
	\edef\ArrowCenter{6}
	\edef\ArrowRight{11}
	\edef\RespX{12}
	% Set the starting Y coordinate
	\edef\Y{0}
	
	% Draw header boxes
%	\node [rectangle,draw,inner sep=5pt,right] at (2,-5) {\textbf{SIDH + MAC construction}};
	\node [rectangle,draw,inner sep=5pt,right] at (\InitX,\Y) {\textbf{UE}};
	\node [rectangle,draw,inner sep=5pt,left] at ((7,\Y) {\textbf{gNB}};
	\node [rectangle,draw,inner sep=5pt,left] at (\RespX,\Y) {\textbf{AuC}};
	\NextLine[2]
	
	\AdversaryAction{$h \getsr \mathbb{Z}_q$}
	\NextLine[1.2]
	\AdversaryAction{$(\cert_{G}^{*},\sigma^{*}_{G}) \gets \SanSig.\Sanit (\cert_{G},\sigma_{G},\mathit{MOD}(g^h), \pk_{sig}^{\AuC},\sk_{san}^{\gNB} )$}
	\NextLine[3]
	
	\AdversaryToClient{\framebox[1.1\width]{$\boldsymbol{M_1}:[\cert_{G}^{*}$, $\sigma^{*}_{G}$, $g^h]$}}{}
	\NextLine[1.2]
	\ClientAction{\textbf{abort if} $1 \neq \SanSig.\Verify(\cert_{G}^{*}, \sigma^{*}_{G}, \pk_{sig}^{\AuC}, \pk_{san}^{\gNB} )$}
	\NextLine[1.2]
	\ClientAction{$r_{id} \getsr \bits{n}$, $u \getsr \mathbb{Z}_q$, $sk_{i},k_s=\KDF((g^h)^{u})$}
	\NextLine[1.2]
	\ClientAction{$\boldsymbol{M_{A_{0}}} \gets \AuthEnc.\Enc\{k_{i},\pid\|r_{id}\|\tid\}$}
	\NextLine[1.2]
	\ClientAction{$\sigma \gets\Sign(\sk_{san}^\UE,\tid\|M_{A_{0}}\|g^{ u})$}
	\NextLine[3]
	
	\ClientToAdversary{\framebox[1.1\width]{$\boldsymbol{M_2}:[\AuthEnc.\Enc\{k_s,M_{A_{0}}\|\tid\},\sigma,g^{ u}]$}}{}
	\NextLine[1.2]
	\AdversaryAction{$\ra{sk_{i},k_s=\KDF((g^u)^{h})}$}
	\NextLine[1.2]
	\AdversaryAction{$M_{A_{0}}\|\tid \gets \AuthEnc.\Dec(k_s,M_{2})$}
	\NextLine[1.2]
	\AdversaryAction{\textbf{abort if} $1 \neq \Verify(\sigma, M_{A_0} \| \tid \|g^{ u},\pk_{san}^{\UE})$}
	\NextLine[1.2]
	\AdversaryAction{{$a \getsr \mathbb{Z}_q$}}
	\NextLine[1.2]
	\AdversaryAction{$(\cert_{G}^{*},\sigma^{*}_{G}) \gets \SanSig.\Sanit (\ra{\cert_{G}},\sigma_{G}, \mathit{MOD}(g^a), \pk_{sig}^{\AuC}, \sk_{san}^{\gNB} )$}
	\NextLine[3]

	\AdversaryToServer{\framebox[1.1\width]{$\boldsymbol{M_3}:[\cert_{G}^{*}$, $\sigma^{*}_{G}$, $g^a]$}}{}
	\NextLine[1.2]
	\ServerAction{\textbf{abort if} $\SanSig.\Verify(\cert_{G},\sigma_{G},\pk_{sig}^{\AuC},\pk_{san}^{\gNB})\neq1$}
	\NextLine[1.2]
	\ServerAction{$b \getsr \mathbb{Z}_q$}
	\NextLine[1.2]
	\ServerAction{$k_s'=\KDF((g^a)^{b})$}
	\NextLine[1.2]
	\ServerAction{$(\cert_{G}^{*},\sigma^{*}_{G}) \gets \SanSig.\Sign (  \ra{\cert_{G}},\sigma_{G},\mathit{MOD}(g^b), \pk_{san}^{\gNB}, \sk_{sig}^{\AuC} )$}
	\NextLine[3]
	
	\ServerToAdversary{\framebox[1.1\width]{$\boldsymbol{M_4}:[\cert_{G}^{*}$, $\sigma^{*}_{G}$, $g^b]$}}{}
	\NextLine[1.2]
	\AdversaryAction{\textbf{abort if} $\SanSig.\Verify(\cert_{G},\sigma_{G},\pk_{sig}^{\AuC},\pk_{san}^{\gNB})\neq1$}
	\NextLine[1.2]
	\AdversaryAction{$k_s'=\KDF((g^b)^{a})$}
	\NextLine[3]
	
	\AdversaryToServer{\framebox[1.1\width]{$\boldsymbol{M_5}:\AuthEnc.\Enc\{k_s',M_{A_{0}}\|\tid\}$}}{}
	\NextLine[1.2]
	\ServerAction{$\AuthEnc.\Dec\{k_s, M_5\}$}
	\NextLine[1.2]
	\ServerAction{$\pid\|r_{id} \gets \AuthEnc.\Dec(\mathit{KEY}[\tid],M_{A_{0}})$}
	\NextLine[1.2]
	\ServerAction{$\tid^*\gets \tid \oplus r_{id}$,$\UID_i\getsr \mathbb{P}$,$\RUID \getsr \mathcal{R}$}
	\NextLine[1.2]
	\ServerAction{$\omega_{U} \gets \mathsf{NonWitCreate}(\sk_{acc},\mathit{RL},X,\UID_i)$}
	\NextLine[1.4]
	\ServerAction{$\mathit{\cert}_{fix}^{U}
 \gets \UID_i\|T_U$}
	\NextLine[1.4]
	\ServerAction{$\sigma_U \gets \SanSig.\Sign(\mathit{\cert}_U,\sk_{sig}^{\AuC},\pk_{san}^{\UE},\mathit{ADM}(\RUID\|\tid^*\|\omega_{U} \|v)$}
	\NextLine[1.2]
	\ServerAction{$\boldsymbol{M_{A_1}}$:$\AuthEnc.\Enc\{k_{i},\sigma_{U}\|\cert_{U}\|\tid^*\}$}
	\NextLine[3]
 
	\ServerToAdversary{\framebox[1.1\width]{$\boldsymbol{M_{6}}:\AuthEnc.\Enc\{k_s', M_{A_1} \}$}}{}
	\NextLine[1]
	\AdversaryAction{$M_{A_1} \gets \AuthEnc.\Dec\{k_s', M_{6} \}$}
	\NextLine[3]
 
	\AdversaryToClient{\framebox[1.1\width]{$\boldsymbol{M_{7}}:\AuthEnc.\Enc\{k_s,M_{A_1}\}$}}{}
	\NextLine[1.2]
	\ClientAction{$\ra{\sigma_{U}\|\cert_{U}\|\tid^* \gets \AuthEnc.\Dec(k_i,\AuthEnc.\Dec(k_s,M_7))}$}
	\NextLine[1.2]
	\ClientAction{\textbf{abort if} $\SanSig.\Verify(\cert_{U},\sigma_{U},\pk_{sig}^{\AuC},\pk_{san}^{\UE})\neq1$}
	\NextLine[1.2]
	\ClientAction{\textbf{Check$\&$Update}
	$\tid*=\tid \oplus r_{id}$}
	\NextLine[1.2]
	\ClientAction{$\mathit{ACK} \gets \AuthEnc.\Enc(k_i,flag\|\tid)$}
	\NextLine[3]
 
    \ClientToAdversary{\framebox[1.1\width]{$\boldsymbol{\mathit{ACK}'} \gets \AuthEnc.\Enc(k_s,\mathit{ACK}\|\tid)$}}{}
    \NextLine[1.2]
	\AdversaryAction{$\ra{\mathit{ACK}\|\tid}\gets \AuthEnc.\Dec(k_s,\mathit{ACK}')$}
	\NextLine[3]
 
	\AdversaryToServer{\framebox[1.1\width]{$\boldsymbol{ACK}" \gets\AuthEnc.\Enc\{ k_s', \ra{\mathit{ACK}},\tid \}$}}{}
	\NextLine[1.2]
	\ServerAction{$ \ra{\mathit{ACK}}\| \tid \gets \AuthEnc.\Dec(k_s', \mathit{ACK}")$}
	\NextLine[1.2]
	\ServerAction{$flag \| \tid \gets \AuthEnc.\Dec(k_i,\ra{\mathit{ACK}})$}
	\NextLine[1.2]
	\ServerAction{\textbf{Check$\&$Update} $\tid=\tid$}
	\end{tikzpicture}
}
	\end{adjustbox}
	\caption{The Initial Authentication protocol of $\UniHand$ Scheme.}
	\label{fig:init-auth-alt}
\end{figure}
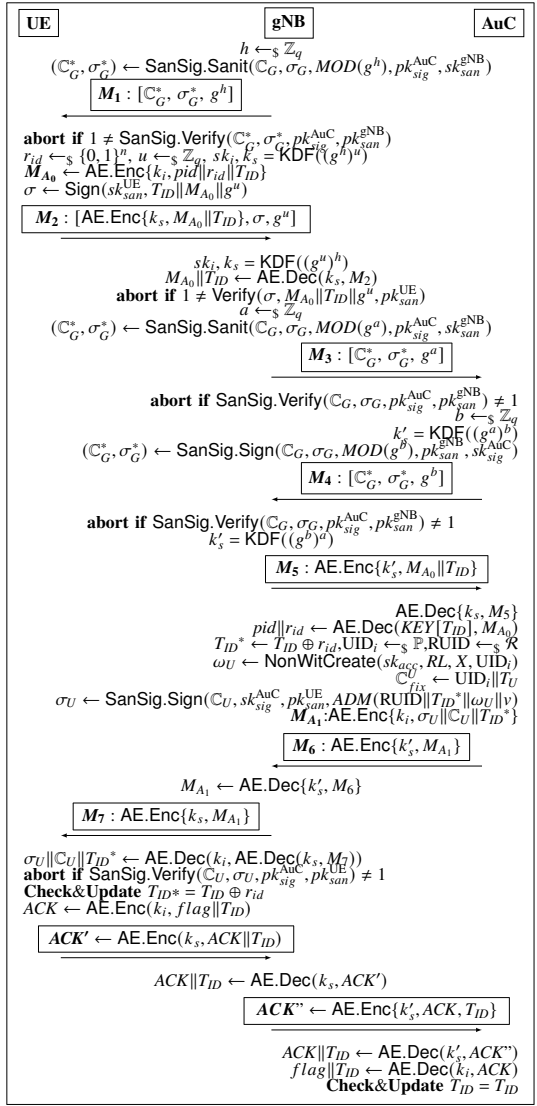

\begin{itemize}[leftmargin=*]
\item[] \textbf{Step 1}:  $\gNB \rightarrow \UE$. $\mathbf{M_{1}}$:[$\ra{\cert_{G}^{*}},\sigma^{*}_{G},g^{h}$].\\
The first step is establishing a secure session key between $\UE$ and $\gNB$ via exchanging signed Diffie–Hellman (DH) values. In this step, the $\gNB$ randomly samples an ephemeral key ($h$) and computes ($g^h$). Then the $\gNB$ utilises the $\SanSig.\Sanit(.)$ algorithm to sanitise their certificate to include the generated DH ephemeral ($\mathit{MOD}(g^h)$) for key integrity and to prevent the known MITM attack on DH. The updated/sanitised certificate with  $(g^h)$ are sent to the $\UE$ via $M_1$.

\item[] \textbf{Step 2}:  $\UE \rightarrow \gNB$. $\mathbf{M_{2}}$:$[\AuthEnc.\Enc\{k_s,M_{A_{0}}\|\tid\},\sigma,g^{ u}]$\\
Upon receiving $M_1$, $\UE$ first verifies the signature using $(\SanSig. \Verify())$ and checks the message integrity (i.e. $g^h \stackrel{?}{=} g^h$). If both verification hold then $\UE$ samples randomly ($r_{id} \& u$) and computes session keys ($sk_i,k_s$). Next $\UE$ encrypts $r_{id}, \pid$ and $\tid$ using $k_i$, where $\pid \& \tid$ is user pseudo identities and $k_i$ is the symmetric long-term key shared between the $\UE$ and $\AuC$ to construct message $M_{A_0}$(to preserve message confidentiality/privacy in case of honest but curious $\gNB$).Then the $\UE$ signs $M_{A_0}$, along with $\tid$ and $g^u$ using user's secret sanitising key ($\sk_{san}^{UE}$). Finally, the $\UE$ composes a message $M_2$ which consists of the signature, ephemeral DH ($g^u$) along with the encryption of $M_{A_0}\|\tid$ (using the session key ($k_s$)), then sends it to $\gNB$. 

\item[] \textbf{Step 3}: $\gNB \rightarrow \AuC$. $\mathbf{M_3}:[\cert_{G}^{*}$, $\sigma^{*}_{G}$, $g^a]$\\
After receiving the message $M_2$, $\gNB$ first computes session keys ($sk_i,k_s$) and decrypt $M_2$. Subsequently, the $\UE$ verifies the signature $\sigma$ and checks the message integrity (i.e. $g^u \stackrel{?}{=} g^u$). If both verification hold, then $\gNB$ randomly samples a temporary key ($a$) and compute ($g^a$). Then the $\gNB$ utilises the $\SanSig.\Sanit(.)$ algorithm to sanitise his/her certificate to include the generated DH ephemeral ($\mathit{MOD}(g^a)$) for key integrity and to prevent the known MITM attack on DH. The updated/sanitised certificate with $(g^a)$ are sent to the $\AuC$ via $M_3$.

\item[] \textbf{Step 4}: $\AuC \rightarrow \gNB$. $\mathbf{M_4}:[\cert_{G}^{*}$, $\sigma^{*}_{G}$, $g^b]$\\
Upon receiving $M_3$, $\AuC$ first verifies the signature using $(\SanSig. \Verify)$ and checks the message integrity (i.e. $g^a \stackrel{?}{=} g^a$). If both verification hold, then $\AuC$ samples randomly ($b$) and computes the session key ($k_s'$). Next the $\AuC$ re-sign $\cert_G^*$ using $\SanSig.\Sign$, which includes the generated DH ephemeral ($\mathit{MOD}(g^b)$). Then the $\AuC$ constructs a message $M_4$ and sends it to $\gNB$.

\item[] \textbf{Step 5}: $\gNB \rightarrow \AuC$. $\mathbf{M_5}:\AuthEnc.\Enc\{k_s',M_{A_{0}}\|\tid\}$\\
After receiving the message $M_4$, $\gNB$ first verifies the certificate $\cert_G^*$ and checks the message integrity (i.e. $g^b \stackrel{?}{=} g^b$). If both verification hold then $\gNB$ computes session key ($k_s'$) to encrypt $\UE s$ information ($M_{A_0}\|\tid$) and send the ciphertext to the $\AuC$.

\item[] \textbf{Step 6}: $\AuC \rightarrow \gNB$. $\mathbf{M_{6}}:\AuthEnc.\Enc\{k_s', M_{A_1} \}$\\
Upon receiving $M_5$, the $\AuC$ decrypts message $M_{5}$ using the session key ($k_s'$) and recovers the long term key $k_i$ via $\tid$ to decrypts $M_{A_0}$, and obtain $(\pid, r_{id}) $. 
The AuC then computes a new temporary user identifier $\tid^*$, and generates a universal user ID $(\UID_i)$, which will be the user's identifier during HOs and in $\RL$. Next, AuC generates a non-membership witness $(\omega_{U}\gets \Nwitcreate(.)$, and specifies the version number ($v$) of $\RL$. AuC creates and signs the $(\cert_{U})$ by generating ``fixed'' part ($\cert_{fix}^{U}$) and ``modifiable'' part $\cert_{mod}^{U}$, where $\cert_{fix}^{U} = \UID_i \| T_{U}$ ($T_{U}$ is a user subscription validity period) and $\cert_{mod}^{U} = \RUID \|\tid^* \|\omega_{U} \|v$ ($\RUID$ a random-user ID). Then AuC signs the entire certificate using $\SanSig.\Sign(.)$ algorithm. AuC then stores $\UID_i$, $\tid_i$ and $\tid^*_i$ (to prevent de-synchronisation) and encrypts user certificate $\cert_U$ and $\sigma_U$ using ($k_i$), to compose message $M_{A_1}$. Finally, the $\AuC$ encrypts $M_{6}$ using the session key and sends it to the $\gNB$.

\item[] \textbf{Step 7}:  $\gNB \rightarrow \UE$. $\mathbf{M_{7}}:\AuthEnc.\Enc\{k_s,M_{A_1}\}$\\
Upon the receipt of the message from $\AuC$, the $\gNB$ decrypt ${M_6}$ and re-encrypt it using ($k_s$), then forwards it to the $\UE$ via $M_7$.

\item[] \textbf{Step 8}:  $\ra{\UE \xrightarrow{ACK'} \gNB \xrightarrow{ACK"} \AuC}$.\\
Upon the reception of $M_7$, first, the $\UE$ decrypts the message using the session key ($k_s$), then using the long-term key ($k_i$). Next, the $\UE$ verifies his/her certificate using $\SanSig.\Verify(.)$ and checks the integrity of $\tid$ (i.e. $\tid^* \stackrel{?}{=} \tid \oplus r_{id}$). If both verification hold, then $\UE$ sends an acknowledgement message to the $\AuC$ to confirm the new temporary identity and delete the old one.

\end{itemize}
\begin{figure}[h!]
	\centering
	\begin{adjustbox}{max width=0.8\textwidth, max height=1\textheight}
		\fbox{
	\begin{tikzpicture}[yscale=-0.55,>=latex]
    \tikzstyle{every node}=[font=\large]
	\edef\InitX{0}
	\edef\ArrowLeft{1}
	\edef\ArrowCenter{6}
	\edef\ArrowRight{11}
	\edef\RespX{12}
	% Set the starting Y coordinate
	\edef\Y{0}
	
	% Draw header boxes
%	\node [rectangle,draw,inner sep=5pt,right] at (2,-5) {\textbf{SIDH + MAC construction}};
    \node [rectangle,draw,inner sep=5pt,right] at (\InitX,\Y) {\textbf{UE} };
	\node [rectangle,draw,inner sep=5pt,left] at (\RespX,\Y) {\textbf{gNB}};
	\NextLine[2]
	\ServerAction{$h \getsr \mathbb{Z}_q$}
	\NextLine
	\ServerAction{$\cert^{G}_{mod}=\EID\|g^{h}$}
	\NextLine
	\ServerAction{$(\cert_{G}^{*},\sigma^{*}_{G}) \gets \SanSig.\Sanit (  \cert^{G}_{fix}\|\cert^{G}_{mod},\sigma_{G}, \pk_{sig}^{\AuC}, \sk_{san}^{\gNB} )$}
	\NextLine[1.5]
	\ServerToClient{\framebox[1.1\width]{$\boldsymbol{M_1}:[\cert_{G}^{*}$, $\sigma^{*}_{G}$, $g^h]$}}{}
	\NextLine
	\ClientAction{\textbf{abort if} $1 \neq \SanSig.\Verify(\cert_{G}^{*}, \sigma^{*}_{G}, \pk_{sig}^{\AuC}, \pk_{san}^{\gNB} )$}
	\NextLine
	\ClientAction{$u \getsr \mathbb{Z}_q$, $\RUID \getsr \mathcal{R}$}
	\NextLine
	\ClientAction{$sk_{i},k_s=\KDF((g^h)^{u})$}
	\NextLine
	\ClientAction{$\cert^{U}_{mod}=\RUID\|\omega_{U} \|v\|g^{u}$}
	\NextLine
	\ClientAction{$(\cert_{U}^{*},\sigma^{*}_{U}) \gets \SanSig.\Sanit (  \cert^{U}_{fix}\|\cert^{U}_{mod},\sigma_{U}, \pk_{sig}^{\AuC}, \sk_{san}^{\UE} )$}
	\NextLine[1.5]
	\ClientToServer{\framebox[1.1\width]{$\boldsymbol{M_2}:[\AuthEnc.\Enc\{k_s,\cert_{U}^{*},\sigma^{*}_{U}\},g^{u}]$}}{}
	\NextLine
	\ServerAction{$\ra{sk_{i},k_s=\KDF((g^u)^{h})}$}
	\NextLine
	\ServerAction{\textbf{abort if} $1 \neq \SanSig.\Verify(\cert_{U}^{*},\sigma^{*}_{U}, \pk_{sig}^{\AuC}, \pk_{san}^{\UE} )$}
	\NextLine
	\ServerAction{\textbf{abort if} $1 \neq \mathit \Verifyacc(RL_v,\omega_U,\UID_i )$}
	\NextLine
	\ServerAction{\textbf{Update} $ [(\omega^{*}_{U}) \getsr \Nwitupdate(RL,RL^{*},x^{*},\UID_i, \omega_U)]$}
	\NextLine[1.5]
   	\ServerToClient{\framebox[1.1\width]{$\boldsymbol{M_{3}}\gets\AuthEnc.\Enc\{k_s,\omega^{*}_{U}\|v*\}$}}{}
	\NextLine
	\ClientAction{$\ra{\omega^{*}_{U}\|v* \gets \AuthEnc.\Dec(k_s,M_{3})}$}
	\NextLine
	\ClientAction{\textbf{Store} $(\omega^{*}_{U}\|v*)$}
	
	\end{tikzpicture}
}
	\end{adjustbox}	\caption{$\UniHand$'s Universal Handover phase.}
	\label{fig:universal-alt}
\end{figure}
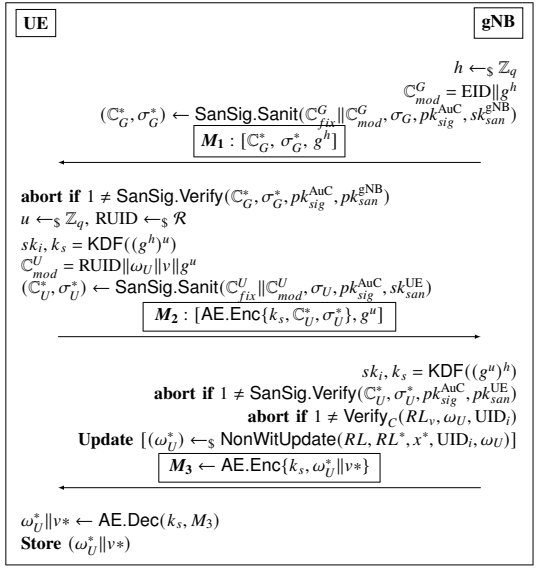

\subsection{Universal HO}
Each roaming user must execute this protocol to secure transit between small cells in 5G. During this protocol, each communicating party mutually authenticates their partner. The Universal HO protocol is described below and illustrated in Figure \ref{fig:universal-alt}.
\begin{itemize}[leftmargin=*]
\item[]\textbf{Step 1:} $\gNB \rightarrow \UE$. $\mathbf{M_{1}}$: [$\ra{\cert_{G}^{*}},\sigma^{*}_{G},g^{h}$].\\
    This step is similar to the initial authentication protocol's first step (step$1$).
    
\item[]\textbf{Step 2:} $\UE \rightarrow \gNB$. $\mathbf{M_2}$:$[\AuthEnc.\Enc\{k_s,\cert_{U}\|\sigma_U\| \omega_U\|v\},g^{u}]$\\
    Upon receiving the message $M_{1}$, UE verifies $\gNB$ certificate $(\cert_{G})$ using the $\SanSig$ verification algorithm $\SanSig.\Verify(\cert_{G}^{*}, \sigma^{*}_{G})$. Then the $\UE$ checks the message integrity (i.e. $g^h \stackrel{?}{=} g^h$). If both verification hold, $\UE$ samples $u$ and $\RUID$, then computes session keys $(sk_{i},k_s)$. Next, $\UE$ updates their certificate (the modifiable part), i.e. $\mathit{MOD}(\RUID\|\omega\|v\|g^{u})$ (preventing replay attacks), where the modifiable part consists of random-user id, non-membership witness, the accumulator version number and Ephemeral DH, respectively. Then $\UE$ sanitises the updated certificate using the sanitising algorithm $\SanSig.\Sanit(.)$, and encrypt it using $k_s$ to compose a message $M_{2}$. Finally, UE sends $M_{2}$ along with $g^{u}$ (Ephemeral DH) to the $\gNB$.

\item [] \textbf{Step 3}: $\gNB \rightarrow UE$ $\mathbf{M_{3}}$:$\gets\AuthEnc.\Enc\{k_s,\omega^{*}_{U}\|v*\}$\\
    Upon receiving the response message $M_{2}$, $\gNB$ generates the session keys $(sk_{i},k_s)$, to decrypt $ M_{2}$. Subsequently, $\gNB$ verifies $\UE$'s certificate using the verification algorithm $\SanSig.\Verify(\cert_{U}, \sigma_{U})$. If SanSig verification holds, $\gNB$ retrieves the accumulator version $v$ and checks if $v_i = v_{RL}$, to see if RL has been updated. If RL has not been updated, $\gNB$ checks whether the UE ($\Verifyacc(\UID_i,..)$) is in the revocation list. Otherwise, if the revocation list has been updated, where $v_i \neq v_{RL}$, $\gNB$ checks whether $\UID$ is added in the later version of the RL or not. If not, $\gNB$ updates the non-membership witness $ (\omega_{U}^{*})$ (where $x^{*}$ is the new revoked UE). Finally, $\gNB$ encrypts and sends $M_{3}$ to UE, which he will maintain for future communications. Details of this protocol are depicted in Figure 3.
\end{itemize}

\section{Security Framework}
Here we formalise the security properties of our UniHand scheme, which follows Bellare-Rogaway \cite{bellare_security_2006} key exchange models. These models essentially capture the security of a key exchange protocol as a game played between a probabilistic polynomial-time (PPT) adversary $\adv$ and a challenger $\challenger$. The adversary wins the game if it either causes a winning event (i.e. breaking authentication or anonymity) or terminates and guesses a challenge bit $b$ (i.e. breaking key indistinguishability). We utilise the Khan et al. framework \cite{khan_identity_2018} to capture notions of \emph{user unlinkability}, and eCK framework \cite{lamacchia_stronger_2007} to capture key indistinguishability ($\KIND$).

\subsection{Execution Environment}
Here we describe the shared execution environment of all security games. Our analysis uses three distinct games that assess different properties of a key exchange protocol: mutual authentication, key indistinguishability and unlinkability.
In our games, the challenger $\challenger$ maintains a single $\AuC$, running a number of instances of the key exchange protocol $\Prot$, and a set of (up to) $\numParties$ 
users $\UE_1, \ldots, \UE_{\numParties}$ 
(representing users communicating with the authentication centre $\AuC$), and systems $\gNB_1, \ldots, \gNB_{\numParties}$ (representing $\gNB$s communicating with the authentication centre), each potentially running up to $\numSessions$ executions of protocol $\Prot$.  We abuse notation and use $\testsess$ to refer to the $s$-th session owned by party $i$, and also as the state maintained by that session. 
We introduce the state maintained by each session:

\begin{itemize}
    \item $\id\in \{ 1,...,\numParties\}$: Index of the session owner.
    \item $\role \in \{\UE,\gNB,\AuC\}$: Role of the session.
    \item $s \in \{ 1,...,\numSessions\}$: Index of the session.
    \item $\sid \in \{\bits{*},\bot\}$: Session identifier, initialised as $\bot$.
    \item $\pid\in \{ 1,...,\numParties,\bot\}$: Partner UE identifier ($\bot$ if $\role =\UE$).
    \item $\gid\in \{ 1,...,\numParties,\bot\}$: Partner $\gNB$'s identifier.
    \item $\msgsent\in \{\bits{*},\bot\}$: Messages sent by the session.
    \item $\msgrecd\in \{\bits{*},\bot\}$: Messages received by the session.
    \item $k_i\in \{\bits{\secpar},\bot\}$: Long-term AuC/UE symmetric key.
    \item $k\in \{\bits{\secpar},\bot\}$: Established session key.
    \item $\alpha\in \{\inprogress, \acceptflag,\bot\}$: Session status.
    \item $it \in \{\bits{*},\bot\}$: Secret internal state of the session.

\end{itemize}
After initialisation, $\adv$ can interact with $\chal$ via adversary queries. We capture a network adversary capable of injecting, modifying, dropping, delaying or deleting messages at will via $\Send$ queries. Our models allow $\adv$ to initialise $\UE$ and $\gNB$ sessions owned by particular parties. Finally, $\adv$ can leak the long-term secrets of sessions via $\Corrupt$ queries, session keys via $\Reveal$ and the internal state of sessions via $\StateReveal$ queries, as described below.

\textbf{Adversary Queries.}
Here, we define queries that represent the behaviours of the adversary $\adv$ during the execution of the experiments. Note that not all queries are available to the adversary in the same game:
\begin{itemize}
    \item $\Create(i,s,\role)$: allows $\adv$ to initialize new $\UE$ and $\gNB$ sessions $\testsess$ such that $\testsess.\id = i$, $\testsess.\role = \role$.
    \item $\Send(m,i,s,\role) \leftarrow m'$:  allows $\adv$ to send message $m$ to a session $\testsess$ where $\testsess.\role = \role$. $\testsess$ processes the message and potentially outputs a message $m'$.
    \item $\CorruptLTK(i)\to k_i$: allows $\adv$ to leak the shared long-term key $k_i$ of $\UE_i$. 
    \item $\CorruptASK(i,\role) \to  (\sk)$: allows $\adv$ to leak the long-term asymmetric keys of a party, where $\role \in \{ \AuC,\gNB,\UE\}$ (for instance, $\CorruptASK(\AuC,0)$ or $\CorruptASK(\gNB,i)$). $\chal$ checks if $\adv$ previously corrupted these secrets, returning $\bot$ if so, otherwise the $\challenger$ returns $\sk_i^{\role}$.
    \item $\StateReveal(i,s,\role)\to {\testsess}$: allows $\adv$ to reveal the internal state of $\testsess$ where $\testsess.\role = \role$.
    \item $\Reveal(i,s, \role)\to k$: allows $\adv$ to reveal the secret session key $k$ computed during session $\testsess$ where $\testsess.\role = \role$.
    \item $\Test(i,s,\role) \to k_b$ (\textbf{Only used in the $\KIND$ security experiment}):  allows $\adv$ to play the $\KIND$ security game. When $\chal$ receives a $\Test(i,s,\role)$ query, if $\Test$ has already been issued, $\testsess.\status = \acceptflag$,  or $\testsess$ is not $\cleanpredicate$, then $\chal$ returns $\bot$. Otherwise, $\chal$ sets $k_0 \gets \testsess.k$, and $k_1 \getsr \bits{\secpar}$, and returns $k_b$ to $\adv$ (where $b$ was sampled by $\chal$ at the beginning of the experiment).
    \item $\Test(s,i,s',i')$ $\to m$ (\textbf{Only used in the $\Unlink$ security experiment}):  allows $\adv$ to play the $\Unlink$ security game. When $\chal$ receives a $\Test(s,i,s',i')$ query, initialises a new session $\session_b$, where $(\session_0=\testsess)$ or $(\session_1=\session^{s'}_{i'}$), $b$ was sampled by $\chal$, and both $\testsess$ and $\session^{s'}_{i'}$ are $\cleanpredicate$. $\Test$ query is only allowed to be issued by $\adv$ if no session $\session.\status \neq \inprogress$ such that $\session.\id = i$. $\chal$ will respond to any $\Send(m,i,s,\UE)$ or $\Send(m,i',s',\UE)$ queries with $\bot$ until $\session_b.\status \neq \inprogress$.
	\item $\SendTest(m) \to (m')$ (\textbf{Only used in the $\Unlink$ security experiment}): allows $\adv$ to send a message $m$ to $\session_b$ after issuing $\Test$. $\challenger$ returns a $\bot$ if $\session_b.\alpha \neq \inprogress$.
\end{itemize}

\subsection{Matching Conversations}
To capture what secrets the adversary is allowed to compromise without trivially breaking the security of our scheme, we need to define how sessions are \textit{partnered}, and whether those sessions are \textit{clean}. On a high level, partnering ensure that we can trace important sessions to other corruptions $\adv$ has made, and cleanness predicates determine which secrets  $\adv$ were not allowed to compromise. Matching conversations are typically used in the BR model \cite{bellare_authenticated_2008}, and the eCK-PFS model relaxes this notion to \textit{origin sessions}. However, these partnering methods inadequately address our setting, where the $\gNB$ essentially acts as a proxy, re-encrypting messages between the $\UE$ and the $\AuC$. Thus, two problems occur: we need to capture the messages that $\UE$ authenticates to the $\AuC$, and we also need to capture the fact that the $\gNB$ sends messages to two parties, neither of which exactly match $\gNB$'s transcript. Our solution is two-fold: we use \textit{matching sessions (identifiers)} to capture the messages authenticated between the $\UE$ and the $\AuC$, and we introduce \textit{matching subsets} to capture the subset of messages authenticated between the $\gNB$ and the $\AuC$ and $\UE$ respectively. 

\begin{definition}[Matching Subset]
Let $S \subseteq T$ denote that all strings $s$ in the set $S$ are substrings of $T$. A session $\testsess$ has a matching subset with another session $\partsess$, if $\partsess.\msgrecd \neq \bot$, $\testsess.\role \neq \partsess.\role$, and if $\partsess.\role = \gNB$ $\testsess.\msgrecd \subseteq \partsess.\msgsent$ and $\testsess.\msgsent \subseteq \partsess.\msgrecd$, and if $\testsess.\role = \gNB$, $\partsess.\msgrecd \subseteq \testsess.\msgsent$ and $\partsess.\msgsent \subseteq \testsess.\msgrecd$.
 \end{definition}

Next, we introduce the notion of \textit{matching sessions}, where the session identifier $\sid$ of both sessions are either equal (or where one is a prefix string of the other).
\begin{definition}[Matching Sessions]
 Let $S \subset T$ denote that a string $S$ is a (potentially equal) prefix of a string $T$. A session $\testsess$ is a matching session of $\partsess$, if $\partsess.\sid  \neq \bot \neq \testsess.\sid$, $\testsess.\role \neq \partsess.\role$ and $\partsess.\sid \subset \testsess.\sid$ or $\testsess.\sid \subset \partsess.\sid$.
 \end{definition}
 
\ifsubmissionversion

%We now turn to define the  $\Unlink$ security game on a high level and point the readers to Appendix (\ref{Full Security Analysis}) for more details on $\MA$, $\KIND$ and $\Unlink$ security games. We provide the cleanness predicates (restricting adversary queries) for $\MA$ and $\Unlink$ security games below.

\fi

We now turn to define each security game for $\MA$, $\KIND$ and $\Unlink$. 

\subsection{Mutual Authentication}
\label{ma-sec}
In this section, we describe the overall goal of $\adv$ in the $\MA$ security game and the queries that $\adv$ has access to. The experiment $\Exp{\MA,\cleanpredicate}{\Prot,\numParties,\numSessions,\adversary}(\secpar)$ is played between a challenger $\challenger$ and an adversary $\adversary$. At the beginning of the experiment, $\chal$ generates long-term asymmetric keys for the $\AuC$ and each user $\UE_i$ and each gNB $\gNB_i$ (where $i \in [\numParties]$) and long-term symmetric keys for each user $\UE_i$, and then interacts with $\adv$ via $\Create,$ $\Send$, $\CorruptLTK$, $\CorruptASK$, $\StateReveal$ and $\Reveal$ queries. $\adv$ wins (and $\Exp{\MA}{\Prot,\numParties,\numSessions,\adversary}(\secpar)$ outputs $1$), if the adversary has caused a $\cleanpredicate$ session to accept (and set $\testsess.\status \gets \acceptflag$) and there either exists no matching subset session $\partsess$, or no matching session $\pi$.

We now turn to describe our cleanness predicates. In the initial authentication protocol, if a session (owned by $\AuC$ or $\UE$) accepts without either a matching subset or a matching session, then $\adv$ only wins if they have not compromised: 
(a) the state of any (at the point of compromise) matches, or
(b) both the long-term shared key of the $\UE$ partner and the long-term asymmetric key of the intended partner, or
(c) the long-term asymmetric secrets of the $\gNB$.

\begin{definition}[Initial authentication cleanness]
	\label{ma-clean, Initial-auth}
	A session $\testsess$ in the $\MA$ experiment described above is $\cleanpredicate_{IA}$ if the following conditions hold:
	\begin{enumerate}
    \item $\StateReveal(i,s,\testsess.\role)$ has not been issued \emph{and} for all sessions $\partsess$ such that $\partsess$ is a matching subset of $\testsess$ $\StateReveal(j,t,\partsess.\role)$ has not been issued \textit{and} for all sessions $\partsess$ such that $\partsess$ is a matching session of $\testsess$ $\StateReveal(j,t,\partsess.\role)$ has not been issued.

    \item If $\testsess.\role \neq \gNB$, and there exists no $(j,t) \in \numParties \times \numSessions$ such that $\partsess$ is a matching session of $\testsess$, $\CorruptLTK(i)$ or \sloppy $\CorruptASK(\testsess.\pid,(\AuC,\UE)\backslash\testsess.\role)$ have not both been issued before $\testsess.\status = \acceptflag$.
    
    \item If there exists no $(j,t) \in \numParties \times \numSessions$ such that $\partsess$ is a matching subset for $\testsess$, and $\testsess.\role \neq \gNB$, $\CorruptASK(\testsess.\pid,\gNB)$ has not been issued before $\testsess.\status = \acceptflag$. Else, if there exists no $(j,t) \in \numParties \times \numSessions$ such that $\partsess$ is a matching subset for $\testsess$, and $\testsess.\role = \gNB$, $\CorruptASK(\testsess.\pid,\UE)$ and $\CorruptASK(\empty,\AuC)$ have not been issued before $\testsess.\status = \acceptflag$.
    
\end{enumerate}
\end{definition}

\begin{definition}[Universal Handover cleanness]
	\label{ma-clean, uni-HO}
	A session $\testsess$ in the $\MA$ experiment described above is $\cleanpredicate_{UH}$ if the following conditions hold:
	\begin{enumerate}
    \item $\StateReveal(i,s,\testsess.\role)$ has not been issued \emph{and} for all sessions $\partsess$ such that $\partsess$ is a matching subset of $\testsess$ $\StateReveal(j,t,\partsess.\role)$ has not been issued.
    \item If there exists no $(j,t) \in \numParties \times \numSessions$ such that $\partsess$ is a matching subset for $\testsess$, $\CorruptASK(\testsess.\pid,(\gNB,\UE)\backslash(\testsess.\role))$ has not been issued before $\testsess.\status = \acceptflag$.
    \end{enumerate}
\end{definition}

In the mutual authentication game, $\adversary$'s goal is to cause a session $\testsess$ to accept without a matching session (i.e. no $AuC$ session that outputs the messages received by $\testsess$) or matching subset (i.e. no $\gNB$ session has output those messages).

We say that a protocol $\Prot$ is $\MA$-secure, if there exist no PPT algorithms $\adv$ that can win the $\MA$ security game against a clean session with a non-negligible advantage. We formalise this notion below.

\begin{definition}[Mutual Authentication Security]
	\label{def:ma-sec}
	Let $\Prot$ be a key exchange protocol, and $\numParties, 
	\numSessions \in \NN$. 
	For a given cleanness predicate $\cleanpredicate$, and a PPT algorithm $\adversary$, we define the advantage of $\adversary$ in the mutual authentication $\MA$ game to be:
	$\Adv{\MA,\cleanpredicate}{\Prot,\numParties,\numSessions,\adversary}(\secpar) = |\Pr[\Exp{\MA,\cleanpredicate}{\Prot,\numParties,\numSessions,\adversary}(\secpar)=1]-\frac{1}{2}|.$
	We say that $\Prot$ is $\MA$-secure if, for all PPT $\adversary$, $\Adv{\MA,\cleanpredicate}{\Prot,\numParties,\numSessions,\adversary}(\secpar)$ is negligible in security parameter $\secpar$.
\end{definition}

\subsection{Key Indistinguishability}
\label{sec:kind-game}
%In the key indistinguishability game, $\adv$ has access to the $\Test$ query: when $\adv$ issues a $\Test(i,s, \role)$ query, they either receive $\testsess.k$ or a random key from the same distribution (based on $b$). $\adv$ eventually terminates, outputs a guess $b'$ and wins (and $\Exp{\KIND, \cleanpredicate}{\Prot,\numParties,\numSessions,\adversary}(\secpar)$ outputs $1$), if $b'=b$.
Here we describe the overall goal of $\adv$ in the key indistinguishability $\KIND$ security game and the queries that $\adv$ has access to. The experiment $\Exp{\KIND,\cleanpredicate}{\Prot,\numParties,\numSessions,\adversary}(\secpar)$ is played between a challenger $\challenger$ and an adversary $\adversary$. At the beginning of the experiment, $\chal$ generates long-term symmetric keys for the $\AuC$ and each user $\UE_i$ and each gNB $\gNB_i$ (where $i \in [\numParties]$), samples a random bit $b \getsr \bits{}$ and then interacts with $\adv$ via queries mentioned below. At some point, $\adv$ issues a $\Test(i,s, \role)$ query and either receives $\testsess.k$ or a random key from the same distribution (based on $b$). $\adv$ eventually terminates, outputs a guess $b'$ and wins (and $\Exp{\KIND, \cleanpredicate}{\Prot,\numParties,\numSessions,\adversary}(\secpar)$ outputs $1$), if $b'=b$.
We say that a protocol $\Prot$ is $\KIND$-secure, if there exist no PPT algorithms $\adv$ that can win the $\KIND$ security game with non-negligible advantage, formalising this notion below.

\begin{definition}[Key Indistinguishability]
	\label{def:kind-sec}
	Let $\Prot$ be a key exchange protocol, and $\numParties, 
	\numSessions \in \NN$. 
	For a given cleanness predicate $\cleanpredicate$, and a PPT algorithm $\adversary$, we define the advantage of $\adversary$ in the key indistinguishability $\KIND$ game to be:
	$\Adv{\KIND,\cleanpredicate}{\Prot,\numParties,\numSessions,\adversary}(\secpar) = |\Pr[\Exp{\KIND,\cleanpredicate}{\Prot,\numParties,\numSessions,\adversary}(\secpar)=1]-\frac{1}{2}|.$
	We say that $\Prot$ is $\KIND$-secure if, for all PPT $\adversary$, $\Adv{\KIND,\cleanpredicate}{\Prot,\numParties,\numSessions,\adversary}(\secpar)$ is negligible in the parameter $\secpar$.
\end{definition}

\iffullversion
We say that a protocol $\Prot$ is $\MA$-secure, if there exist no PPT algorithms $\adv$ that can win the $\MA$ security game against a clean session with a non-negligible advantage. We formalise this notion below.

\begin{definition}[Mutual Authentication Security]
	\label{def:ma-sec}
	Let $\Prot$ be a key exchange protocol, and $\numParties, 
	\numSessions \in \NN$. 
	For a given cleanness predicate $\cleanpredicate$, and a PPT algorithm $\adversary$, we define the advantage of $\adversary$ in the mutual authentication $\MA$ game to be:
	$\Adv{\MA,\cleanpredicate}{\Prot,\numParties,\numSessions,\adversary}(\secpar) = |\Pr[\Exp{\MA,\cleanpredicate}{\Prot,\numParties,\numSessions,\adversary}(\secpar)=1]-\frac{1}{2}|.$
	We say that $\Prot$ is $\MA$-secure if, for all PPT $\adversary$, $\Adv{\MA,\cleanpredicate}{\Prot,\numParties,\numSessions,\adversary}(\secpar)$ is negligible in security parameter $\secpar$.
\end{definition}

\subsection{Key Indistinguishability}
\label{sec:kind-game}
Here we describe the overall goal of $\adv$ in the key indistinguishability $\KIND$ security game and the queries that $\adv$ has access to. The experiment $\Exp{\KIND,\cleanpredicate}{\Prot,\numParties,\numSessions,\adversary}(\secpar)$ is played between a challenger $\challenger$ and an adversary $\adversary$. At the beginning of the experiment, $\chal$ generates long-term symmetric keys for the $\AuC$ and each user $\UE_i$ and each gNB $\gNB_i$ (where $i \in [\numParties]$), samples a random bit $b \getsr \bits{}$ and then interacts with $\adv$ via queries mentioned below. At some point, $\adv$ issues a $\Test(i,s, \role)$ query and either receive $\testsess.k$ or a random key from the same distribution (based on $b$). $\adv$ eventually terminates, outputs a guess $b'$ and wins (and $\Exp{\KIND, \cleanpredicate}{\Prot,\numParties,\numSessions,\adversary}(\secpar)$ outputs $1$), if $b'=b$.

\textbf{Adversary Queries.}
In the $\KIND$ game, $\adv$ has access $\Create$, $\Send$, $\CorruptLTK$, $\CorruptASK$, $\Reveal$, $\StateReveal$ and $\Test$ queries described above. We now turn to define our cleanness predicate for the key-indistinguishability game.

\fi
\begin{definition}[cleanness predicate]
	\label{def:KIND-clean}
	A session $\testsess$ in the $\KIND$ experiment described above is $\cleanpredicate_{IA} \&$ $\cleanpredicate_{UH}$ if the following conditions hold:
	\begin{enumerate}
	\item $\Reveal({i},{s},\role)$ has not been issued, and if a matching session $\partsess$ exists, $\Reveal(j,t,\partsess.\role)$ has not been issued.
    \item The query $\StateReveal({i},{s},\role)$ has not been issued \emph{and} for all $j,t$ such that $\partsess$  has a matching subset with  $\testsess$, \sloppy $\StateReveal(j,t,\partsess.\role)$ has not been issued.
    %condition 4 capturing KCI
    \item If there is no $(j,t) \in \numParties \times \numSessions$ such that $\partsess$ is a matching subset for $\testsess$, $\CorruptLTK(i)$  and $\CorruptASK(\testsess.\pid)$ have not been both issued before $\testsess.\status = \acceptflag$.
\end{enumerate}
\end{definition}
\iffullversion

We say that a protocol $\Prot$ is $\KIND$-secure, if there exist no PPT algorithms $\adv$ that can win the $\KIND$ security game with non-negligible advantage, formalising this notion below.

\begin{definition}[Key Indistinguishability]
	\label{def:kind-sec}
	Let $\Prot$ be a key exchange protocol, and $\numParties, 
	\numSessions \in \NN$. 
	For a given cleanness predicate $\cleanpredicate$, and a PPT algorithm $\adversary$, we define the advantage of $\adversary$ in the key indistinguishability $\KIND$ game to be:
	$\Adv{\KIND,\cleanpredicate}{\Prot,\numParties,\numSessions,\adversary}(\secpar) = |\Pr[\Exp{\KIND,\cleanpredicate}{\Prot,\numParties,\numSessions,\adversary}(\secpar)=1]-\frac{1}{2}|.$
	We say that $\Prot$ is $\KIND$-secure if, for all PPT $\adversary$, $\Adv{\KIND,\cleanpredicate}{\Prot,\numParties,\numSessions,\adversary}(\secpar)$ is negligible in the parameter $\secpar$.
\end{definition}
\fi
\subsection{Unlinkability }
\label{UNlink-sec}
In this section, we describe the overall goal of $\adv$ in the $\Unlink$ security game and the queries that $\adv$ has access to. The experiment $\Exp{\Unlink, \cleanpredicate}{\Prot,\numParties,\numSessions,\adversary}(\secpar)$ is played between a challenger $\challenger$ and an adversary $\adversary$. At the beginning of the experiment, $\chal$ generates long-term symmetric keys for the $\AuC$ and each user $\UE_i$ and each gNB $\gNB_i$ (where $i \in [\numParties]$), samples a random bit $b \getsr \bits{}$ and then interacts with $\adv$ via queries mentioned below. $\adv$ wins (and $\Exp{\Unlink, \cleanpredicate}{\Prot,\numParties,\numSessions,\adversary}(\secpar)$ outputs $1$), if $\adv$ terminates and outputs a guess bit $b'$ such that $b'=b$.

\textbf{Adversary Queries}
In the $\KIND$ game, $\adv$ has access $\Create$, $\Send$, $\CorruptLTK$, $\CorruptASK$, $\Reveal$, $\StateReveal$, $\Test$ and $\SendTest$ queries described above. Unlike in the $\KIND$ game, $\Test$ in $\Unlink$ allows the adversary to initialise one of two sessions (depending on a bit $b$ sampled by the challenger), and $\SendTest$, which allows the adversary to interact with that session without revealing which party owns it. We now turn to define our cleanness predicate for the unlinkability game.

\begin{definition}[Cleanness predicate]
	\label{unlink-clean}
	A session $\testsess$ in the $\Unlink$ experiment is $\cleanpredicate$ if the following conditions hold:
	\begin{enumerate}
    \item The query $\StateReveal({i},{s},\role)$ has not been issued \emph{and} for all $j,t$ such that $\partsess$ is a matching session (or has a matching subset) with  $\testsess$, \sloppy $\StateReveal(j,t,\partsess.\role)$ has not been issued.
    \item If there is no $(j,t) \in \numParties \times \numSessions$ such that $\partsess$ is a matching subset for $\testsess$, $\CorruptASK(\testsess.\pid, (\gNB,\UE)\backslash\testsess.\role)$ has not been issued before $\testsess.\status = \acceptflag$.

\end{enumerate}
\end{definition}

We say that a protocol $\Prot$ is $\Unlink$-secure, if there exist no PPT algorithms $\adv$ that can win the $\Unlink$ security game with non-negligible advantage, which we formalise below.

\begin{definition}[Unlinkability]
	\label{def:unlink-sec}
	Let $\Prot$ be a key exchange protocol, and $\numParties, 
	\numSessions \in \NN$. 
	For a given cleanness predicate $\cleanpredicate$, and a PPT algorithm $\adversary$, we define the advantage of $\adversary$ in the unlinkability $\Unlink$ game to be:
	$\Adv{\Unlink,\cleanpredicate}{\Prot,\numParties,\numSessions,\adversary}(\secpar) = |\Pr[\Exp{\Unlink,\cleanpredicate}{\Prot,\numParties,\numSessions,\adversary}(\secpar)=1]-\frac{1}{2}|.$
	We say that $\Prot$ is $\Unlink$-secure if, for all PPT $\adversary$, $\Adv{\Unlink,\cleanpredicate}{\Prot,\numParties,\numSessions,\adversary}(\secpar)$ is negligible in the parameter $\secpar$.
\end{definition}

\section{Security Analysis}
\label{Formal Security Analysis}

In this section, we formally prove the security of our protocols. We begin by demonstrating that our protocols achieve mutual authentication, then show that the session keys output from the secure handover scheme are indistinguishable from random, and finally show that external attackers cannot link sessions owned by the same party. 

\subsection{Mutual Authentication security}
Here we formally analyse the $\MA$-security of the $\UniHand$ scheme.

\subsubsection{\texorpdfstring{$\MA$}{Lg}-security of Initial Authentication}
We begin by showing that the Initial Authentication protocol achieves mutual authentication.
%%%%%%%%%%%%%%%%%%%%%%%%%%%%%%%%%%%%%%%%%%%%%%%%%%%%
%%%%%%%%%%%%%%%% MA SECURITY %%%%%%%%%%%%%%%%%%%%%%%
%%%%%%%%%%%%% INITIAL AUTHENTICATION %%%%%%%%%%%%%%%
%%%%%%%%%%%%%%%%%%%%%%%%%%%%%%%%%%%%%%%%%%%%%%%%%%%%

\begin{theorem}
\label{theorem: MA sec- initial auth}
{\textbf{$\MA$-security of Initial Authentication}}. Initial Authentication depicted in Figure \ref{fig:init-auth-alt} is $\MA$-secure under the cleanness predicate $\cleanpredicate_{IA}$ in Definition \ref{ma-clean, Initial-auth}. For any PPT algorithm $\adv$, $\Adv{\MA,\cleanpredicate_{IA}}{\Prot,\numParties,\numSessions,\adversary}(\secpar)$ is negligible assuming that the $\SanSig$, $\Enc$ and $\KDF$ schemes achieve $\EUFCMA$, $\AuthEnc$ and $\KDF$ security and under the $\DDH$ assumption.
\end{theorem}

\begin{pro}
Our proof is divided into three cases, denoted by $\Adv{\MA,\cleanpredicate, c_1}{\Prot,\numParties,\numSessions,\adversary}(\secpar)$, $\Adv{\MA,\cleanpredicate, c_2}{\Prot,\numParties,\numSessions,\adversary}(\secpar)$ and $\Adv{\MA,\cleanpredicate, c_3}{\Prot,\numParties,\numSessions,\adversary}(\secpar)$
    
We then bound the advantage of $\adv$ winning the game under certain assumptions to $\Adv{\MA,\cleanpredicate_{IA}}{\Prot,\numParties,\numSessions,\adversary}(\secpar)\leq(\Adv{\MA,\cleanpredicate, c_1}{\Prot,\numParties,\numSessions,\adversary}(\secpar)+\Adv{\MA,\cleanpredicate, c_2}{\Prot,\numParties,\numSessions,\adversary}(\secpar)+\Adv{\MA,\cleanpredicate, c_3}{\Prot,\numParties,\numSessions,\adversary}(\secpar))$. 
\ifsubmissionversion
Due to space constraints, we give only a proof sketch and point readers to \cite{authors_unihand_2023}\footnote[1]{The full version of the security analysis and the security framework is available in the Supplementary Material.} for the full proof details.
\fi

\textbf{Case 1:} We assume the first $\cleanpredicate$ session $\testsess$  to accept without a matching session or subset sets $\testsess.\role =\UE$. 
Now we begin by bounding the advantage of $\adv$ in Case 1.1.

%%%%%%%%%%%%%%%%%%%%%%%%%%%%%%%%%%%%%%%%%%%%%%%%%%%%
%%%%%%%%%%%%%%%% MA SECURITY %%%%%%%%%%%%%%%%%%%%%%%
%%%%%%%%%%%%% INITIAL AUTHENTICATION %%%%%%%%%%%%%%%
%%%%%%%%%%%%%%%%% CASE 1.1 %%%%%%%%%%%%%%%%%%%%%%%%%

\textbf{Case 1.1:} According to the definition of this case, $\UE$ either accepts messages $\mathbf{M_1}$ and $\mathbf{M_7}$ without a matching subset (i.e. without honest $\gNB$ partner), or $\mathbf{M_{A_1}}$ without a matching session identifier (i.e. without honest $\AuC$).  In this subcase $\adv$ cannot corrupt the long-term $\UE$ symmetric secret or the $\testsess.\pid_{\gNB}$ asymmetric key.

\ifsubmissionversion
On a high level, we show that $\adv$ cannot inject DH public keys between the $\gNB$ and the $\UE$ due to the $\gNB$ signatures over these values, based on the $\EUFCMA$ security of the $\SanSig$ scheme. As the secret DH output is used to derive keys for the authenticated encryption scheme securely, and all other messages exchanged between the $\gNB$ and the $\UE$ are encrypted, $\adv$ cannot modify messages sent between the $\gNB$ and the $\UE$ due to the $\AuthEnc$ security of the $\Enc$ scheme. Similarly, since all messages sent between the $\AuC$ and the $\UE$ are encrypted under the $\UE$ long-term symmetric key (and the $\UE$ long-term key cannot be compromised by $\adv$, $\adv$ cannot modify messages sent between the $\gNB$ and the $\UE$ due to the $\AuthEnc$ security of the $\Enc$ scheme.
\fi

\iffullversion

Here we provide the details of the security games:
\textbf{Game 0}: This is the original mutual authentication experiment described in \ref{ma-sec}: \[\Adv{\MA,\cleanpredicate,\mathsf{C1.1}}{\Prot,\numParties,\numSessions,\adversary}(\secpar) \leq \Adv{}{G_0}\]

\textbf{Game 1} : In this game, we guess the index $(i,s) \in \numParties \times \numSessions$ of the first $\UE$ session that accepts without a matching session or subset, introducing a factor of $\numParties \times \numSessions$ in $\adv$'s advantage: \[\Adv{}{G_0} \leq \numParties \cdot\numSessions \Adv{}{G_{1}}.\]

\textbf{Game 2} : In this game, we guess the index $j \in \numParties$ of the $\gNB$ party such that $\testsess.\pid$, introducing a factor of $\numParties \times \numSessions$ in $\adv$'s advantage. \[\Adv{}{G_1} \leq \numParties \Adv{}{G_{2}}.\]

\textbf{Game 3}: Here we introduce an abort event, where $\chal$ aborts if $\testsess$ receives a message $\mathbf{M_1}$ without setting $\testsess.\status \gets \rejectflag$ but $\mathbf{M_1}$ was not output by a session owned by $\testsess.\pid$. We do so by defining a reduction $\bdv_1$ that initialises a $\SanSig$ challenger $\chal_{\SanSig}$, that outputs $\pk_{sig}^{\challenger}$ and  $\pk_{san}^{\challenger}$, which we embed into the $\AuC$'s $\pk_{sig}^{\AuC}$  and $\gNB$'s $\pk_{san}^{\gNB}$ respectively. Anytime $\AuC$ or $\gNB$ needs to generate a signature over a message $m$, $\bdv_1$ instead queries $\chal$ with $m$. Now, if $\testsess$ receives a message $\mathbf{M_1}$ without setting $\testsess.\status \gets \rejectflag$ but $\mathbf{M_1}$ was not output by a session owned by $\testsess.\pid$, then $\adv$ must have produced a message $\mathbf{M_1} = C^*_G,\sigma^*_G,g^h$ such that $\SanSig.\Verify(C^*_G,\sigma^*_G,\pk_{sig}^{\chal},\pk_{san}^{\chal})=1$, which is a valid signature forgery. $\bdv_1$ responds to $\chal_{\SanSig}$ with $C^*_G,\sigma^*_G$ and triggers the abort event. 
Thus, the probability that $\bdv_1$ triggers the abort event is bounded by the $\EUFCMA$ security of $\SanSig$: \[\Adv{}{G_2} \leq \Adv{}{G_3}+\Adv{\EUFCMA}{\bdv_1, \SanSig}.\]

\textbf{Game 4} : In this game, we guess the index $t \in \numSessions$ of the $\gNB$ session $\partsess$ that output $\mathbf{M_1}$ received by $\testsess$, introducing a factor of $\numSessions$ in $\adv$'s advantage. \[\Adv{}{G_3} \leq \numParties \cdot\numSessions \Adv{}{G_{4}}.\]

\textbf{Game 5}: Here we introduce another abort event that triggers if $\adv$ sends a Diffie-Hellman public keyshare $g^h$ to the session $\testsess$, i.e. session $\testsess$ receives $g^h$ that was not output from a $\gNB$ session, but instead from $\adv$. Since this trigger event requires the signature $\sigma^*_G$ in $\mathbf{M_1}$ to verify over $g^h$, and by \textbf{Game 3} we already abort if $\sigma^*_G$ comes from $\adv$, it follows that \[\Adv{}{G_4} \leq \Adv{}{G_5}.\]

\textbf{Game 6}: In this game, we replace $g^{hu}$ computed honestly in $\testsess$ with a uniformly random and independent value $\hat{g^{hu}}$. We do so by defining a reduction $\bdv_2$ that initialises a $\DDH$ challenger $\chal_{\DDH}$, and replaces $g^u$, $g^h$ and $g^{hu}$ computed by $\testsess$ and $\partsess$ with the outputs of $\chal_{\DDH}$, $g^a$, $g^b$, $g^c$. We note that if the bit $b$ sampled by $\chal_{\DDH}$ is 1, then $c = ab$ and we are in \textbf{Game 5}; otherwise, $c \getsr \Zq$ and we are in \textbf{Game 6}. Any $\adv$ that can distinguish \textbf{Game 5} from \textbf{Game 6} can break the $\DDH$ assumption. Thus:  \[\Adv{}{G_5} \leq \Adv{}{G_6}+ \Adv{G,g,q}{\bdv_2 \DDH}.\]

\textbf{Game 7}: In this game we replace the session and encryption keys $sk_i, k_s$ with uniformly random values $\hat{sk_i}, \hat{k_s}$. We do so by defining a reduction $\bdv_3$ that interacts with a $\KDF$ challenger $\chal_{\KDF}$, querying $\chal_{\KDF}$ with $\hat{g^{uh}}$ and replacing the computation of $sk_i,k_s$ in $\testsess$ and $\partsess$ with the outputs from the $\chal_{\KDF}$ $\hat{sk_i}, \hat{k_s}$. Since $sk_i, k_s \gets \KDF(\hat{g^{uh}})$ and by \textbf{Game 6} $\hat{g^{uh}}$ is already uniformly random and independent, this change is sound. Any $\adversary$ that can distinguish \textbf{Game 6} from \textbf{Game 7} can be used to break $\KDF$ security of the $\KDF$ scheme. Thus:  \[\Adv{}{G_6} \leq \Adv{}{G_7}+ \Adv{\KDF}{\bdv_3,\KDF}.\]

\textbf{Game 8}: In this game, we introduce an abort event that triggers if $\testsess$ decrypts $\mathbf{M_7}$ (keyed by $\hat{k_s}$), but $\mathbf{M_7}$ was not output by $\partsess$. We do so by defining a reduction $\bdv_4$ that initialises an $\Enc$ challenger $\chal_{\AuthEnc}$, which $\bdv_4$ queries when $\testsess$ needs to encrypt or decrypt with $\hat{k_s}$. The abort event only triggers if $\adv$ can produce a valid ciphertext that decrypts under $\hat{k_s}$, and we can submit $\mathbf{M_7}$ to $\chal_{\AuthEnc}$, breaking the $\AuthEnc$ security of the $\Enc$ scheme. By \textbf{Game 7} $\hat{k_s}$ is already uniformly random and independent, and this replacement is sound. Any $\adversary$ that can trigger the abort event can be used by $\bdv_4$ to break the $\AuthEnc$ security of $\Enc$. This implies: \[\Adv{}{G_7} \leq \Adv{}{G_8}+ \Adv{\AuthEnc}{\bdv_4,\Enc}.\]

\textbf{Game 9}: In this game, we introduce an abort event that triggers if $\testsess$ decrypts $\mathbf{M_{A_1}}$ (keyed by $k_i$), but $\mathbf{M_{A_1}}$ was not output by a session owned by $\AuC$. We do so by defining a reduction $\bdv_5$ that initialises an $\Enc$ challenger $\chal_{\AuthEnc}$, which $\bdv_5$ queries whenever $\chal$ needs to encrypt or decrypt with $k_i$. The abort event only triggers if $\adv$ can produce a valid ciphertext that decrypts under $k_i$, and we can submit $\mathbf{M_{A_1}}$ to $\chal_{\AuthEnc}$, breaking the $\AuthEnc$ security of the $\Enc$ scheme. By the definition of the $\MA$ security game, $k_i$ is uniformly randomly sampled at the beginning of the experiment, and this replacement is sound. Any $\adversary$ that can trigger the abort event can be used by $\bdv_5$ to break the $\AuthEnc$ security of $\Enc$. This implies: \[\Adv{}{G_8} \leq \Adv{}{G_9}+ \Adv{\AuthEnc}{\bdv_5,\Enc}.\]

\textbf{Game 10}: In this game, the session $\testsess$ will only accept $\mathbf{M_1}$ and $\mathbf{M_7}$ from $\gNB$ and $\mathbf{M_{A_1}}$ from $\AuC$ if they were generated honestly by $\gNB$ and $\AuC$ sessions, and thus $\testsess$ has both matching sessions and matching subsets. Thus the advantage of $\adv$ in winning the $\MA$-security experiment is negligible. 
    \[\Adv{}{G_{10}}=0.
    \]
\fi

We turn to bound the advantage of $\adv$ in \textbf{Case 1.2}.

%%%%%%%%%%%%%%%%%%%%%%%%%%%%%%%%%%%%%%%%%%%%%%%%%%%%
%%%%%%%%%%%%%%%% MA SECURITY %%%%%%%%%%%%%%%%%%%%%%%
%%%%%%%%%%%%% INITIAL AUTHENTICATION %%%%%%%%%%%%%%%
%%%%%%%%%%%%%%%%% CASE 1.2 %%%%%%%%%%%%%%%%%%%%%%%%%

\textbf{Case 1.2:} According to the definition of this case, $\UE$ either accepts messages $\mathbf{M_1}$ and $\mathbf{M_7}$ without a matching subset (i.e. without honest $\gNB$ partner), or $\mathbf{M_{A_1}}$ without a matching session identifier (i.e. without honest $\AuC$).  In this subcase, $\adv$ cannot corrupt the long-term $\AuC$ asymmetric key or the $\testsess.\pid_{\gNB}$ asymmetric key.

\ifsubmissionversion
On a high-level, we show that $\adv$ cannot inject DH public keys between the $\gNB$ and the $\UE$ due to the $\gNB$ signatures over these values, based on the $\EUFCMA$ security of the $\SanSig$ scheme. As the secret DH output is used to derive keys for the authenticated encryption scheme securely, and all other messages exchanged between the $\gNB$ and the $\UE$ are encrypted, $\adv$ cannot modify messages sent between the $\gNB$ and the $\UE$ due to the $\AuthEnc$ security of the $\Enc$ scheme. Similarly, we show that $\adv$ cannot inject DH public keys between the $\gNB$ and the $\AuC$ due to the $\gNB$ signatures and the $\AuC$ signatures over these values, based on the $\EUFCMA$ security of the $\SanSig$ scheme. As the secret DH output is used to derive keys for the authenticated encryption scheme securely, and all other messages exchanged between the $\gNB$ and the $\AuC$ are encrypted, $\adv$ cannot modify messages sent between the $\gNB$ and the $\AuC$ due to the $\AuthEnc$ security of the $\Enc$ scheme. As the $\gNB$ proxies ciphertexts between the $\UE$ and the $\AuC$, this means that $\adv$ cannot modify any messages between $\UE$ and $\AuC$.
\fi

\iffullversion

Here we provide the details of the security games:\\
\textbf{Game 0}: This is the original mutual authentication experiment described in \cref{ma-sec}: \[\Adv{\MA,\cleanpredicate,\mathsf{C1.2}}{\Prot,\numParties,\numSessions,\adversary}(\secpar) \leq \Adv{}{G_0}\]

\textbf{Game 1} : In this game, we guess the index $(i,s) \in \numParties \times \numSessions$ of the first $\UE$ session that accepts without a matching session or subset, introducing a factor of $\numParties \times \numSessions$ in $\adv$'s advantage: \[\Adv{}{G_0} \leq \numParties \cdot\numSessions \Adv{}{G_{1}}.\]

\textbf{Game 2}: Here we introduce an abort event, where $\chal$ aborts if $\testsess$ receives a message $\mathbf{M_1}$ without setting $\testsess.\status \gets \rejectflag$ but $\mathbf{M_1}$ was not output by a session owned by $\testsess.\pid$. We do so by defining a reduction $\bdv_6$ that initialises a $\SanSig$ challenger $\chal_{\SanSig}$, that outputs $\pk_{sig}^{\challenger}$ and  $\pk_{san}^{\challenger}$, which we embed into the $\AuC$'s $\pk_{sig}^{\AuC}$  and $\gNB$'s $\pk_{san}^{\gNB}$ respectively. Anytime $\AuC$ or $\gNB$ needs to generate a signature over a message $m$, $\bdv_1$ instead queries $\chal$ with $m$. Now, if $\testsess$ receives a message $\mathbf{M_1}$ without setting $\testsess.\status \gets \rejectflag$ but $\mathbf{M_1}$ was not output by a session owned by $\testsess.\pid$, then $\adv$ must have produced a message $\mathbf{M_1} = C^*_G,\sigma^*_G,g^h$ such that $\SanSig.\Verify(C^*_G,\sigma^*_G,\pk_{sig}^{\chal},\pk_{san}^{\chal})=1$, which is a valid signature forgery. $\bdv_6$ responds to $\chal_{\SanSig}$ with $C^*_G,\sigma^*_G$ and triggers the abort event. 
Thus, the probability that $\bdv_6$ triggers the abort event is bounded by the $\EUFCMA$ security of $\SanSig$: \[\Adv{}{G_1} \leq \Adv{}{G_2}+\Adv{\EUFCMA}{\bdv_6, \SanSig}.\]

\textbf{Game 3} : In this game, we guess the index $(j,t) \in \numParties \times \numSessions$ of the $\gNB$ session $\partsess$ that output $\mathbf{M_1}$ received by $\testsess$, introducing a factor of $\numParties \times \numSessions$ in $\adv$'s advantage. \[\Adv{}{G_2} \leq \numParties \cdot\numSessions \Adv{}{G_{3}}.\]

\textbf{Game 4}: Here we introduce another abort event that triggers if $\adv$ sends a Diffie-Hellman public keyshare $g^h$ to the session $\testsess$, i.e. session $\testsess$ receives $g^h$ that was not output from a $\gNB$ session, but instead from $\adv$. Since this trigger event requires the signature $\sigma^*_G$ in $\mathbf{M_1}$ to verify over $g^h$, and by \textbf{Game 2} we already abort if $\sigma^*_G$ comes from $\adv$, it follows that \[\Adv{}{G_3} \leq \Adv{}{G_4}.\]

\textbf{Game 5}: In this game, we replace $g^{hu}$ computed honestly in $\testsess$ with a uniformly random and independent value $\hat{g^{hu}}$. We do so by defining a reduction $\bdv_7$ that initialises a $\DDH$ challenger $\chal_{\DDH}$, and replaces $g^u$, $g^h$ and $g^{hu}$ computed by $\testsess$ and $\partsess$ with the outputs of $\chal_{\DDH}$, $g^a$, $g^b$, $g^c$. We note that if the bit $b$ sampled by $\chal_{\DDH}$ is 1, then $c = ab$, and we are in \textbf{Game 4}. Otherwise, $c \getsr \Zq$, and we are in \textbf{Game 5}. Any $\adv$ that can distinguish \textbf{Game 4} from \textbf{Game 5} can break the $\DDH$ assumption. Thus:  \[\Adv{}{G_4} \leq \Adv{}{G_5}+ \Adv{G,g,q}{\bdv_7 \DDH}.\]

\textbf{Game 6}: In this game we replace the session and encryption keys $sk_i, k_s$ with uniformly random values $\hat{sk_i}, \hat{k_s}$. We do so by defining a reduction $\bdv_8$ that interacts with a $\KDF$ challenger $\chal_{\KDF}$, querying $\chal_{\KDF}$ with $\hat{g^{uh}}$ and replacing the computation of $sk_i,k_s$ in $\testsess$ and $\partsess$ with the outputs from the $\chal_{\KDF}$ $\hat{sk_i}, \hat{k_s}$. Since $sk_i, k_s \gets \KDF(\hat{g^{uh}})$ and by \textbf{Game 5} $\hat{g^{uh}}$ is already uniformly random and independent, this change is sound. Any $\adversary$ that can distinguish \textbf{Game 5} from \textbf{Game 6} can be used to break $\KDF$ security of the $\KDF$ scheme. Thus:  \[\Adv{}{G_5} \leq \Adv{}{G_6}+ \Adv{\KDF}{\bdv_8,\KDF}.\]

\textbf{Game 7}: In this game, we introduce an abort event that triggers if $\testsess$ decrypts $\mathbf{M_7}$ (keyed by $\hat{k_s}$), but $\mathbf{M_7}$ was not output by $\partsess$. We do so by defining a reduction $\bdv_9$ that initialises an $\Enc$ challenger $\chal_{\AuthEnc}$, which $\bdv_9$ queries when $\testsess$ needs to encrypt or decrypt with $\hat{k_s}$. The abort event only triggers if $\adv$ can produce a valid ciphertext that decrypts under $\hat{k_s}$, and we can submit $\mathbf{M_7}$ to $\chal_{\AuthEnc}$, breaking the $\AuthEnc$ security of the $\Enc$ scheme. By \textbf{Game 6} $\hat{k_s}$ is already uniformly random and independent, and this replacement is sound. Any $\adversary$ that can trigger the abort event can be used by $\bdv_9$ to break the $\AuthEnc$ security of $\Enc$. This implies: \[\Adv{}{G_6} \leq \Adv{}{G_7}+ \Adv{\AuthEnc}{\bdv_9,\Enc}.\]

\textbf{Game 8}: Here we introduce an abort event, where $\chal$ aborts if $\testsess$ receives a message $\mathbf{M_{A_1}}$ without setting $\testsess.\status \gets \rejectflag$ but $\mathbf{M_{A_1}}$ was not output by a session owned by $\AuC$. We do so by defining a reduction $\bdv_{10}$ that initialises a $\SanSig$ challenger $\chal_{\SanSig}$, that outputs $\pk_{sig}^{\challenger}$ and  $\pk_{san}^{\challenger}$, which we embed into the $\AuC$'s $\pk_{sig}^{\AuC}$  and $\UE_i$'s $\pk_{san}^{\UE}$ respectively. Anytime $\AuC$ or $\UE$ needs to generate a signature over a message $m$, $\bdv_{10}$ instead queries $\chal$ with $m$. Now, if $\testsess$ receives a message $\mathbf{M_{A_1}}$ without setting $\testsess.\status \gets \rejectflag$ but $\mathbf{M_{A_1}}$ was not output by a session owned by $\AuC$, then $\adv$ must have produced a message $\mathbf{M_{A_1}} = \AuthEnc.\Enc(k_i,\sigma_U\|C_U\|T_{ID}^*\|\omega\|v)$ such that $\SanSig.\Verify(C_U,\sigma_U,\pk_{sig}^{\chal},\pk_{san}^{\chal})=1$, which is a valid signature forgery. $\bdv_{10}$ responds to $\chal_{\SanSig}$ with $C_U,\sigma_U$ and triggers the abort event. 
Thus, the probability that $\bdv_{10}$ triggers the abort event is bounded by the $\EUFCMA$ security of $\SanSig$: \[\Adv{}{G_7} \leq \Adv{}{G_8}+\Adv{\EUFCMA}{\bdv_{10}, \SanSig}.\]

\textbf{Game 9}: In this game, the session $\testsess$ will only accept $\mathbf{M_1}$ and $\mathbf{M_7}$ from $\gNB$ and $\mathbf{M_{A_1}}$ from $\AuC$ if they have matching subsets and sessions. Thus the advantage of $\adv$ in winning the $\MA$-security experiment is negligible. 
    \[\Adv{}{G_9}=0.
    \]
\fi

%%%%%%%%%%%%%%%%%%%%%%%%%%%%%%%%%%%%%%%%%%%%%%%%%%%%
%%%%%%%%%%%%%%%% MA SECURITY %%%%%%%%%%%%%%%%%%%%%%%
%%%%%%%%%%%%% INITIAL AUTHENTICATION %%%%%%%%%%%%%%%
%%%%%%%%%%%%%%%%% CASE 2.1 %%%%%%%%%%%%%%%%%%%%%%%%%

We turn to bound the advantage of $\adv$ in \textbf{Case 2}, where We assume the first $\cleanpredicate$ session $\testsess$ to accept without a matching session or subset sets $\testsess.\role =\AuC$. 

\textbf{Case 2.1:} According to the definition of this case, $\AuC$ either accepts messages $\mathbf{M_3}$, $\mathbf{M_5}$ or $\mathbf{ACK''}$ without a matching subset (i.e. without honest $\gNB$ partner), or $\mathbf{M_{A_0}}$, $ACK$ without a matching session identifier (i.e. without honest $\UE$).  In this subcase $\adv$ cannot corrupt the long-term $\UE$ symmetric key or the $\testsess.\pid_{\gNB}$ asymmetric key.

\ifsubmissionversion
On a high level, we show that $\adv$ cannot inject DH public keys between the $\gNB$ and the $\UE$ due to the $\gNB$ signatures over these values, based on the $\EUFCMA$ security of the $\SanSig$ scheme. As the secret DH output is used to derive keys for the authenticated encryption scheme securely, and all other messages exchanged between the $\gNB$ and the $\UE$ are encrypted, $\adv$ cannot modify messages sent between the $\gNB$ and the $\UE$ due to the $\AuthEnc$ security of the $\Enc$ scheme. Similarly, we show that $\adv$ cannot inject DH public keys between the $\gNB$ and the $\AuC$ due to the $\gNB$ signatures and the $\AuC$ signatures over these values, based on the $\EUFCMA$ security of the $\SanSig$ scheme. As the secret DH output is used to derive keys for the authenticated encryption scheme securely, and all other messages exchanged between the $\gNB$ and the $\AuC$ are encrypted, $\adv$ cannot modify messages sent between the $\gNB$ and the $\AuC$ due to the $\AuthEnc$ security of the $\Enc$ scheme. As the $\gNB$ proxies ciphertexts between the $\UE$ and the $\AuC$, this means that $\adv$ cannot modify any messages between $\UE$ and $\AuC$.

\fi

\iffullversion
\textbf{Case 2.1:} According to the definition of this case, $\AuC$ either accepts messages $\mathbf{M_3}$, $\mathbf{M_5}$ or $\mathbf{ACK''}$ without a matching subset (i.e. without honest $\gNB$ partner), or $\mathbf{M_{A_0}}$, $ACK$ without a matching session identifier (i.e. without honest $\UE$).  In this subcase $\adv$ cannot corrupt the long-term $\UE$ symmetric key or the $\testsess.\pid_{\gNB}$ asymmetric key.
Here we provide the security analysis:

\textbf{Game 0}: This is the original mutual authentication game described in \ref{ma-sec}: \[\Adv{\MA,\cleanpredicate,\mathsf{C2.1}}{\Prot,\numParties,\numSessions,\adversary}(\secpar) \leq \Adv{}{G_0}.\]

\textbf{Game 1} : In this game, we guess the index $s \in \numSessions$ of the first $\AuC$ session that accepts without a matching session or subset, introducing a factor of $\numSessions$ in $\adv$'s advantage: \[\Adv{}{G_0} \leq \numSessions \Adv{}{G_{1}}.\]

\textbf{Game 2}: Here we introduce an abort event, where $\chal$ aborts if $\testsess$ receives a message $\mathbf{M_3}$ without setting $\testsess.\status \gets \rejectflag$ but $\mathbf{M_3}$ was not output by a session owned by $\testsess.\pid_{\gNB}$. We do so by defining a reduction $\bdv_{1}$ that initialises a $\SanSig$ challenger $\chal_{\SanSig}$, that outputs $\pk_{sig}^{\challenger}$ and  $\pk_{san}^{\challenger}$, which we embed into the $\AuC$'s $\pk_{sig}^{\AuC}$  and $\gNB$'s $\pk_{san}^{\gNB}$ respectively. Anytime $\gNB$ needs to generate a signature over a message $m$, $\bdv_{1}$ instead queries $\chal$ with $m$. Now, if $\testsess$ receives a message $\mathbf{M_3}$ without setting $\testsess.\status \gets \rejectflag$ but $\mathbf{M_3}$ (resp. $\mathbf{M_5}$ was not output by a session owned by $\testsess.\pid_{\gNB}$, then $\adv$ must have produced a message $\mathbf{M_3} = C^*_G,\sigma^*_G,g^h$ such that $\SanSig.\Verify(C^*_G,\sigma^*_G,\pk_{sig}^{\chal},\pk_{san}^{\chal})=1$, which is a valid signature forgery. $\bdv_{1}$ responds to $\chal_{\SanSig}$ with $C^*_G,\sigma^*_G$ and triggers the abort event. 
Thus, the probability that $\bdv_{1}$ triggers the abort event is bounded by the $\EUFCMA$ security of $\SanSig$: \[\Adv{}{G_1} \leq \Adv{}{G_2}+\Adv{\EUFCMA}{\bdv_{1},\SanSig}.\]

\textbf{Game 3} : In this game, we guess the index $(j,t) \in \numParties \times \numSessions$ of the $\gNB$ session $\partsess$ that output $\mathbf{M_3}$ received by $\testsess$, introducing a factor of $\numParties \times \numSessions$ in $\adv$'s advantage. \[\Adv{}{G_2} \leq \numParties \cdot\numSessions \Adv{}{G_{3}}.\]

\textbf{Game 4}: Here we introduce another abort event that triggers if $\adv$ sends a Diffie-Hellman public keyshare $g^a$ to the session $\testsess$, i.e. session $\testsess$ receives $g^a$ that was not output from a $\gNB$ session, but instead from $\adv$. Since this trigger event requires the signature $\sigma^*_G$ in $\mathbf{M_3}$ to verify over $g^a$, and by \textbf{Game 2} we already abort if $\sigma^*_G$ comes from $\adv$, it follows that \[\Adv{}{G_3} \leq \Adv{}{G_4}.\]

\textbf{Game 5}: In this game, we replace $g^{ab}$ computed honestly in $\testsess$ with a uniformly random and independent value $\hat{g^{ab}}$. We do so by defining a reduction $\bdv_{2}$ that initialises a $\DDH$ challenger $\chal_{\DDH}$, and replaces $g^a$, $g^b$ and $g^{ab}$ computed by $\testsess$ and $\partsess$ with the outputs of $\chal_{\DDH}$. We note that if the bit $b$ sampled by $\chal_{\DDH}$ is 1, then $\hat{g^{ab}} =g^{ab}$ and we are in \textbf{Game 4}, otherwise $\hat{ab} \getsr \Zq$ and we are in \textbf{Game 5}. Any $\adv$ that can distinguish \textbf{Game 4} from \textbf{Game 5} can break the $\DDH$ assumption. Thus:  \[\Adv{}{G_4} \leq \Adv{}{G_5}+ \Adv{G,g,q}{\bdv_{2} \DDH}.\]

\textbf{Game 6}: In this game we replace the session and encryption keys $k_s'$ with uniformly random values $\hat{k_s}'$. We do so by defining a reduction $\bdv_{3}$ that interacts with a $\KDF$ challenger $\chal_{\KDF}$, querying $\chal_{\KDF}$ with $\hat{g^{ab}}$ and replacing the computation of $k_s'$ in $\testsess$ and $\partsess$ with the outputs from the $\chal_{\KDF}$ $\hat{k_s}'$. Since $k_s \gets \KDF(\hat{g^{ab}})$ and by \textbf{Game 5} $\hat{g^{ab}}$ is already uniformly random and independent, this change is sound. Any $\adversary$ that can distinguish \textbf{Game 5} from \textbf{Game 6} can be used to break $\KDF$ security of the $\KDF$ scheme. Thus:  \[\Adv{}{G_5} \leq \Adv{}{G_6}+ \Adv{\KDF}{\bdv_{3},\KDF}.\]

\textbf{Game 7}: In this game, we introduce an abort event that triggers if $\testsess$ decrypts $\mathbf{M_5}$ (keyed by $\hat{k_s}'$), but $\mathbf{M_5}'$ was not output by $\partsess$. We do so by defining a reduction $\bdv_{4}$ that initialises an $\Enc$ challenger $\chal_{\AuthEnc}$, which $\bdv_{4}$ queries when $\testsess$ needs to encrypt or decrypt with $\hat{k_s}'$. The abort event only triggers if $\adv$ can produce a valid ciphertext that decrypts under $\hat{k_s}'$, and we can submit $\mathbf{M_5}$ to $\chal_{\AuthEnc}$, breaking the $\AuthEnc$ security of the $\Enc$ scheme. By \textbf{Game 6} $\hat{k_s}'$ is already uniformly random and independent, and this replacement is sound. Any $\adversary$ that can trigger the abort event can be used by $\bdv_{4}$ to break the $\AuthEnc$ security of $\Enc$. This implies: \[\Adv{}{G_6} \leq \Adv{}{G_7}+ \Adv{\AuthEnc}{\bdv_{4},\Enc}.\]

\textbf{Game 8} : In this game, we guess the index $i \in \numParties$ of the $\UE_i$ key $k_i$ $\testsess$ uses to decrypt $\mathbf{M_{A_0}}$, introducing a factor of $\numParties$ in $\adv$'s advantage: \[\Adv{}{G_7} \leq \numParties \Adv{}{G_{8}}.\]

\textbf{Game 9}: In this game, we introduce an abort event that triggers if $\testsess$ decrypts $\mathbf{M_{A_0}}$, $\mathbf{ACK}$ (all keyed by $k_i$), but $\mathbf{M_{A_0}}$ or $\mathbf{ACK}$ was not output by an honest $\UE$ session. We do so by defining a reduction $\bdv_{5}$ that initialises an $\Enc$ challenger $\chal_{\AuthEnc}$, which $\bdv_{5}$ queries when $\bdv_{5}$ needs to encrypt or decrypt with $k_i$. The abort event only triggers if $\adv$ can produce a valid ciphertext that decrypts under $k_i$, and we can submit $\mathbf{M_{A_0}}$ or $\mathbf{ACK}$ to $\chal_{\AuthEnc}$, breaking the $\AuthEnc$ security of the $\Enc$ scheme. By the definition of the execution environment, $k_i$ is generated uniformly random and independent,o and by the definition of \textbf{Case 2.1} $\adv$ cannot issue $\CorruptLTK(i,UE)$ and this replacement is sound. Any $\adversary$ that can trigger the abort event can be used by $\bdv_{5}$ to break the $\AuthEnc$ security of $\Enc$. This implies: \[\Adv{}{G_8} \leq \Adv{}{G_9}+ \Adv{\AuthEnc}{\bdv_{5},\Enc}.\]
 
     %\item [] 
\textbf{Game 10}: In this game, the $\testsess$ only accepts $\mathbf{M_{3}}$, $\mathbf{M_5}$ and $\mathbf{ACK''}$ from an matching subset $\gNB$ and $\mathbf{M_{A_0}}$, $\mathbf{ACK}$ from an matching session $\UE$. Thus, summing the probabilities, we find that the $\adv$ has a negligible advantage in winning the $\MA$-security experiment.: 
        \[\Adv{}{G_9}=0 \]
\fi

We turn to bound the advantage of $\adv$ in \textbf{Case 2.2}.

%%%%%%%%%%%%%%%%%%%%%%%%%%%%%%%%%%%%%%%%%%%%%%%%%%%%
%%%%%%%%%%%%%%%% MA SECURITY %%%%%%%%%%%%%%%%%%%%%%%
%%%%%%%%%%%%% INITIAL AUTHENTICATION %%%%%%%%%%%%%%%
%%%%%%%%%%%%%%%%% CASE 2.2 %%%%%%%%%%%%%%%%%%%%%%%%%

\textbf{Case 2.2:} According to the definition of this case, $\AuC$ either accepts messages $\mathbf{M_3}$, $\mathbf{M_5}$ or $\mathbf{ACK''}$ without a matching subset (i.e. without honest $\gNB$ partner), or $\mathbf{M_{A_0}}$, $ACK$ without a matching session identifier (i.e. without honest $\UE$).  In this subcase, $\adv$ cannot corrupt the long-term asymmetric keys of $\UE$ or the $\testsess.\pid_{\gNB}$ asymmetric key.

\ifsubmissionversion
On a high level, we show that $\adv$ cannot inject DH public keys between the $\gNB$ and the $\UE$ due to the $\gNB$ and $\UE$ signatures over these values, based on the $\EUFCMA$ security of the $\SanSig$ scheme. As the secret DH output is used to derive keys for the authenticated encryption scheme securely, and all other messages exchanged between the $\gNB$ and the $\UE$ are encrypted, $\adv$ cannot modify messages sent between the $\gNB$ and the $\UE$ due to the $\AuthEnc$ security of the $\Enc$ scheme. Similarly, we show that $\adv$ cannot inject DH public keys between the $\gNB$ and the $\AuC$ due to the $\gNB$ signatures over these values, based on the $\EUFCMA$ security of the $\SanSig$ scheme. As the secret DH output is used to derive keys for the authenticated encryption scheme securely, and all other messages exchanged between the $\gNB$ and the $\AuC$ are encrypted, $\adv$ cannot modify messages sent between the $\gNB$ and the $\AuC$ due to the $\AuthEnc$ security of the $\Enc$ scheme. As the $\gNB$ proxies ciphertexts between the $\UE$ and the $\AuC$, this means that $\adv$ cannot modify any messages between $\UE$ and $\AuC$.
\fi

\iffullversion
Here we provide the security analysis:

\textbf{Game 0}: This is the original mutual authentication game described in \ref{ma-sec}: \[\Adv{\MA,\cleanpredicate,\mathsf{C2.1}}{\Prot,\numParties,\numSessions,\adversary}(\secpar) \leq \Adv{}{G_0}.\]

\textbf{Game 1} : In this game, we guess the index $s \in \numSessions$ of the first $\AuC$ session that accepts without a matching session or subset, introducing a factor of $\numSessions$ in $\adv$'s advantage: \[\Adv{}{G_0} \leq \numSessions \Adv{}{G_{1}}.\]

\textbf{Game 2}: Here we introduce an abort event, where $\chal$ aborts if $\testsess$ receives a message $\mathbf{M_3}$ without setting $\testsess.\status \gets \rejectflag$ but $\mathbf{M_3}$ was not output by a session owned by $\testsess.\pid_{\gNB}$. We do so by defining a reduction $\bdv_{1}$ that initialises a $\SanSig$ challenger $\chal_{\SanSig}$, that outputs $\pk_{sig}^{\challenger}$ and  $\pk_{san}^{\challenger}$, which we embed into the $\AuC$'s $\pk_{sig}^{\AuC}$  and $\gNB$'s $\pk_{san}^{\gNB}$ respectively. Anytime $\gNB$ needs to generate a signature over a message $m$, $\bdv_{1}$ instead queries $\chal$ with $m$. Now, if $\testsess$ receives a message $\mathbf{M_3}$ without setting $\testsess.\status \gets \rejectflag$ but $\mathbf{M_3}$ (resp. $\mathbf{M_5}$ was not output by a session owned by $\testsess.\pid_{\gNB}$, then $\adv$ must have produced a message $\mathbf{M_3} = C^*_G,\sigma^*_G,g^h$ such that $\SanSig.\Verify(C^*_G,\sigma^*_G,\pk_{sig}^{\chal},\pk_{san}^{\chal})=1$, which is a valid signature forgery. $\bdv_{1}$ responds to $\chal_{\SanSig}$ with $C^*_G,\sigma^*_G$ and triggers the abort event. 
Thus, the probability that $\bdv_{1}$ triggers the abort event is bounded by the $\EUFCMA$ security of $\SanSig$: \[\Adv{}{G_1} \leq \Adv{}{G_2}+\Adv{\EUFCMA}{\bdv_{1},\SanSig}.\]

\textbf{Game 3} : In this game, we guess the index $(j,t) \in \numParties \times \numSessions$ of the $\gNB$ session $\partsess$ that output $\mathbf{M_3}$ received by $\testsess$, introducing a factor of $\numParties \times \numSessions$ in $\adv$'s advantage. \[\Adv{}{G_2} \leq \numParties \cdot\numSessions \Adv{}{G_{3}}.\]

\textbf{Game 4}: Here we introduce another abort event that triggers if $\adv$ sends a Diffie-Hellman public keyshare $g^a$ to the session $\testsess$, i.e. session $\testsess$ receives $g^a$ that was not output from a $\gNB$ session, but instead from $\adv$. Since this trigger event requires the signature $\sigma^*_G$ in $\mathbf{M_3}$ to verify over $g^a$, and by \textbf{Game 2} we already abort if $\sigma^*_G$ comes from $\adv$, it follows that \[\Adv{}{G_3} \leq \Adv{}{G_4}.\]

\textbf{Game 5}: In this game, we replace $g^{ab}$ computed honestly in $\testsess$ with a uniformly random and independent value $\hat{g^{ab}}$. We do so by defining a reduction $\bdv_{2}$ that initialises a $\DDH$ challenger $\chal_{\DDH}$, and replaces $g^a$, $g^b$ and $g^{ab}$ computed by $\testsess$ and $\partsess$ with the outputs of $\chal_{\DDH}$. We note that if the bit $b$ sampled by $\chal_{\DDH}$ is 1, then $\hat{g^{ab}} =g^{ab}$ and we are in \textbf{Game 4}, otherwise $\hat{ab} \getsr \Zq$ and we are in \textbf{Game 5}. Any $\adv$ that can distinguish \textbf{Game 4} from \textbf{Game 5} can break the $\DDH$ assumption. Thus:  \[\Adv{}{G_4} \leq \Adv{}{G_5}+ \Adv{G,g,q}{\bdv_{2} \DDH}.\]

\textbf{Game 6}: In this game we replace the encryption keys $k_s'$ with uniformly random values $\hat{k_s}'$. We do so by defining a reduction $\bdv_{3}$ that interacts with a $\KDF$ challenger $\chal_{\KDF}$, querying $\chal_{\KDF}$ with $\hat{g^{ab}}$ and replacing the computation of $k_s'$ in $\testsess$ and $\partsess$ with the outputs from the $\chal_{\KDF}$ $\hat{k_s}'$. Since $k_s \gets \KDF(\hat{g^{ab}})$ and by \textbf{Game 5} $\hat{g^{ab}}$ is already uniformly random and independent, this change is sound. Any $\adversary$ that can distinguish \textbf{Game 5} from \textbf{Game 6} can be used to break $\KDF$ security of the $\KDF$ scheme. Thus:  \[\Adv{}{G_5} \leq \Adv{}{G_6}+ \Adv{\KDF}{\bdv_{3},\KDF}.\]

\textbf{Game 7}: In this game, we introduce an abort event that triggers if $\testsess$ decrypts $\mathbf{M_5}$ (keyed by $\hat{k_s}'$), but $\mathbf{M_5}'$ was not output by $\partsess$. We do so by defining a reduction $\bdv_{4}$ that initialises an $\Enc$ challenger $\chal_{\AuthEnc}$, which $\bdv_{4}$ queries when $\testsess$ needs to encrypt or decrypt with $\hat{k_s}'$. The abort event only triggers if $\adv$ can produce a valid ciphertext that decrypts under $\hat{k_s}'$, and we can submit $\mathbf{M_5}$ to $\chal_{\AuthEnc}$, breaking the $\AuthEnc$ security of the $\Enc$ scheme. By \textbf{Game 6} $\hat{k_s}'$ is already uniformly random and independent, and this replacement is sound. Any $\adversary$ that can trigger the abort event can be used by $\bdv_{4}$ to break the $\AuthEnc$ security of $\Enc$. This implies: \[\Adv{}{G_6} \leq \Adv{}{G_7}+ \Adv{\AuthEnc}{\bdv_{4},\Enc}.\]

\textbf{Game 8} : In this game, we guess the index $l \in \numParties$ of the $\UE_l$ party $\testsess$ communicates with, such that $\testsess.\pid_{\UE}=l$, introducing a factor of $\numParties$ in $\adv$'s advantage: \[\Adv{}{G_7} \leq \numParties \Adv{}{G_{8}}.\]

\textbf{Game 9}: Here we introduce an abort event, where $\chal$ aborts if $\partsess$ receives a message $\mathbf{M_2}$, but $\mathbf{M_2}$ was not output by a session owned by $\testsess.\pid_{\UE}$. We do so by defining a reduction $\bdv_{5}$ that initialises a $\mathbf{Signature}$ challenger $\chal_{\mathbf{Sign}}$, that outputs $\pk_{sig}^{\challenger}$, which we embed into the $\UE_l$'s $\pk_{sig}^{\UE}$. Anytime $\UE$ needs to generate a signature over a message $m$, $\bdv_{5}$ instead queries $\chal$ with $m$. Now, if $\partsess$ receives a message $\mathbf{M_2}$ where $\mathbf{M_2}$ was not output by a session owned by $\testsess.\pid_{\UE}$, then $\adv$ must have produced a message $T_{ID}\|\mathbf{M_{A_0}}\|g^u$ such that $\mathbf{Sign}.\Verify(T_{ID}\|\mathbf{M_{A_0}}\|g^u,\sigma,\pk_{sig}^{\chal})=1$, which is a valid signature forgery. $\bdv_{5}$ responds to $\chal_{\mathbf{Sign}}$ with $T_{ID}\|\mathbf{M_{A_0}}\|g^u,\sigma$ and triggers the abort event. 
Thus, the probability that $\bdv_{5}$ triggers the abort event is bounded by the $\EUFCMA$ security of $\mathbf{Sign}$: \[\Adv{}{G_8} \leq \Adv{}{G_9}+\Adv{\EUFCMA}{\bdv_{5},\mathbf{Sign}}.\]

\textbf{Game 10} : In this game, we guess the index $v \in \numParties$ of the $\UE_l$ session that $\partsess$ receives $\mathbf{M_2}$ from, introducing a factor of $\numSessions$ in $\adv$'s advantage: \[\Adv{}{G_9} \leq \numParties \Adv{}{G_{10}}.\]

\textbf{Game 11}: In this game, we replace $g^{uh}$ computed honestly in $\partsess$ and $\session^{l}_{v}$ with a uniformly random and independent value $\hat{g^{uh}}$. We do so by defining a reduction $\bdv_{6}$ that initialises a $\DDH$ challenger $\chal_{\DDH}$, and replaces $g^u$, $g^h$ and $g^{uh}$ computed by $\testsess$ and $\partsess$ with the outputs of $\chal_{\DDH}$, $g^a$, $g^b$, $g^c$. We note that if the bit $b$ sampled by $\chal_{\DDH}$ is 1, then $\hat{g^{uh}} =g^{ab}$ and we are in \textbf{Game 10}, otherwise $c \getsr \Zq$ and we are in \textbf{Game 11}. Any $\adv$ that can distinguish \textbf{Game 10} from \textbf{Game 11} can break the $\DDH$ assumption. Thus:  \[\Adv{}{G_{10}} \leq \Adv{}{G_{11}}+ \Adv{G,g,q}{\bdv_{6} \DDH}.\]

\textbf{Game 12}: In this game we replace the session and encryption keys $sk_i, k_s$ with uniformly random values $\hat{sk_i}, \hat{k_s}$ in $\partsess$ and $\session^{l}_{v}$. We do so by defining a reduction $\bdv_7$ that interacts with a $\KDF$ challenger $\chal_{\KDF}$, querying $\chal_{\KDF}$ with $\hat{g^{uh}}$ and replacing the computation of $sk_i,k_s$ in $\session^l_v$ and $\partsess$ with the outputs from the $\chal_{\KDF}$ $\hat{sk_i}, \hat{k_s}$. Since $sk_i, k_s \gets \KDF(\hat{g^{uh}})$ and by \textbf{Game 11} $\hat{g^{uh}}$ is already uniformly random and independent, this change is sound. Any $\adversary$ that can distinguish \textbf{Game 11} from \textbf{Game 12} can be used to break $\KDF$ security of the $\KDF$ scheme. Thus:  \[\Adv{}{G_{11}} \leq \Adv{}{G_{12}}+ \Adv{\KDF}{\bdv_7,\KDF}.\]

\textbf{Game 13}: In this game, we introduce an abort event that triggers if $\partsess$ decrypts $\mathbf{ACK'}$ (keyed by $\hat{k_s}$), but $\mathbf{ACK'}$ was not output by $\session^{l}_{v}$. We do so by defining a reduction $\bdv_8$ that initialises an $\Enc$ challenger $\chal_{\AuthEnc}$, which $\bdv_8$ queries when $\testsess$ needs to encrypt or decrypt with $\hat{k_s}$. The abort event only triggers if $\adv$ can produce a valid ciphertext that decrypts under $\hat{k_s}$, and we can submit $\mathbf{ACK'}$ to $\chal_{\AuthEnc}$, breaking the $\AuthEnc$ security of the $\Enc$ scheme. By \textbf{Game 12} $\hat{k_s}$ is already uniformly random and independent and this replacement is sound. Any $\adversary$ that can trigger the abort event can be used by $\bdv_8$ to break the $\AuthEnc$ security of $\Enc$. This implies: \[\Adv{}{G_{12}} \leq \Adv{}{G_{13}}+ \Adv{\AuthEnc}{\bdv_8,\Enc}.\]

\textbf{Game 14}: In this game, all messages sent to $\partsess$ come from an $\UE$ session without modification. Similarly, $\testsess$ only accepts $\mathbf{M_3},\mathbf{M_5}, \mathbf{ACK''}$ from an matching subset session $\partsess$. Thus, it follows that all messages added to the session identifier by $\testsess$ come from some matching $\UE$ session $\session^{l}_{v}$. Thus we find that the $\adv$ has a negligible advantage in winning the $\MA$-security experiment.: 
        \[\Adv{}{G_14}=0
        \]
\fi

We turn to bound the advantage of $\adv$ in \textbf{Case 3}, where assume that the first $\cleanpredicate$ session $\testsess$  to accept without a matching session or a matching subset is a $\gNB$ (i.e. $\testsess.\role =\gNB$). 

%%%%%%%%%%%%%%%%%%%%%%%%%%%%%%%%%%%%%%%%%%%%%%%%%%%%
%%%%%%%%%%%%%%%% MA SECURITY %%%%%%%%%%%%%%%%%%%%%%%
%%%%%%%%%%%%% INITIAL AUTHENTICATION %%%%%%%%%%%%%%%
%%%%%%%%%%%%%%%%%%% CASE 3 %%%%%%%%%%%%%%%%%%%%%%%%%

\textbf{Case 3:} According to the definition of this case, $\gNB$ either accepts messages $\mathbf{M_2}$, $\mathbf{M_4}$, $\mathbf{M_6}$, or $\mathbf{ACK''}$, without a matching subset (i.e. without honest $\UE$ or $\AuC$ partners).  Note that in this case, $\adv$ cannot corrupt the asymmetric long-term keys of either the $\AuC$ or the $\testsess.\pid$.

\ifsubmissionversion
On a high level, we show that $\adv$ cannot inject DH public keys from the $\UE$ to the $\gNB$ due to the $\UE$ signatures over these values, based on the $\EUFCMA$ security of the $\SanSig$ scheme. As the secret DH output is used to derive keys for the authenticated encryption scheme securely, and all other messages exchanged from the $\UE$ to the $\gNB$ are encrypted, $\adv$ cannot modify these messages due to the $\AuthEnc$ security of the $\Enc$ scheme. Similarly, we show that $\adv$ cannot inject DH public keys from the $\AuC$ to the $\gNB$ due to the $\AuC$ signatures over these values, based on the $\EUFCMA$ security of the $\SanSig$ scheme. As the secret DH output is used to derive keys for the authenticated encryption scheme securely, and all other messages exchanged from the $\UE$ to the $\gNB$ are encrypted, $\adv$ cannot modify these messages due to the $\AuthEnc$ security of the $\Enc$ scheme. this means that $\adv$ cannot modify any messages between $\UE$ and $\AuC$.
\fi

\iffullversion
Here we describe the security analysis:

\textbf{Game 0}: This is the original mutual authentication game described in \ref{ma-sec}: \[\Adv{\MA,\cleanpredicate,\mathsf{C3}}{\Prot,\numParties,\numSessions,\adversary}(\secpar) \leq \Adv{}{G_0}.\]

\textbf{Game 1} : In this game, we guess the index $(i,s) \in \numParties 
\times \numSessions$ of the first $\gNB$ session that accepts without a matching subset, introducing a factor of $\numParties \times \numSessions$ in $\adv$'s advantage: \[\Adv{}{G_0} \leq \numParties \cdot\numSessions \Adv{}{G_{1}}.\]

\textbf{Game 2}: Here we introduce an abort event, where $\chal$ aborts if $\testsess$ receives a message $\mathbf{M_4}$ without setting $\testsess.\status \gets \rejectflag$ but $\mathbf{M_4}$ was not output by a session owned by $\testsess.\pid_{\AuC}$. We do so by defining a reduction $\bdv_{1}$ that initialises a $\SanSig$ challenger $\chal_{\SanSig}$, that outputs $\pk_{sig}^{\challenger}$ and  $\pk_{san}^{\challenger}$, which we embed into the $\AuC$'s $\pk_{sig}^{\AuC}$  and $\gNB$'s $\pk_{san}^{\gNB}$ respectively. Anytime $\AuC$ needs to generate a signature over a message $m$, $\bdv_{1}$ instead queries $\chal$ with $m$. Now, if $\testsess$ receives a message $\mathbf{M_4}$ without setting $\testsess.\status \gets \rejectflag$ but $\mathbf{M_4}$ was not output by a session owned by $\testsess.\pid_{\AuC}$, then $\adv$ must have produced a message $\mathbf{M_4} = C^*_G,\sigma^*_G,g^b$ such that $\SanSig.\Verify(C^*_G,\sigma^*_G,\pk_{sig}^{\chal},\pk_{san}^{\chal})=1$, which is a valid signature forgery. $\bdv_{1}$ responds to $\chal_{\SanSig}$ with $C^*_G,\sigma^*_G$ and triggers the abort event. 
Thus, the probability that $\bdv_{1}$ triggers the abort event is bounded by the $\EUFCMA$ security of $\SanSig$: \[\Adv{}{G_1} \leq \Adv{}{G_2}+\Adv{\EUFCMA}{\bdv_{1},\SanSig}.\]

\textbf{Game 3} : In this game, we guess the index $j \in \numSessions$ of the $\AuC$ session $\partsess$ that output $\mathbf{M_6}$ received by $\testsess$, introducing a factor of $\numSessions$ in $\adv$'s advantage. \[\Adv{}{G_2} \leq \numSessions \Adv{}{G_{3}}.\]

\textbf{Game 4}: Here we introduce another abort event that triggers if $\adv$ sends a Diffie-Hellman public keyshare $g^b$ to the session $\testsess$, i.e. session $\testsess$ receives $g^b$ that was not output from a $\AuC$ session, but instead from $\adv$. Since this trigger event requires the signature $\sigma^*_G$ in $\mathbf{M_4}$ to verify over $g^b$, and by \textbf{Game 2} we already abort if $\sigma^*_G$ comes from $\adv$, it follows that \[\Adv{}{G_3} \leq \Adv{}{G_4}.\]

\textbf{Game 5}: In this game, we replace $g^{ab}$ computed honestly in $\testsess$ with a uniformly random and independent value $\hat{g^{ab}}$. We do so by defining a reduction $\bdv_{2}$ that initialises a $\DDH$ challenger $\chal_{\DDH}$, and replaces $g^a$, $g^b$ and $g^{ab}$ computed by $\testsess$ and $\partsess$ with the outputs of $\chal_{\DDH}$. We note that if the bit $b$ sampled by $\chal_{\DDH}$ is 1, then $\hat{g^{ab}} =g^{ab}$ and we are in \textbf{Game 4}, otherwise $\hat{ab} \getsr \Zq$ and we are in \textbf{Game 5}. Any $\adv$ that can distinguish \textbf{Game 4} from \textbf{Game 5} can break the $\DDH$ assumption. Thus:  \[\Adv{}{G_4} \leq \Adv{}{G_5}+ \Adv{G,g,q}{\bdv_{2} \DDH}.\]

\textbf{Game 6}: In this game we replace the encryption keys $k_s'$ with uniformly random values $\hat{k_s}'$. We do so by defining a reduction $\bdv_{3}$ that interacts with a $\KDF$ challenger $\chal_{\KDF}$, querying $\chal_{\KDF}$ with $\hat{g^{ab}}$ and replacing the computation of $k_s'$ in $\testsess$ and $\partsess$ with the outputs from the $\chal_{\KDF}$ $\hat{k_s}'$. Since $k_s \gets \KDF(\hat{g^{ab}})$ and by \textbf{Game 5} $\hat{g^{ab}}$ is already uniformly random and independent, this change is sound. Any $\adversary$ that can distinguish \textbf{Game 5} from \textbf{Game 6} can be used to break $\KDF$ security of the $\KDF$ scheme. Thus:  \[\Adv{}{G_5} \leq \Adv{}{G_6}+ \Adv{\KDF}{\bdv_{3},\KDF}.\]

\textbf{Game 7}: In this game, we introduce an abort event that triggers if $\testsess$ decrypts $\mathbf{M_6}$ (keyed by $\hat{k_s}'$), but $\mathbf{M_6}'$ was not output by $\partsess$. We do so by defining a reduction $\bdv_{4}$ that initialises an $\Enc$ challenger $\chal_{\AuthEnc}$, which $\bdv_{4}$ queries when $\testsess$ needs to encrypt or decrypt with $\hat{k_s}'$. The abort event only triggers if $\adv$ can produce a valid ciphertext that decrypts under $\hat{k_s}'$, and we can submit $\mathbf{M_6}$ to $\chal_{\AuthEnc}$, breaking the $\AuthEnc$ security of the $\Enc$ scheme. By \textbf{Game 6} $\hat{k_s}'$ is already uniformly random and independent, and this replacement is sound. Any $\adversary$ that can trigger the abort event can be used by $\bdv_{4}$ to break the $\AuthEnc$ security of $\Enc$. This implies: \[\Adv{}{G_6} \leq \Adv{}{G_7}+ \Adv{\AuthEnc}{\bdv_{4},\Enc}.\]

\textbf{Game 8} : In this game, we guess the index $l \in \numParties$ of the $\UE_l$ party $\testsess$ communicates with, such that $\testsess.\pid_{\UE}=l$, introducing a factor of $\numParties$ in $\adv$'s advantage: \[\Adv{}{G_7} \leq \numParties \Adv{}{G_{8}}.\]

\textbf{Game 9}: Here we introduce an abort event, where $\chal$ aborts if $\partsess$ receives a message $\mathbf{M_2}$, but $\mathbf{M_2}$ was not output by a session owned by $\testsess.\pid_{\UE}$. We do so by defining a reduction $\bdv_{5}$ that initialises a $\mathbf{Signature}$ challenger $\chal_{\mathbf{Sign}}$, that outputs $\pk_{sig}^{\challenger}$, which we embed into the $\UE_l$'s $\pk_{sig}^{\UE}$. Anytime $\UE$ needs to generate a signature over a message $m$, $\bdv_{5}$ instead queries $\chal$ with $m$. Now, if $\partsess$ receives a message $\mathbf{M_2}$ where $\mathbf{M_2}$ was not output by a session owned by $\testsess.\pid_{\UE}$, then $\adv$ must have produced a message $T_{ID}\|\mathbf{M_{A_0}}\|g^u$ such that $\mathbf{Sign}.\Verify(T_{ID}\|\mathbf{M_{A_0}}\|g^u,\sigma,\pk_{sig}^{\chal})=1$, which is a valid signature forgery. $\bdv_{5}$ responds to $\chal_{\mathbf{Sign}}$ with $T_{ID}\|\mathbf{M_{A_0}}\|g^u,\sigma$ and triggers the abort event. 
Thus, the probability that $\bdv_{5}$ triggers the abort event is bounded by the $\EUFCMA$ security of $\mathbf{Sign}$: \[\Adv{}{G_8} \leq \Adv{}{G_9}+\Adv{\EUFCMA}{\bdv_{5},\mathbf{Sign}}.\]

\textbf{Game 10} : In this game, we guess the index $v \in \numParties$ of the $\UE_l$ session that $\partsess$ receives $\mathbf{M_2}$ from, introducing a factor of $\numSessions$ in $\adv$'s advantage: \[\Adv{}{G_9} \leq \numParties \Adv{}{G_{10}}.\]

\textbf{Game 11}: In this game, we replace $g^{uh}$ computed honestly in $\testsess$ and $\session^{l}_{v}$ with a uniformly random and independent value $\hat{g^{uh}}$. We do so by defining a reduction $\bdv_{6}$ that initialises a $\DDH$ challenger $\chal_{\DDH}$, and replaces $g^u$, $g^h$ and $g^{uh}$ computed by $\testsess$ and $\session^{l}_{v}$ with the outputs of $\chal_{\DDH}$, $g^a$, $g^b$, $g^c$. We note that if the bit $b$ sampled by $\chal_{\DDH}$ is 1, then $\hat{g^{uh}} =g^{ab}$ and we are in \textbf{Game 10}, otherwise $c \getsr \Zq$ and we are in \textbf{Game 11}. Any $\adv$ that can distinguish \textbf{Game 10} from \textbf{Game 11} can break the $\DDH$ assumption. Thus:  \[\Adv{}{G_{10}} \leq \Adv{}{G_{11}}+ \Adv{G,g,q}{\bdv_{6} \DDH}.\]

\textbf{Game 12}: In this game we replace the session and encryption keys $sk_i, k_s$ with uniformly random values $\hat{sk_i}, \hat{k_s}$ in $\testsess$ and $\session^{l}_{v}$. We do so by defining a reduction $\bdv_7$ that interacts with a $\KDF$ challenger $\chal_{\KDF}$, querying $\chal_{\KDF}$ with $\hat{g^{uh}}$ and replacing the computation of $sk_i,k_s$ in $\session^l_v$ and $\partsess$ with the outputs from the $\chal_{\KDF}$ $\hat{sk_i}, \hat{k_s}$. Since $sk_i, k_s \gets \KDF(\hat{g^{uh}})$ and by \textbf{Game 11} $\hat{g^{uh}}$ is already uniformly random and independent, this change is sound. Any $\adversary$ that can distinguish \textbf{Game 11} from \textbf{Game 12} can be used to break $\KDF$ security of the $\KDF$ scheme. Thus:  \[\Adv{}{G_{11}} \leq \Adv{}{G_{12}}+ \Adv{\KDF}{\bdv_7,\KDF}.\]

\textbf{Game 13}: In this game, we introduce an abort event that triggers if $\testsess$ decrypts $\mathbf{ACK'}$ (keyed by $\hat{k_s}$), but $\mathbf{ACK'}$ was not output by $\session^{l}_{v}$. We do so by defining a reduction $\bdv_8$ that initialises an $\Enc$ challenger $\chal_{\AuthEnc}$, which $\bdv_8$ queries when $\testsess$ needs to encrypt or decrypt with $\hat{k_s}$. The abort event only triggers if $\adv$ can produce a valid ciphertext that decrypts under $\hat{k_s}$, and we can submit $\mathbf{ACK'}$ to $\chal_{\AuthEnc}$, breaking the $\AuthEnc$ security of the $\Enc$ scheme. By \textbf{Game 12} $\hat{k_s}$ is already uniformly random and independent, and this replacement is sound. Any $\adversary$ that can trigger the abort event can be used by $\bdv_8$ to break the $\AuthEnc$ security of $\Enc$. This implies: \[\Adv{}{G_{12}} \leq \Adv{}{G_{13}}+ \Adv{\AuthEnc}{\bdv_8,\Enc}.\]

\textbf{Game 14}: In this game, all messages sent to $\testsess$ come from an $\UE$ or $\AuC$ session without modification. Thus, it follows that all messages sent to $\testsess$ come from some matching subset $\UE$ session $\session^{l}_{v}$ or a matching subset $\AuC$ session $\partsess$, and thus we find that the $\adv$ has negligible advantage in winning the $\MA$-security experiment: 
        \[\Adv{}{G_{14}}=0
        \]
\fi
\end{pro}

%%%%%%%%%%%%%%%%%%%%%%%%%%%%%%%%%%%%%%%%%%%%%%%%%%%%
%%%%%%%%%%%%%%%% MA SECURITY %%%%%%%%%%%%%%%%%%%%%%%
%%%%%%%%%%%%% UNIVERSAL HANDOVER %%%%%%%%%%%%%%%%%%%
%%%%%%%%%%%%%%%%%%%%%%%%%%%%%%%%%%%%%%%%%%%%%%%%%%%%

\subsubsection{\texorpdfstring{$\MA$}{Lg}- security of Universal Handover}
\label{MA of Universal Handover}
This section formally analyses the $\MA$-security of the Universal Handover protocol.

\begin{theorem}
\label{theorem: MA sec- HO}
{\textbf{$\mathbf{\MA}$-security of the Universal Handover}}. Universal Handover depicted in Figure \ref{fig:universal-alt} is $\MA$-secure under the cleanness predicate in Definition \ref{ma-clean, uni-HO}. For any PPT algorithm $\adv$ against the $\MA$ experiment, $\Adv{\MA,\cleanpredicate}{\Prot,\numParties,\numSessions,\adversary}(\secpar)$ is negligible assuming the EUFCMA security of $\SanSig$, Auth security of $\AuthEnc$, the KDF security of $\KDF$ and the $\DDH$ assumption.
\end{theorem}
\begin{pro}
Our proof is divided into two cases  $\Adv{\MA,\cleanpredicate, c_1}{\Prot,\numParties,\numSessions,\adversary}(\secpar)$ and $\Adv{\MA,\cleanpredicate, c_2}{\Prot,\numParties,\numSessions,\adversary}(\secpar)$.
\begin{enumerate}
 \item\textbf{ Case 1:} The test session $\testsess$, where $\testsess.\rho =\UE$ accepts messages $\mathbf{M_1}$ and $\mathbf{M_3}$ without an honest matching $\gNB$ partner (no matching subset). \ifsubmissionversion
The analysis of \textbf{Case 1} proceeds similarly to \textbf{Case 1.1} of the initial authentication protocol and thus, for brevity, we omit the proof of this case, which can be found in \cite{authors_unihand_2023}\footnote[1]{The full version of the security analysis and the security framework is available in the Supplementary Material.}.
\fi
   \item\textbf{ Case 2:} The test session $\testsess$, where $\testsess.\rho =\gNB$ accepts message $\mathbf{M_2}$ without an honest matching $\UE$ partner (no matching subset). \ifsubmissionversion
The analysis of \textbf{Case 2} proceeds similarly to \textbf{Case 3} of the initial authentication protocol, and thus, for brevity, here we omit the proof of this case, which can be found in \cite{authors_unihand_2023}\footnotemark[1]{}.
\fi
 
\end{enumerate}

We then bound the advantage of $\adv$ winning the game under certain assumptions to $\Adv{\MA,\cleanpredicate}{\Prot,\numParties,\numSessions,\adversary}(\secpar)\leq(\Adv{\MA,\cleanpredicate, C_1}{\Prot,\numParties,\numSessions,\adversary}(\secpar)+\Adv{\MA,\cleanpredicate, C_2}{\Prot,\numParties,\numSessions,\adversary}(\secpar))$.

\iffullversion

%%%%%%%%%%%%%%%%%%%%%%%%%%%%%%%%%%%%%%%%%%%%%%%%%%%%
%%%%%%%%%%%%%%%% MA SECURITY %%%%%%%%%%%%%%%%%%%%%%%
%%%%%%%%%%%%% UNIVERSAL HANDOVER %%%%%%%%%%%%%%%%%%%
%%%%%%%%%%%%%%%%%%% CASE 1 %%%%%%%%%%%%%%%%%%%%%%%%%

\textbf{Case 1:} According to the definition of this case, $\UE$ accepts messages $\mathbf{M_1}$ and $\mathbf{M_3}$ without an honest matching partner ($\gNB$) (no matching subset). In this case, the $\adv$ cannot corrupt the $\testsess.\pid_{\gNB}$ asymmetric key.
\fi

\iffullversion
Here we provide the security analysis:
\textbf{Game 0}: This is the original mutual authentication game in Definition \ref{ma-sec}: \[\Adv{\MA,\cleanpredicate,\mathsf{C1}}{\Prot,\numParties,\numSessions,\adversary}(\secpar) \leq \Adv{}{G_0}.\]

\textbf{Game 1} : In this game, we guess the index $(i,s) \in \numParties \times \numSessions$ of the first $\UE$ session that accepts without a matching subset, introducing a factor of $\numParties \times \numSessions$ in $\adv$'s advantage: \[\Adv{}{G_0} \leq \numParties \cdot\numSessions\cdot \Adv{}{G_{1}}.\]

\textbf{Game 2} : In this game, we guess the index $j \in \numParties$ of the $\gNB$ party such that $\testsess.\pid = j$, introducing a factor of $\numParties$ in $\adv$'s advantage. \[\Adv{}{G_1} \leq \numParties \Adv{}{G_{2}}.\]

\textbf{Game 3}: Here we introduce an abort event, where $\chal$ aborts if $\testsess$ receives a message $\mathbf{M_1}$ without setting $\testsess.\status \gets \rejectflag$ but $\mathbf{M_1}$ was not output by a session owned by $\testsess.\pid$. We do so by defining a reduction $\bdv_1$ that initialises a $\SanSig$ challenger $\chal_{\SanSig}$, that outputs $\pk_{sig}^{\challenger}$ and  $\pk_{san}^{\challenger}$, which we embed into the $\AuC$'s $\pk_{sig}^{\AuC}$  and $\gNB$'s $\pk_{san}^{\gNB}$ respectively. Anytime $\gNB$ needs to generate a signature over a message $m$, $\bdv_1$ instead queries $\chal$ with $m$. Now, if $\testsess$ receives a message $\mathbf{M_1}$ without setting $\testsess.\status \gets \rejectflag$ but $\mathbf{M_1}$ was not output by a session owned by $\testsess.\pid$, then $\adv$ must have produced a message $\mathbf{M_1} = C^*_G,\sigma^*_G,g^h$ such that $\SanSig.\Verify(C^*_G,\sigma^*_G,\pk_{sig}^{\chal},\pk_{san}^{\chal})=1$, which is a valid signature forgery. $\bdv_1$ responds to $\chal_{\SanSig}$ with $C^*_G,\sigma^*_G$ and triggers the abort event. 
Thus, the probability that $\bdv_1$ triggers the abort event is bounded by the $\EUFCMA$ security of $\SanSig$: \[\Adv{}{G_2} \leq \Adv{}{G_3}+\Adv{\EUFCMA}{\bdv_1, \SanSig}.\]

\textbf{Game 4} : In this game, we guess the index $t \in \numSessions$ of the $\gNB$ session $\partsess$ that output $\mathbf{M_1}$ received by $\testsess$, introducing a factor of $\numSessions$ in $\adv$'s advantage. \[\Adv{}{G_3} \leq \numSessions \cdot\Adv{}{G_{4}}.\]

\textbf{Game 5}: Here we introduce another abort event that triggers if $\adv$ sends a Diffie-Hellman public keyshare $g^h$ to the session $\testsess$, i.e. session $\testsess$ receives $g^h$ that was not output from a $\gNB$ session, but instead from $\adv$. Since this trigger event requires the signature $\sigma^*_G$ in $\mathbf{M_1}$ to verify over $g^h$, and by \textbf{Game 3} we already abort if $\sigma^*_G$ comes from $\adv$, it follows that \[\Adv{}{G_4} \leq \Adv{}{G_5}.\]

\textbf{Game 6}: In this game, we replace $g^{hu}$ computed honestly in $\testsess$ with a uniformly random and independent value $\hat{g^{hu}}$. We do so by defining a reduction $\bdv_2$ that initialises a $\DDH$ challenger $\chal_{\DDH}$, and replaces $g^u$, $g^h$ and $g^{hu}$ computed by $\testsess$ and $\partsess$ with the outputs of $\chal_{\DDH}$, $g^a$, $g^b$, $g^c$. We note that if the bit $b$ sampled by $\chal_{\DDH}$ is 1, then $c = ab$ and we are in \textbf{Game 5}, otherwise, $c \getsr \Zq$ and we are in \textbf{Game 6}. Any $\adv$ that can distinguish \textbf{Game 5} from \textbf{Game 6} can break the $\DDH$ assumption. Thus:  \[\Adv{}{G_5} \leq \Adv{}{G_6}+ \Adv{G,g,q}{\bdv_2 \DDH}.\]

\textbf{Game 7}: In this game we replace the session and encryption keys $sk_i, k_s$ with uniformly random values $\hat{sk_i}, \hat{k_s}$. We do so by defining a reduction $\bdv_3$ that interacts with a $\KDF$ challenger $\chal_{\KDF}$, querying $\chal_{\KDF}$ with $\epsilon$ and replacing the computation of $sk_i,k_s$ in $\testsess$ and $\partsess$ with the outputs from the $\chal_{\KDF}$ $\hat{sk_i}, \hat{k_s}$. Since $sk_i, k_s \gets \KDF(\hat{g^{uh}})$ and by \textbf{Game 6} $\hat{g^{uh}}$ is already uniformly random and independent, this change is sound. Any $\adversary$ that can distinguish \textbf{Game 6} from \textbf{Game 7} can be used to break $\KDF$ security of the $\KDF$ scheme. Thus:  \[\Adv{}{G_6} \leq \Adv{}{G_7}+ \Adv{\KDF}{\bdv_3,\KDF}.\]

\textbf{Game 8}: In this game, we introduce an abort event that triggers if $\testsess$ decrypts $\mathbf{M_3}$ (keyed by $\hat{k_s}$), but $\mathbf{M_3}$ was not output by $\partsess$. We do so by defining a reduction $\bdv_4$ that initialises an $\Enc$ challenger $\chal_{\AuthEnc}$, which $\bdv_4$ queries when $\testsess$ needs to encrypt or decrypt with $\hat{k_s}$. The abort event only triggers if $\adv$ can produce a valid ciphertext that decrypts under $\hat{k_s}$, and we can submit $\mathbf{M_3}$ to $\chal_{\AuthEnc}$, breaking the $\AuthEnc$ security of the $\Enc$ scheme. By \textbf{Game 7} $\hat{k_s}$ is already uniformly random and independent and this replacement is sound. Any $\adversary$ that can trigger the abort event can be used by $\bdv_4$ to break the $\AuthEnc$ security of $\Enc$. This implies: \[\Adv{}{G_7} \leq \Adv{}{G_8}+ \Adv{\AuthEnc}{\bdv_4,\Enc}.\]

     %\item [] 
\textbf{Game 9}: In this game, the $\testsess$ only accepts $\mathbf{M_1}$, $\mathbf{M_3}$ from an honest matching partner. Thus, summing the probabilities we find that the $\adv$ has negligible advantage in winning the $\MA$-security experiment: 
\[\Adv{}{G_9}=0\]
\fi
\iffullversion
\textbf{Case 2:} According to the definition of this case, $\gNB$ accepts messages $\mathbf{M_2}$ without an honest matching partner ($\UE$) (no matching subset). Note that in this case, $\adv$ cannot corrupt the asymmetric long-term keys of the $\testsess.\pid$.
\fi

\iffullversion

Here we analyse \textbf{Case 2}:

\textbf{Game 0}: This is the original mutual authentication game described in Section \ref{ma-sec}: \[\Adv{\MA,\cleanpredicate,\mathsf{C2}}{\Prot,\numParties,\numSessions,\adversary}(\secpar) \leq \Adv{}{G_0}.\]

\textbf{Game 1} : In this game, we guess the index $(i,s) \in \numParties 
\times \numSessions$ of the first $\gNB$ session that accepts without a matching subset, introducing a factor of $\numParties \times \numSessions$ in $\adv$'s advantage: \[\Adv{}{G_0} \leq \numParties \cdot\numSessions \Adv{}{G_{1}}.\]

\textbf{Game 2} : In this game, we guess the index $j \in \numParties$ of the $\UE$ party such that $\testsess.\pid = j$, introducing a factor of $\numParties$ in $\adv$'s advantage. \[\Adv{}{G_1} \leq \numParties \Adv{}{G_{2}}.\]

\textbf{Game 3}: Here we introduce an abort event, where $\chal$ aborts if $\testsess$ receives a message $\mathbf{M_2}$ without setting $\testsess.\status \gets \rejectflag$ but $\mathbf{M_2}$ was not output by a session owned by $\testsess.\pid_{\UE}$. We do so by defining a reduction $\bdv_{1}$ that initialises a $\SanSig$ challenger $\chal_{\SanSig}$, that outputs $\pk_{sig}^{\challenger}$ and  $\pk_{san}^{\challenger}$, which we embed into the $\AuC$'s $\pk_{sig}^{\AuC}$  and $\UE$'s $\pk_{san}^{\UE}$ respectively. Anytime $\UE$ needs to generate a signature over a message $m$, $\bdv_{1}$ instead queries $\chal$ with $m$. Now, if $\testsess$ receives a message $\mathbf{M_2}$ without setting $\testsess.\status \gets \rejectflag$ but $\mathbf{M_2}$ was not output by a session owned by $\testsess.\pid_{\UE}$, then $\adv$ must have produced a message $\mathbf{M_2} = \cert_{U}\|\sigma_U\| \omega_U\|v$ such that $\SanSig.\Verify(\cert^*_{U},\sigma^*_U,\pk_{sig}^{\chal},\pk_{san}^{\chal})=1$, which is a valid signature forgery. $\bdv_{1}$ responds to $\chal_{\SanSig}$ with $\cert^*_{U},\sigma^*_U$ and triggers the abort event. 
Thus, the probability that $\bdv_{1}$ triggers the abort event is bounded by the $\EUFCMA$ security of $\SanSig$: \[\Adv{}{G_2} \leq \Adv{}{G_3}+\Adv{\EUFCMA}{\bdv_{1},\SanSig}.\]

\textbf{Game 4} : In this game, we guess the index $j \in \numSessions$ of the $\UE$ session that $\testsess$ receives $\mathbf{M_2}$ from, introducing a factor of $\numSessions$ in $\adv$'s advantage: \[\Adv{}{G_3} \leq \numSessions \Adv{}{G_{4}}.\]

\textbf{Game 5}: Here we introduce another abort event that triggers if $\adv$ sends a Diffie-Hellman public keyshare $g^u$ to the session $\testsess$, i.e. session $\testsess$ receives $g^u$ that was not output from a $\UE$ session, but instead from $\adv$. Since this trigger event requires the signature $\sigma^*_U$ in $\mathbf{M_2}$ to verify over $g^u$, and by \textbf{Game 2} we already abort if $\sigma^*_U$ comes from $\adv$, it follows that \[\Adv{}{G_4} \leq \Adv{}{G_5}.\]

\textbf{Game 6}: In this game, we replace $g^{uh}$ computed honestly in $\testsess$ and $\partsess$ with a uniformly random and independent value $\hat{g^{uh}}$. We do so by defining a reduction $\bdv_{2}$ that initialises a $\DDH$ challenger $\chal_{\DDH}$, and replaces $g^u$, $g^h$ and $g^{uh}$ computed by $\testsess$ and $\partsess$ with the outputs of $\chal_{\DDH}$, $g^a$, $g^b$, $g^c$. We note that if the bit $b$ sampled by $\chal_{\DDH}$ is 1, then $\hat{g^{uh}} =g^{ab}$ and we are in \textbf{Game 5}, otherwise $c \getsr \Zq$ and we are in \textbf{Game 6}. Any $\adv$ that can distinguish \textbf{Game 5} from \textbf{Game 6} can break the $\DDH$ assumption. Thus:  \[\Adv{}{G_{5}} \leq \Adv{}{G_{6}}+ \Adv{G,g,q}{\bdv_{2} \DDH}.\]

\textbf{Game 7}: In this game we replace the session and encryption keys $sk_i, k_s$ with uniformly random values $\hat{sk_i}, \hat{k_s}$ in $\testsess$ and $\partsess$. We do so by defining a reduction $\bdv_3$ that interacts with a $\KDF$ challenger $\chal_{\KDF}$, querying $\chal_{\KDF}$ with $\hat{g^{uh}}$ and replacing the computation of $sk_i,k_s$ in $\partsess$ and $\testsess$ with the outputs from the $\chal_{\KDF}$ $\hat{sk_i}, \hat{k_s}$. Since $sk_i, k_s \gets \KDF(\hat{g^{uh}})$ and by \textbf{Game 6} $\hat{g^{uh}}$ is already uniformly random and independent, this change is sound. Any $\adversary$ that can distinguish \textbf{Game 6} from \textbf{Game 7} can be used to break $\KDF$ security of the $\KDF$ scheme. Thus:  \[\Adv{}{G_{6}} \leq \Adv{}{G_{7}}+ \Adv{\KDF}{\bdv_3,\KDF}.\]

\textbf{Game 8}: In this game, we introduce an abort event that triggers if $\testsess$ decrypts $\mathbf{M_2}$ (keyed by $\hat{k_s}$), but $\mathbf{M_2}$ was not output by $\partsess$. We do so by defining a reduction $\bdv_4$ that initialises an $\Enc$ challenger $\chal_{\AuthEnc}$, which $\bdv_4$ queries when $\testsess$ needs to encrypt or decrypt with $\hat{k_s}$. The abort event only triggers if $\adv$ can produce a valid ciphertext that decrypts under $\hat{k_s}$, and we can submit $\mathbf{M_2}$ to $\chal_{\AuthEnc}$, breaking the $\AuthEnc$ security of the $\Enc$ scheme. By \textbf{Game 7} $\hat{k_s}$ is already uniformly random and independent and this replacement is sound. Any $\adversary$ that can trigger the abort event can be used by $\bdv_4$ to break the $\AuthEnc$ security of $\Enc$. This implies: \[\Adv{}{G_{7}} \leq \Adv{}{G_{8}}+ \Adv{\AuthEnc}{\bdv_4,\Enc}.\]

     %\item [] 
\textbf{Game 9}: In this game, the $\testsess$ only accepts $\mathit{M_2}$ from an honest matching partner. Thus, summing the probabilities we find that the $\adv$ has negligible advantage in winning the $\MA$-security experiment: 
\[\Adv{}{G_9}=0\]
\fi

\end{pro}

%%%%%%%%%%%%%%%%%%%%%%%%%%%%%%%%%%%%%%%%%%%%%%%%%%%%
%%%%%%%%%%%%%%%% KIND SECURITY %%%%%%%%%%%%%%%%%%%%%
%%%%%%%%%%%%% INITIAL AUTHENTICATION %%%%%%%%%%%%%%%
%%%%%%%%%%%%%%%%%%%%%%%%%%%%%%%%%%%%%%%%%%%%%%%%%%%%

\subsection{Key Indistinguishability}
Here we prove the key indistinguishability of our protocols.
\ifsubmissionversion 
Due to space constraints, we encapsulate both protocols into Theorem \ref{thm:kind-both}.

\begin{theorem}{\textbf{$\mathbf{\KIND}$-security of $\UniHand$}}\label{thm:kind-both}. $\UniHand$ scheme depicted in Figure \ref{fig:init-auth-alt} and Figure \ref{fig:universal-alt}  are $\KIND$-secure under the cleanness predicate (Definition \ref{def:KIND-clean}). For any PPT algorithm $\adv$ against the $\KIND$ experiment, $\Adv{\KIND,\cleanpredicate}{\Prot,\numParties,\numSessions,\adversary}(\secpar)$ is negligible assuming the EUFCMA security of $\SanSig$, Auth security of $\AuthEnc$, the KDF security of $\KDF$ and the $\DDH$ assumption.
\end{theorem}

Our proof is divided into two cases $\Adv{\KIND,\cleanpredicate, c_1}{\Prot,\numParties,\numSessions,\adversary}(\secpar)$ and $\Adv{\KIND,\cleanpredicate, c_2}{\Prot,\numParties,\numSessions,\adversary}(\secpar)$:

\begin{enumerate}[label*=\arabic*.] 
    \item\textbf{Case 1:} The test session $\testsess$ accepts messages without an honest matching session (or matching subset). 
     
    \item \textbf{Case 2:} The test session $\testsess$ accepts messages with a matching session (and subset).
\end{enumerate}

We then bound the advantage of $\adv$ winning the game under certain assumptions to $\Adv{\KIND,\cleanpredicate}{\Prot,\numParties,\numSessions,\adversary}(\secpar)\leq(\Adv{\KIND,\cleanpredicate, C_1}{\Prot,\numParties,\numSessions,\adversary}(\secpar)+\Adv{\KIND,\cleanpredicate, C_2}{\Prot,\numParties,\numSessions,\adversary}(\secpar)$.

\ifsubmissionversion

On a high-level, in \textbf{Case 1} we show that $\adv$'s advantage in causing a session to accept without a matching session (or subset) to be equal to $\adv$ breaking the MA-security of $\UniHand$. In \textbf{Case 2} we show that since the Diffie-Hellman values are exchanged honestly, the DH outputs can be replaced with uniformly random values in $\testsess$, which are in turn used to derive the session key. Hence, the session key is indistinguishable from a uniformly random value from the same distribution (i.e. indistinguishable from the output of a $\Test$ query), and $\adv$ has no advantage in guessing $b$.
\fi

\iffullversion

\subsubsection{$\KIND$-security of Initial Authentication}
This section formally analyses the $\KIND$-security of the Initial Authentication protocol.

\begin{theorem}{\textbf{$\mathbf{\KIND}$-security of Initial Authentication}}. Initial Authentication depicted in Figure \ref{fig:init-auth-alt} is $\KIND$-secure under the cleanness predicate defined in definition \ref{ma-clean, Initial-auth} . For any PPT algorithm $\adv$ against the $\KIND$ experiment, $\Adv{\KIND,\cleanpredicate}{\Prot,\numParties,\numSessions,\adversary}(\secpar)$ is negligible assuming the KDF security of $\KDF$ and the $\DDH$ assumption.
\end{theorem}
\begin{pro}
Our proof is divided into two cases, donated by $\Adv{\KIND,\cleanpredicate_{IA}, C_1}{\Prot,\numParties,\numSessions,\adversary}(\secpar)$ and $\Adv{\KIND,\cleanpredicate_{IA}, C_2}{\Prot,\numParties,\numSessions,\adversary}(\secpar)$ 

\begin{enumerate}[label*=\arabic*.] 
    \item\textbf{Case 1:} The test session $\testsess$ accepts without a matching subset. 
     
    \item \textbf{Case 2:} The test session $\testsess$ accepts with a matching subset.
\end{enumerate}

We then bound the advantage of $\adv$ winning the game under certain assumptions to $\Adv{\KIND,\cleanpredicate_{IA}}{\Prot,\numParties,\numSessions,\adversary}(\secpar)\leq(\Adv{\KIND,\cleanpredicate_{IA}, C_1}{\Prot,\numParties,\numSessions,\adversary}(\secpar)+\Adv{\KIND,\cleanpredicate_{IA}, C_2}{\Prot,\numParties,\numSessions,\adversary}(\secpar)$.

%%%%%%%%%%%%%%%%%%%%%%%%%%%%%%%%%%%%%%%%%%%%%%%%%%%%
%%%%%%%%%%%%%%%% KIND SECURITY %%%%%%%%%%%%%%%%%%%%%
%%%%%%%%%%%%% INITIAL AUTHENTICATION %%%%%%%%%%%%%%%
%%%%%%%%%%%%%%%%%%% CASE 1 %%%%%%%%%%%%%%%%%%%%%%%%%

\textbf{Case 1:} $\testsess$ accepts without a matching session (or subset). 
Here we describe the analysis of \textbf{Case 1}:
\textbf{Game 0}: This is the original key indistinguishability experiment in\textbf{ Appendix \ref{sec:kind-game}}: \[\Adv{\KIND,\cleanpredicate_{IA}, C_1}{\Prot,\numParties,\numSessions,\adversary}(\secpar) \leq Adv_{G_{0}}.\]

\textbf{Game 1}: In this game, we introduce an abort event that triggers if $\adv$ issues a query $\Test(i,s,\role)$ such that $\testsess$ accepts without a matching session or subset. This is exactly equal to the $\MA$ security experiment, and thus we have :\[\Adv{}{G_{0}} \leq \Adv{}{G_{1}} + \Adv{\MA,\cleanpredicate_{IA}}{\Prot,\numParties,\numSessions,\adversary}(\secpar).\]
    
Since, by \textbf{Case 1}, $\testsess$ has no matching session (or subset), and by \textbf{Game 1}, we abort if $\testsess$ accepts without matching session, it follows that $\adv$ can never query $\Test(i,s\role)$ and thus the $\KIND$ game proceeds identically regardless of the bit $b$ sampled by $\challenger$. Thus
    \[\Adv{}{G_{1}}=0.\]

\textbf{Case 2 :} $\testsess$ accepts with a matching session and subset. 

%%%%%%%%%%%%%%%%%%%%%%%%%%%%%%%%%%%%%%%%%%%%%%%%%%%%
%%%%%%%%%%%%%%%% KIND SECURITY %%%%%%%%%%%%%%%%%%%%%
%%%%%%%%%%%%% INITIAL AUTHENTICATION %%%%%%%%%%%%%%%
%%%%%%%%%%%%%%%%%%% CASE 2 %%%%%%%%%%%%%%%%%%%%%%%%%

First we recall that cleanness predicate \ref{def:KIND-clean} prevents the $\adv$ from issuing a $\Reveal(i,s,\testsess.\role)$ query $\testsess$ (and to any session $\partsess$ such that $\partsess$ is a matching session or subset with $\testsess$), nor can it issue a $\StateReveal(i,s,\testsess.\role)$, nor to any session $\partsess$ such that $\partsess$ is a matching session or subset with $\testsess$. We proceed via the following sequence of games.

\textbf{Game $0$}: This is the original key indistinguishability experiment in Appendix \ref{sec:kind-game}: \[\Adv{\KIND,\cleanpredicate_{IA}, C_2}{\Prot,\numParties,\numSessions,\adversary}(\secpar) \leq Adv_{G_{0}}.\]

\textbf{Game 1}: In this game, we guess the index $(i,s) \in \numParties 
\times \numSessions$, and abort if $\adv$ issues a $\Test(i^*,s^*,\session^{s^*}_{i^*})$ query such that $i\neq i^*$ and $s \neq s^*$. This introduces the following bound: \[\Adv{}{G_{0}} \leq \numSessions \cdot \numParties \cdot \Adv{}{G_{1}}.\]

\textbf{Game 2}: In this game, we guess the index $(j,t) \in \numParties 
\times \numSessions$, and abort if $\partsess$ is not the matching subset of $\testsess$, which must exist by \textbf{Case 2}. This introduces the following bound: \[\Adv{}{G_{1}} \leq \numSessions \cdot \numParties \cdot \Adv{}{G_{2}}.\]

\textbf{Game 3}: In this game, we replace $g^{uh}$ computed honestly in $\testsess$ and $\partsess$ with a uniformly random and independent value $\hat{g^{uh}}$. We do so by defining a reduction $\bdv_{1}$ that initialises a $\DDH$ challenger $\chal_{\DDH}$, and replaces $g^u$, $g^h$ and $g^{uh}$ computed by $\testsess$ and $\partsess$ with the outputs of $\chal_{\DDH}$, $g^a$, $g^b$, $g^c$. We note that if the bit $b$ sampled by $\chal_{\DDH}$ is 1, then $\hat{g^{uh}} =g^{ab}$ and we are in \textbf{Game 2}, otherwise $c \getsr \Zq$ and we are in \textbf{Game 3}. Any $\adv$ that can distinguish \textbf{Game 2} from \textbf{Game 3} can break the $\DDH$ assumption. Thus: \[\Adv{}{G_{3}} \leq \Adv{}{G_{4}}+ \Adv{G,g,q}{\bdv_{1} \DDH}.\]
     
\textbf{Game 4}: In this game we replace the session and encryption keys $sk_i, k_s$ with uniformly random values $\hat{sk_i}, \hat{k_s}$ in $\testsess$ and $\partsess$. We do so by defining a reduction $\bdv_2$ that interacts with a $\KDF$ challenger $\chal_{\KDF}$, querying $\chal_{\KDF}$ with $\hat{g^{uh}}$ and replacing the computation of $sk_i,k_s$ in $\partsess$ and $\testsess$ with the outputs from the $\chal_{\KDF}$ $\hat{sk_i}, \hat{k_s}$. Since $sk_i, k_s \gets \KDF(\hat{g^{uh}})$ and by \textbf{Game 3} $\hat{g^{uh}}$ is already uniformly random and independent, this change is sound. Any $\adversary$ that can distinguish \textbf{Game 3} from \textbf{Game 4} can be used to break $\KDF$ security of the $\KDF$ scheme. Thus:  \[\Adv{}{G_{3}} \leq \Adv{}{G_{4}}+ \Adv{\KDF}{\bdv_1,\KDF}.\]

Here we emphasise that as a result of these changes, the session key $\hat{sk_i}$ is now uniformly random and independent of the protocol execution regardless of the bit $b$ sampled by $\chal$, thus $\adv$ has no advantage in guessing the bit $b$:
\[\Adv{}{G_{4}}=0.\]
\end{pro}

%%%%%%%%%%%%%%%%%%%%%%%%%%%%%%%%%%%%%%%%%%%%%%%%%%%%
%%%%%%%%%%%%%%%% KIND SECURITY %%%%%%%%%%%%%%%%%%%%%
%%%%%%%%%%%%% UNIVERSAL HANDOVER %%%%%%%%%%%%%%%%%%%
%%%%%%%%%%%%%%%%%%%%%%%%%%%%%%%%%%%%%%%%%%%%%%%%%%%%

\subsubsection{$\KIND$-security of Universal Handover}
Here we formally analyse the $\KIND$-security of the Universal Handover protocol.

\begin{theorem}{\textbf{$\mathbf{\KIND}$-security of Universal Handover}}. Universal Handover depicted in Figure \ref{fig:universal-alt} is $\KIND$-secure under the cleanness predicate in Definition \ref{def:KIND-clean}. For any PPT algorithm $\adv$ against the $\KIND$ experiment, $\Adv{\KIND,\cleanpredicate_{UH}}{\Prot,\numParties,\numSessions,\adversary}(\secpar)$ is negligible assuming the EUFCMA security of $\SanSig$, Auth security of $\AuthEnc$, the KDF security of $\KDF$ and the $\DDH$ assumption.
\end{theorem}

\begin{pro}
Our proof is divided into two cases, denoted by $\Adv{\KIND,\cleanpredicate, C_1}{\Prot,\numParties,\numSessions,\adversary}(\secpar)$ and $\Adv{\KIND,\cleanpredicate, C_2}{\Prot,\numParties,\numSessions,\adversary}(\secpar)$ 
\begin{enumerate}[label*=\arabic*.] 
    \item\textbf{ Case 1:} The test session $\testsess$ accepts messages without a matching subset. 
     
    \item \textbf{Case 2:} The test session $\testsess$ accepts messages with a matching subset.
\end{enumerate}

We then bound the advantage of $\adv$ winning the game under certain assumptions to $\Adv{\KIND,\cleanpredicate}{\Prot,\numParties,\numSessions,\adversary}(\secpar)\leq(\Adv{\KIND,\cleanpredicate, C_1}{\Prot,\numParties,\numSessions,\adversary}(\secpar)+\Adv{\KIND,\cleanpredicate, C_2}{\Prot,\numParties,\numSessions,\adversary}(\secpar)$.

We begin by treating \textbf{Case 1}.

%%%%%%%%%%%%%%%%%%%%%%%%%%%%%%%%%%%%%%%%%%%%%%%%%%%%
%%%%%%%%%%%%%%%% KIND SECURITY %%%%%%%%%%%%%%%%%%%%%
%%%%%%%%%%%%% UNIVERSAL HANDOVER %%%%%%%%%%%%%%%%%%%
%%%%%%%%%%%%%%%%%%% CASE 1 %%%%%%%%%%%%%%%%%%%%%%%%%

\textbf{Case 1:} $\testsess$ accepts without a matching subset. 
Here we describe the analysis of \textbf{Case 1}:
\textbf{Game 0}: This is the original key indistinguishability experiment in Appendix \ref{sec:kind-game}: \[\Adv{\KIND,\cleanpredicate_{UH}, C_1}{\Prot,\numParties,\numSessions,\adversary}(\secpar) \leq Adv_{G_{0}}.\]

\textbf{Game 1}: In this game, we introduce an abort event that triggers if $\adv$ issues a query $\Test(i,s,\role)$ such that $\testsess$ accepts without a matching session or subset. This is exactly equal to the $\MA$ security experiment, and thus we have :\[\Adv{}{G_{0}} \leq \Adv{}{G_{1}} + \Adv{\MA,\cleanpredicate_{UH}}{\Prot,\numParties,\numSessions,\adversary}(\secpar).\]
    
Since, by \textbf{Case 1}, $\testsess$ has no matching session (or subset), and by \textbf{Game 1}, we abort if $\testsess$ accepts without matching session, it follows that $\adv$ can never query $\Test(i,s\role)$ and thus the $\KIND$ game proceeds identically regardless of the bit $b$ sampled by $\challenger$. Thus
    \[\Adv{}{G_{1}}=0.\]

We now turn to \textbf{Case 2}.

%%%%%%%%%%%%%%%%%%%%%%%%%%%%%%%%%%%%%%%%%%%%%%%%%%%%
%%%%%%%%%%%%%%%% KIND SECURITY %%%%%%%%%%%%%%%%%%%%%
%%%%%%%%%%%%% UNIVERSAL HANDOVER %%%%%%%%%%%%%%%%%%%
%%%%%%%%%%%%%%%%%%% CASE 2 %%%%%%%%%%%%%%%%%%%%%%%%%

\textbf{Case 2:} $\testsess$ accepts with a matching subset. 

First we recall that cleanness predicate \ref{def:KIND-clean} prevents the $\adv$ from issuing a $\Reveal(i,s,\testsess.\role)$ query $\testsess$ (and to any session $\partsess$ such that $\partsess$ is a matching session or subset with $\testsess$), nor can it issue a $\StateReveal(i,s,\testsess.\role)$, nor to any session $\partsess$ such that $\partsess$ is a matching session or subset with $\testsess$. We proceed via the following sequence of games.

\textbf{Game $0$}: This is the original key indistinguishability experiment in Appendix \ref{sec:kind-game}: \[\Adv{\KIND,\cleanpredicate_{UH}, C_2}{\Prot,\numParties,\numSessions,\adversary}(\secpar) \leq Adv_{G_{0}}.\]

\textbf{Game 1}: In this game, we guess the index $(i,s) \in \numParties 
\times \numSessions$, and abort if $\adv$ issues a $\Test(i^*,s^*,\session^{s^*}_{i^*})$ query such that $i\neq i^*$ and $s \neq s^*$. This introduces the following bound: \[\Adv{}{G_{0}} \leq \numSessions \cdot \numParties \cdot \Adv{}{G_{1}}.\]

\textbf{Game 2}: In this game, we guess the index $(j,t) \in \numParties 
\times \numSessions$, and abort if $\partsess$ is not the matching subset of $\testsess$, which must exist by \textbf{Case 2}. This introduces the following bound: \[\Adv{}{G_{1}} \leq \numSessions \cdot \numParties \cdot \Adv{}{G_{2}}.\]

\textbf{Game 3}: In this game, we replace $g^{uh}$ computed honestly in $\testsess$ and $\partsess$ with a uniformly random and independent value $\hat{g^{uh}}$. We do so by defining a reduction $\bdv_{1}$ that initialises a $\DDH$ challenger $\chal_{\DDH}$, and replaces $g^u$, $g^h$ and $g^{uh}$ computed by $\testsess$ and $\partsess$ with the outputs of $\chal_{\DDH}$, $g^a$, $g^b$, $g^c$. We note that if the bit $b$ sampled by $\chal_{\DDH}$ is 1, then $\hat{g^{uh}} =g^{ab}$ and we are in \textbf{Game 2}, otherwise $c \getsr \Zq$ and we are in \textbf{Game 3}. Any $\adv$ that can distinguish \textbf{Game 2} from \textbf{Game 3} can break the $\DDH$ assumption. Thus: \[\Adv{}{G_{3}} \leq \Adv{}{G_{4}}+ \Adv{G,g,q}{\bdv_{1} \DDH}.\]
     
\textbf{Game 4}: In this game we replace the session and encryption keys $sk_i, k_s$ with uniformly random values $\hat{sk_i}, \hat{k_s}$ in $\testsess$ and $\partsess$. We do so by defining a reduction $\bdv_2$ that interacts with a $\KDF$ challenger $\chal_{\KDF}$, querying $\chal_{\KDF}$ with $\hat{g^{uh}}$ and replacing the computation of $sk_i,k_s$ in $\partsess$ and $\testsess$ with the outputs from the $\chal_{\KDF}$ $\hat{sk_i}, \hat{k_s}$. Since $sk_i, k_s \gets \KDF(\hat{g^{uh}})$ and by \textbf{Game 3} $\hat{g^{uh}}$ is already uniformly random and independent, this change is sound. Any $\adversary$ that can distinguish \textbf{Game 3} from \textbf{Game 4} can be used to break $\KDF$ security of the $\KDF$ scheme. Thus:  \[\Adv{}{G_{3}} \leq \Adv{}{G_{4}}+ \Adv{\KDF}{\bdv_1,\KDF}.\]

Here we emphasise that as a result of these changes, the session key $\hat{sk_i}$ is now uniformly random and independent of the protocol execution regardless of the bit $b$ sampled by $\chal$, thus $\adv$ has no advantage in guessing the bit $b$:
\[\Adv{}{G_{4}}=0.\]
\end{pro}
\fi

\subsection{Unlinkability}
Here we discuss the unlinkability of $\UniHand$. We consider a strong notion of anonymity where $\adv$ wins simply by linking a so-called ``test'' session to another protocol execution - since it knows the identities of all other protocol executions, this allows de-anonymisation. Thus, our framework captures user anonymity, user confidentiality and untraceability. 

\ifsubmissionversion Due to space constraints, we include the proof sketch of $\Unlink$-security for the initial authentication protocol since the universal handover analysis follows similarly (and more simply). For full details of each proof, refer to \cite{authors_unihand_2023}\footnotemark[1]{}. 

%%%%%%%%%%%%%%%%%%%%%%%%%%%%%%%%%%%%%%%%%%%%%%%%%%%%
%%%%%%%%%%%%%%%% UNLINK SECURITY %%%%%%%%%%%%%%%%%%%
%%%%%%%%%%%%% INITIAL AUTHENTICATION %%%%%%%%%%%%%%%
%%%%%%%%%%%%%%%%%%%%%%%%%%%%%%%%%%%%%%%%%%%%%%%%%%%%

\begin{theorem}
\label{theorem: UNlink sec- initial auth.}
{\textbf{$\Unlink$-security of $\UniHand$}}.  $\UniHand$ depicted in Figure \ref{fig:init-auth-alt} and Figure\ref{fig:universal-alt} is unlinkable under the cleanness predicate in Definition \ref{unlink-clean}. For any PPT algorithm $\adv$ against the $\Unlink$ experiment, $\Adv{\Unlink,\cleanpredicate}{\Prot,\numParties,\numSessions,\adversary}(\secpar)$ is negligible assuming the EUFCMA security of $\SanSig$, Conf security of $\AuthEnc$, the KDF security of $\KDF$ and the $\DDH$ assumption.
\end{theorem}

\ifsubmissionversion
On a high level, we show that $\adv$ cannot inject DH public keys from the $\UE$ to the $\gNB$ due to the $\UE$ signatures over these values, based on the $\EUFCMA$ security of the $\SanSig$ scheme. As the secret DH output is used to derive keys for the authenticated encryption scheme securely, and all other messages exchanged from the $\UE$ to the $\gNB$ are encrypted, $\adv$ cannot break the confidentiality of the messages and hence cannot guess the identity of the intended user. Similarly, we show that $\adv$ cannot inject DH public keys from the $\AuC$ to the $\gNB$ due to the $\AuC$ signatures over these values, based on the $\EUFCMA$ security of the $\SanSig$ scheme. As the secret DH output is used to derive keys for the authenticated encryption scheme securely, and all other messages exchanged from the $\UE$ to the $\gNB$ are encrypted, $\adv$ cannot learn plaintext messages and hence cannot guess the identity of the intended user.  
\fi

\fi

\iffullversion
\subsubsection{$\Unlink$-security of Initial Authentication}
\label{UNlink of Initial Authentication}
This section formally analyses the $\Unlink$-security of the Initial Authentication protocol.
\begin{theorem}
\label{theorem: UNlink sec- initial auth.}
{\textbf{$\Unlink$-security of Initial Authentication}}.  The Initial Authentication depicted in Figure \ref{fig:init-auth-alt} is unlinkable under the cleanness predicate in Definition \ref{unlink-clean}. For any PPT algorithm $\adv$ against the $\Unlink$ experiment, $\Adv{\Unlink,\cleanpredicate}{\Prot,\numParties,\numSessions,\adversary}(\secpar)$ is negligible assuming the EUFCMA security of $\SanSig$, Conf security of $\AuthEnc$, the KDF security of $\KDF$ and the $\DDH$ assumption.
\end{theorem}

%%%%%%%%%%%%%%%%%%%%%%%%%%%%%%%%%%%%%%%%%%%%%%%%%%%%
%%%%%%%%%%%%%%%% UNLINK SECURITY %%%%%%%%%%%%%%%%%%%
%%%%%%%%%%%%% INITIAL AUTHENTICATION %%%%%%%%%%%%%%%
%%%%%%%%%%%%%%%%%%%%%%%%%%%%%%%%%%%%%%%%%%%%%%%%%%%%

Our proof is divided into two cases, denoted by $\Adv{\KIND,\cleanpredicate, C_1}{\Prot,\numParties,\numSessions,\adversary}(\secpar)$ and $\Adv{\KIND,\cleanpredicate, C_2}{\Prot,\numParties,\numSessions,\adversary}(\secpar)$ 
\begin{enumerate}[label*=\arabic*.] 
    \item\textbf{ Case 1:} The test session $\session_b$ (such that $\adv$ issues $\Test(i,s,i^*,s^*)$) accepts messages without a matching subset. 
     
    \item \textbf{Case 2:} The test session $\session_b$ (such that $\adv$ issues $\Test(i,s,i^*,s^*)$) accepts messages with a matching subset.
\end{enumerate}

We then bound the advantage of $\adv$ winning the game under certain assumptions to $\Adv{\Unlink,\cleanpredicate}{\Prot,\numParties,\numSessions,\adversary}(\secpar)\leq(\Adv{\Unlink,\cleanpredicate, C_1}{\Prot,\numParties,\numSessions,\adversary}(\secpar)+\Adv{\Unlink,\cleanpredicate, C_2}{\Prot,\numParties,\numSessions,\adversary}(\secpar)$.

We begin by treating \textbf{Case 1}.

%%%%%%%%%%%%%%%%%%%%%%%%%%%%%%%%%%%%%%%%%%%%%%%%%%%%
%%%%%%%%%%%%%%%% UNLINK SECURITY %%%%%%%%%%%%%%%%%%%
%%%%%%%%%%%%% INITIAL AUTHENTICATION %%%%%%%%%%%%%%%
%%%%%%%%%%%%%%%%%%% CASE 1 %%%%%%%%%%%%%%%%%%%%%%%%%

\textbf{Case 1:} $\session_b$ accepts without a matching subset. 
Here we describe the analysis of \textbf{Case 1}:
\textbf{Game 0}: This is the original unlinkability experiment in Appendix \ref{UNlink-sec}: \[\Adv{\Unlink,\cleanpredicate_{IA}, C_1}{\Prot,\numParties,\numSessions,\adversary}(\secpar) \leq Adv_{G_{0}}.\]

\textbf{Game 1}: In this game, we introduce an abort event that triggers if $\adv$ issues a query $\Test(i,s,i^*,s^*)$ and $\session_b$ accepts without a matching session or subset. This is exactly equal to the $\MA$ security experiment, and thus we have :\[\Adv{}{G_{0}} \leq \Adv{}{G_{1}} + \Adv{\MA,\cleanpredicate_{IA}}{\Prot,\numParties,\numSessions,\adversary}(\secpar).\]
    
Since, by \textbf{Case 1}, $\session$ has no matching session (or subset). By \textbf{Game 1}, we abort if $\session_b$ accepts without matching session, it follows that $\adv$ can never terminate and output a guess bit $b'$ and thus the $\Unlink$ game proceeds identically regardless of the bit $b$ sampled by $\challenger$. Thus
    \[\Adv{}{G_{1}}=0.\]

We now turn to \textbf{Case 2}.

%%%%%%%%%%%%%%%%%%%%%%%%%%%%%%%%%%%%%%%%%%%%%%%%%%%%
%%%%%%%%%%%%%%%% UNLINK SECURITY %%%%%%%%%%%%%%%%%%%
%%%%%%%%%%%%% INITIAL AUTHENTICATION %%%%%%%%%%%%%%%
%%%%%%%%%%%%%%%%%%% CASE 2 %%%%%%%%%%%%%%%%%%%%%%%%%

\textbf{Case 2:} $\testsess$ accepts with a matching subset. 

First we recall that cleanness predicate Definition \ref{def:KIND-clean} prevents the $\adv$ from issuing a $\StateReveal(i,s,\testsess.\role)$, nor to any session $\partsess$ such that $\partsess$ is a matching session or subset with $\testsess$. We proceed via the following sequence of games.

\textbf{Game 0}: This is the original unlinkability experiment in Definition \ref{def:unlink-sec}: \[\Adv{\Unlink,\cleanpredicate_{UH}}{\Prot,\numParties,\numSessions,\adversary}(\secpar) \leq \Adv{}{G_0}.\]
    
\textbf{Game 1} : In this game, we guess the index $(i,s) \in \numParties \times \numSessions$ of the $\session_b$ session, introducing a factor of $\numParties \times \numSessions$ in $\adv$'s advantage: \[\Adv{}{G_0} \leq \numParties \cdot\numSessions \Adv{}{G_{1}}.\]

\textbf{Game 2}: Here we introduce an abort event, where $\chal$ aborts if $\session_b$ receives a message $\mathbf{M_1}$ without setting $\session_b.\status \gets \rejectflag$ but $\mathbf{M_1}$ was not output by a session owned by $\session_b.\pid$. We do so by defining a reduction $\bdv_1$ that initialises a $\SanSig$ challenger $\chal_{\SanSig}$, that outputs $\pk_{sig}^{\challenger}$ and  $\pk_{san}^{\challenger}$, which we embed into the $\AuC$'s $\pk_{sig}^{\AuC}$  and $\gNB$'s $\pk_{san}^{\gNB}$ respectively. Anytime $\AuC$ or $\gNB$ needs to generate a signature over a message $m$, $\bdv_1$ instead queries $\chal$ with $m$. Now, if $\session_b$ receives a message $\mathbf{M_1}$ without setting $\session_b.\status \gets \rejectflag$ but $\mathbf{M_1}$ was not output by a session owned by $\session_b.\pid$, then $\adv$ must have produced a message $\mathbf{M_1} = C^*_G,\sigma^*_G,g^h$ such that $\SanSig.\Verify(C^*_G,\sigma^*_G,\pk_{sig}^{\chal},\pk_{san}^{\chal})=1$, which is a valid signature forgery. $\bdv_1$ responds to $\chal_{\SanSig}$ with $C^*_G,\sigma^*_G$ and triggers the abort event. 
Thus, the probability that $\bdv_1$ triggers the abort event is bounded by the $\EUFCMA$ security of $\SanSig$: \[\Adv{}{G_1} \leq \Adv{}{G_2}+\Adv{\EUFCMA}{\bdv_1, \SanSig}.\]

\textbf{Game 3}: In this game, we guess the index $(j,t) \in \numParties \times \numSessions$ of the $\gNB$ session $\partsess$ that output $\mathbf{M_1}$ received by $\session_b$, introducing a factor of $\numParties \times \numSessions$ in $\adv$'s advantage. \[\Adv{}{G_2} \leq \numParties \cdot\numSessions \Adv{}{G_{3}}.\]

\textbf{Game 4}: Here we introduce another abort event that triggers if $\adv$ sends a Diffie-Hellman public keyshare $g^h$ to the session $\session_b$, i.e. session $\session_b$ receives $g^h$ that was not output from a $\gNB$ session, but instead from $\adv$. Since this trigger event requires the signature $\sigma^*_G$ in $\mathbf{M_1}$ to verify over $g^h$, and by \textbf{Game 2} we already abort if $\sigma^*_G$ comes from $\adv$, it follows that \[\Adv{}{G_3} \leq \Adv{}{G_4}.\]

\textbf{Game 5}: In this game, we replace $g^{hu}$ computed honestly in $\session_b$ with a uniformly random and independent value $\hat{g^{hu}}$. We do so by defining a reduction $\bdv_2$ that initialises a $\DDH$ challenger $\chal_{\DDH}$, and replaces $g^u$, $g^h$ and $g^{hu}$ computed by $\session_b$ and $\partsess$ with the outputs of $\chal_{\DDH}$, $g^a$, $g^b$, $g^c$. We note that if the bit $b$ sampled by $\chal_{\DDH}$ is 1, then $c = ab$ and we are in \textbf{Game 4}. Otherwise, $c \getsr \Zq$ and we are in \textbf{Game 5}. Any $\adv$ that can distinguish \textbf{Game 4} from \textbf{Game 5} can break the $\DDH$ assumption. Thus:  \[\Adv{}{G_4} \leq \Adv{}{G_5}+ \Adv{G,g,q}{\bdv_2 \DDH}.\]

\textbf{Game 6}: In this game we replace the session and encryption keys $sk_i, k_s$ with uniformly random values $\hat{sk_i}, \hat{k_s}$. We do so by defining a reduction $\bdv_3$ that interacts with a $\KDF$ challenger $\chal_{\KDF}$, querying $\chal_{\KDF}$ with $\hat{g^{uh}}$ and replacing the computation of $sk_i,k_s$ in $\session_b$ and $\partsess$ with the outputs from the $\chal_{\KDF}$ $\hat{sk_i}, \hat{k_s}$. Since $sk_i, k_s \gets \KDF(\hat{g^{uh}})$ and by \textbf{Game 5} $\hat{g^{uh}}$ is already uniformly random and independent, this change is sound. Any $\adversary$ that can distinguish \textbf{Game 5} from \textbf{Game 6} can be used to break $\KDF$ security of the $\KDF$ scheme. Thus:  \[\Adv{}{G_5} \leq \Adv{}{G_6}+ \Adv{\KDF}{\bdv_3,\KDF}.\]

\textbf{Game 7}: In this game, we replace the plaintexts in messages $\mathbf{M_2}$, $\mathbf{M_7}$ and $\mathbf{M_8}$ with uniformly random values of the same length. We do so by defining a reduction $\bdv_4$ that initialises an $\Enc$ challenger $\chal_{\Conf}$, which $\bdv_4$ queries $(p,p^*)$ (where $p$ is the original plaintext message and $p^*\getsr \bits{|p|}$) when it needs to encrypt with $\hat{k_s}$. By \textbf{Game 6} $\hat{k_s}$ is already uniformly random and independent, and this replacement is sound. If the bit $b$ sampled by $\chal_{\Conf}$ is 0, then we are in \textbf{Game 6}, otherwise, we are in \textbf{Game 7}. Any $\adversary$ that can distinguish between  \textbf{Game 6} and \textbf{Game 7} can be used by $\bdv_4$ to break the $\Conf$ security of $\Enc$. This implies: \[\Adv{}{G_6} \leq \Adv{}{G_7}+ \Adv{\AuthEnc}{\bdv_4,\Enc}.\]

\textbf{Game 8}: In this game, we guess the index $s \in \numSessions$ of the $\AuC$ session (which we denote $\session_{\AuC}$ that is a matching session of $\session_b$, introducing a factor of $\numSessions$ in $\adv$'s advantage: \[\Adv{}{G_7} \leq \numSessions \Adv{}{G_{8}}.\]

\textbf{Game 9}: Here we introduce an abort event, where $\chal$ aborts if $\partsess$ receives a message $\mathbf{M_4}$ without setting $\partsess.\status \gets \rejectflag$ but $\mathbf{M_4}$ was not output by a session owned by $\AuC$. We do so by defining a reduction $\bdv_5$ that initialises a $\SanSig$ challenger $\chal_{\SanSig}$, that outputs $\pk_{sig}^{\challenger}$ and  $\pk_{san}^{\challenger}$, which we embed into the $\AuC$'s $\pk_{sig}^{\AuC}$  and $\gNB$'s $\pk_{san}^{\gNB}$ respectively. Anytime $\AuC$ or $\gNB$ needs to generate a signature over a message $m$, $\bdv_5$ instead queries $\chal$ with $m$. Now, if $\partsess$ receives a message $\mathbf{M_4}$ without setting $\partsess.\status \gets \rejectflag$ but $\mathbf{M_4}$ was not output by a session owned by $\AuC$, then $\adv$ must have produced a message $\mathbf{M_4} = C^*_G,\sigma^*_G,g^b$ such that $\SanSig.\Verify(C^*_G,\sigma^*_G,\pk_{sig}^{\chal},\pk_{san}^{\chal})=1$, which is a valid signature forgery. $\bdv_5$ responds to $\chal_{\SanSig}$ with $C^*_G,\sigma^*_G$ and triggers the abort event. 
Thus, the probability that $\bdv_5$ triggers the abort event is bounded by the $\EUFCMA$ security of $\SanSig$: \[\Adv{}{G_8} \leq \Adv{}{G_9}+\Adv{\EUFCMA}{\bdv_5, \SanSig}.\]

\textbf{Game 10}: In this game, we replace $g^{ab}$ computed honestly in $\partsess$ and $\session_{\AuC}$ with a uniformly random and independent value $\hat{g^{ab}}$. We do so by defining a reduction $\bdv_{6}$ that initialises a $\DDH$ challenger $\chal_{\DDH}$, and replaces $g^a$, $g^b$ and $g^{ab}$ computed by $\partsess$ and $\sessiong_{\AuC}$ with the outputs of $\chal_{\DDH}$. We note that if the bit $b$ sampled by $\chal_{\DDH}$ is 1, then $\hat{g^{ab}} =g^{ab}$ and we are in \textbf{Game 9}, otherwise $\hat{ab} \getsr \Zq$ and we are in \textbf{Game 10}. Any $\adv$ that can distinguish \textbf{Game 9} from \textbf{Game 10} can break the $\DDH$ assumption. Thus:  \[\Adv{}{G_9} \leq \Adv{}{G_{10}}+ \Adv{G,g,q}{\bdv_{6} \DDH}.\]

\textbf{Game 11}: In this game we replace the session and encryption keys $k_s'$ with uniformly random values $\hat{k_s}'$. We do so by defining a reduction $\bdv_{7}$ that interacts with a $\KDF$ challenger $\chal_{\KDF}$, querying $\chal_{\KDF}$ with $\hat{g^{ab}}$ and replacing the computation of $k_s'$ in $\partsess$ and $\session_{\AuC}$ with the outputs from the $\chal_{\KDF}$ $\hat{k_s}'$. Since $k_s \gets \KDF(\hat{g^{ab}})$ and by \textbf{Game 10} $\hat{g^{ab}}$ is already uniformly random and independent, this change is sound. Any $\adversary$ that can distinguish \textbf{Game 9} from \textbf{Game 10} can be used to break $\KDF$ security of the $\KDF$ scheme. Thus:  \[\Adv{}{G_9} \leq \Adv{}{G_{10}}+ \Adv{\KDF}{\bdv_{7},\KDF}.\]

\textbf{Game 12}: In this game, we replace the plaintexts in messages $\mathbf{M_5}$, $\mathbf{M_6}$ and $\mathtbf{ACK''}$ with uniformly random values of the same length. We do so by defining a reduction $\bdv_8$ that initialises an $\Enc$ challenger $\chal_{\Conf}$, which $\bdv_8$ queries with $(p,p^*)$ (where $p$ is the original plaintext message and $p^*\getsr \bits{|p|}$) when it needs to encrypt with $\hat{k_s}'$. By \textbf{Game 11} $\hat{k_s}'$ is already uniformly random and independent, and this replacement is sound. If the bit $b$ sampled by $\chal_{\Conf}$ is 0, then we are in \textbf{Game 11}, otherwise, we are in \textbf{Game 12}. Any $\adversary$ that can distinguish between  \textbf{Game 11} and \textbf{Game 12} can be used by $\bdv_8$ to break the $\Conf$ security of $\Enc$. This implies: \[\Adv{}{G_{11}} \leq \Adv{}{G_{12}}+ \Adv{\Conf}{\bdv_8,\Enc}.\]
     
Now we highlight that all plaintext messages sent across the network to and from $\session_b$ and its matching session and subsets are uniformly random and independent of the bit $b$ sampled  by the challenger. Thus it follows that $\adv$ has no advantage in guessing the bit $b$, and summing the probabilities $\adv$ has a negligible advantage in winning the $\Unlink$ game. Thus: 
\[\Adv{}{G_{12}}=0.\]

\subsubsection{$\Unlink$-security of Universal Handover}
\label{UNlink of Universal Handover}
This section formally analyses the $\Unlink$-security of the Universal Handover protocol.
\begin{theorem}
\label{theorem: UNlink sec- HO}
{\textbf{$\Unlink$-security of Universal Handover}}.  The Universal Handover depicted in Figure \ref{fig:universal-alt} is unlinkable under the cleanness predicate in Definition \ref{unlink-clean}. For any PPT algorithm $\adv$ against the $\Unlink$ experiment, $\Adv{\Unlink,\cleanpredicate}{\Prot,\numParties,\numSessions,\adversary}(\secpar)$ is negligible assuming the EUFCMA security of $\SanSig$, Conf security of $\AuthEnc$, the KDF security of $\KDF$ and the $\DDH$ assumption.
\end{theorem}

\begin{pro}  

%%%%%%%%%%%%%%%%%%%%%%%%%%%%%%%%%%%%%%%%%%%%%%%%%%%%
%%%%%%%%%%%%%%%% UNLINK SECURITY %%%%%%%%%%%%%%%%%%%
%%%%%%%%%%%%% UNIVERSAL HANDOVER %%%%%%%%%%%%%%%%%%%
%%%%%%%%%%%%%%%%%%%%%%%%%%%%%%%%%%%%%%%%%%%%%%%%%%%%

Our proof is divided into two cases, denoted by $\Adv{\Unlink,\cleanpredicate_{UH}, C_1}{\Prot,\numParties,\numSessions,\adversary}(\secpar)$ and $\Adv{\Unlink,\cleanpredicate_{UH}, C_2}{\Prot,\numParties,\numSessions,\adversary}(\secpar)$ 
\begin{enumerate}[label*=\arabic*.] 
    \item\textbf{ Case 1:} The test session $\session_b$ (such that $\adv$ issues $\Test(i,s,i^*,s^*)$) accepts messages without a matching subset. 
     
    \item \textbf{Case 2:} The test session $\session_b$ (such that $\adv$ issues $\Test(i,s,i^*,s^*)$) accepts messages with a matching subset.
\end{enumerate}

We then bound the advantage of $\adv$ winning the game under certain assumptions to $\Adv{\Unlink,\cleanpredicate}{\Prot,\numParties,\numSessions,\adversary}(\secpar)\leq(\Adv{\Unlink,\cleanpredicate, C_1}{\Prot,\numParties,\numSessions,\adversary}(\secpar)+\Adv{\Unlink,\cleanpredicate, C_2}{\Prot,\numParties,\numSessions,\adversary}(\secpar)$.

We begin by treating \textbf{Case 1}.

%%%%%%%%%%%%%%%%%%%%%%%%%%%%%%%%%%%%%%%%%%%%%%%%%%%%
%%%%%%%%%%%%%%%% UNLINK SECURITY %%%%%%%%%%%%%%%%%%%
%%%%%%%%%%%%% UNIVERSAL HANDOVER %%%%%%%%%%%%%%%%%%%
%%%%%%%%%%%%%%%%%%% CASE 1 %%%%%%%%%%%%%%%%%%%%%%%%%

\textbf{Case 1:} $\session_b$ accepts without a matching subset. 
Here we describe the analysis of \textbf{Case 1}:
\textbf{Game 0}: This is the original unlinkability experiment in Appendix \ref{UNlink-sec}: \[\Adv{\Unlink,\cleanpredicate, C_1}{\Prot,\numParties,\numSessions,\adversary}(\secpar) \leq Adv_{G_{0}}.\]

\textbf{Game 1}: In this game, we introduce an abort event that triggers if $\adv$ issues a query $\Test(i,s,i^*,s^*)$ and $\session_b$ accepts without a matching session or subset. This is exactly equal to the $\MA$ security experiment, and thus we have :\[\Adv{}{G_{0}} \leq \Adv{}{G_{1}} + \Adv{\MA,\cleanpredicate_{IA}}{\Prot,\numParties,\numSessions,\adversary}(\secpar).\]
    
Since, by \textbf{Case 1}, $\session$ has no matching session (or subset), and by \textbf{Game 1}, we abort if $\session_b$ accepts without matching session, it follows that $\adv$ can never terminate and output a guess bit $b'$ and thus the $\Unlink$ game proceeds identically regardless of the bit $b$ sampled by $\challenger$. Thus
    \[\Adv{}{G_{1}}=0.\]

We now turn to \textbf{Case 2}.

%%%%%%%%%%%%%%%%%%%%%%%%%%%%%%%%%%%%%%%%%%%%%%%%%%%%
%%%%%%%%%%%%%%%% UNLINK SECURITY %%%%%%%%%%%%%%%%%%%
%%%%%%%%%%%%% INITIAL AUTHENTICATION %%%%%%%%%%%%%%%
%%%%%%%%%%%%%%%%%%% CASE 2 %%%%%%%%%%%%%%%%%%%%%%%%%

\textbf{Case 2:} $\testsess$ accepts with a matching subset. 

First, we recall that cleanness predicate Definition \ref{def:unlink-sec} prevents the $\adv$ from issuing a $\StateReveal(i,s,\testsess.\role)$, nor to any session $\partsess$ such that $\partsess$ is a matching subset with $\testsess$. We proceed via the following sequence of games.

\textbf{Game 0}: This is the original unlinkability experiment in Definition \ref{def:unlink-sec}: \[\Adv{\Unlink,\cleanpredicate_{UH}}{\Prot,\numParties,\numSessions,\adversary}(\secpar) \leq \Adv{}{G_0}.\]
    
\textbf{Game 1} : In this game, we guess the index $(i,s) \in \numParties \times \numSessions$ of the $\session_b$ session, introducing a factor of $\numParties \times \numSessions$ in $\adv$'s advantage: \[\Adv{}{G_0} \leq \numParties \cdot\numSessions \Adv{}{G_{1}}.\]

\textbf{Game 2}: Here we introduce an abort event, where $\chal$ aborts if $\session_b$ receives a message $\mathbf{M_1}$ without setting $\session_b.\status \gets \rejectflag$ but $\mathbf{M_1}$ was not output by a session owned by $\session_b.\pid$. We do so by defining a reduction $\bdv_1$ that initialises a $\SanSig$ challenger $\chal_{\SanSig}$, that outputs $\pk_{sig}^{\challenger}$ and  $\pk_{san}^{\challenger}$, which we embed into the $\AuC$'s $\pk_{sig}^{\AuC}$  and $\gNB$'s $\pk_{san}^{\gNB}$ respectively. Anytime $\gNB$ needs to generate a signature over a message $m$, $\bdv_1$ instead queries $\chal$ with $m$. Now, if $\session_b$ receives a message $\mathbf{M_1}$ without setting $\session_b.\status \gets \rejectflag$ but $\mathbf{M_1}$ was not output by a session owned by $\session_b.\pid$, then $\adv$ must have produced a message $\mathbf{M_1} = C^*_G,\sigma^*_G,g^h$ such that $\SanSig.\Verify(C^*_G,\sigma^*_G,\pk_{sig}^{\chal},\pk_{san}^{\chal})=1$, which is a valid signature forgery. $\bdv_1$ responds to $\chal_{\SanSig}$ with $C^*_G,\sigma^*_G$ and triggers the abort event. 
Thus, the probability that $\bdv_1$ triggers the abort event is bounded by the $\EUFCMA$ security of $\SanSig$: \[\Adv{}{G_1} \leq \Adv{}{G_2}+\Adv{\EUFCMA}{\bdv_1, \SanSig}.\]

\textbf{Game 3}: In this game, we guess the index $(j,t) \in \numParties \times \numSessions$ of the $\gNB$ session $\partsess$ that output $\mathbf{M_1}$ received by $\session_b$, introducing a factor of $\numParties \times \numSessions$ in $\adv$'s advantage. \[\Adv{}{G_2} \leq \numParties \cdot\numSessions \Adv{}{G_{3}}.\]

\textbf{Game 4}: Here we introduce another abort event that triggers if $\adv$ sends a Diffie-Hellman public keyshare $g^h$ to the session $\session_b$, i.e. session $\session_b$ receives $g^h$ that was not output from a $\gNB$ session, but instead from $\adv$. Since this trigger event requires the signature $\sigma^*_G$ in $\mathbf{M_1}$ to verify over $g^h$, and by \textbf{Game 2} we already abort if $\sigma^*_G$ comes from $\adv$, it follows that \[\Adv{}{G_3} \leq \Adv{}{G_4}.\]

\textbf{Game 5}: In this game, we replace $g^{hu}$ computed honestly in $\session_b$ with a uniformly random and independent value $\hat{g^{hu}}$. We do so by defining a reduction $\bdv_2$ that initialises a $\DDH$ challenger $\chal_{\DDH}$, and replaces $g^u$, $g^h$ and $g^{hu}$ computed by $\session_b$ and $\partsess$ with the outputs of $\chal_{\DDH}$, $g^a$, $g^b$, $g^c$. We note that if the bit $b$ sampled by $\chal_{\DDH}$ is 1, then $c = ab$ and we are in \textbf{Game 4}. Otherwise, $c \getsr \Zq$ and we are in \textbf{Game 5}. Any $\adv$ that can distinguish \textbf{Game 4} from \textbf{Game 5} can break the $\DDH$ assumption. Thus:  \[\Adv{}{G_4} \leq \Adv{}{G_5}+ \Adv{G,g,q}{\bdv_2 \DDH}.\]

\textbf{Game 6}: In this game we replace the session and encryption keys $sk_i, k_s$ with uniformly random values $\hat{sk_i}, \hat{k_s}$. We do so by defining a reduction $\bdv_3$ that interacts with a $\KDF$ challenger $\chal_{\KDF}$, querying $\chal_{\KDF}$ with $\hat{g^{uh}}$ and replacing the computation of $sk_i,k_s$ in $\session_b$ and $\partsess$ with the outputs from the $\chal_{\KDF}$ $\hat{sk_i}, \hat{k_s}$. Since $sk_i, k_s \gets \KDF(\hat{g^{uh}})$ and by \textbf{Game 5} $\hat{g^{uh}}$ is already uniformly random and independent, this change is sound. Any $\adversary$ that can distinguish \textbf{Game 5} from \textbf{Game 6} can be used to break $\KDF$ security of the $\KDF$ scheme. Thus:  \[\Adv{}{G_5} \leq \Adv{}{G_6}+ \Adv{\KDF}{\bdv_3,\KDF}.\]

\textbf{Game 7}: In this game, we replace the plaintexts in messages $\mathbf{M_2}$, $\mathbf{M_7}$ and $\mathbf{M_8}$ with uniformly random values of the same length. We do so by defining a reduction $\bdv_4$ that initialises an $\Enc$ challenger $\chal_{\Conf}$, which $\bdv_4$ queries $(p,p^*)$ (where $p$ is the original plaintext message and $p^*\getsr \bits{|p|}$) when it needs to encrypt with $\hat{k_s}$. By \textbf{Game 6} $\hat{k_s}$ is already uniformly random and independent, and this replacement is sound. If the bit $b$ sampled by $\chal_{\Conf}$ is 0, then we are in \textbf{Game 6}, otherwise, we are in \textbf{Game 7}. Any $\adversary$ that can distinguish between  \textbf{Game 6} and \textbf{Game 7} can be used by $\bdv_4$ to break the $\Conf$ security of $\Enc$. This implies: \[\Adv{}{G_6} \leq \Adv{}{G_7}+ \Adv{\AuthEnc}{\bdv_4,\Enc}.\]
     
Now we highlight that all plaintext messages sent across the network to and from $\session_b$ and its matching subset are uniformly random and independent of the bit $b$ sampled  by the challenger. Thus it follows that $\adv$ has no advantage in guessing the bit $b$, and summing the probabilities $\adv$ has a negligible advantage in winning the $\Unlink$ game. Thus: 
\[\Adv{}{G_{7}}=0.\]

\end{pro}

\fi

\section{Discussion}
\label{Discussion}

In this section, we first compare our proposed scheme $\UniHand$ with existing state-of-the-art AKA protocols proposed for 5G \cite{3rd_generation_partnership_project_3gpp_security_2020},\cite{cao_cppha_2021}, \cite{fan_rehand_2020} and \cite{koutsos_5g-aka_2019} in terms of security features and then describe the computational and communication cost of the proposed $\UniHand$ scheme. Table \ref{tab:sec.Features} show that the existing schemes cannot simultaneously ensure all essential security features (described in Section \ref{Design Goals}). In particular, no previous scheme is KCI-resilient, which is important in preventing impersonation attacks due to the leakage of long-term keys. Similarly, managing the significant number of subscribers in the 5G network is always essential. However, apart from $\UniHand$, only  \cite{fan_rehand_2020} manages subscription revocation. The revocation mechanism used in \cite{fan_rehand_2020} is based on Nyberg’s one-way accumulator \cite{nyberg_fast_1996} that is a static base accumulator (i.e. regeneration of the accumulator is required for every addition/deletion to the revocation list), hence introducing computation overhead and negatively affecting the network’s efficiency. 
\ra{From Table \ref{tab:sec.Features}, we can also see that none of the previous schemes has considered \textit{fully} protecting previous communication in the presence of an adversary with a compromised long-term key (PFS) or supported KEF for fair secret key generation other than RUSH protocol \cite{zhang_robust_2021}. Both properties are essential to maintain the security of the 5G mobile network in case of long-term key corruption or trusted third-party corruption. For example, the protocol presented in \cite{zhang_robust_2021} utilises the authentication protocol of the conventional 5G AKA and extends it to propose a new handover protocol with the assistance of Blockchain. The proposed handover protocol in \cite{zhang_robust_2021} supports PFS but their initial authentication, which is based on the conventional 5G AKA, can not support PFS; thus, we can say that the protocol presented in \cite{zhang_robust_2021} can partially address the PFS in the overall scheme. Similarly, the protocol of \cite{zhang_robust_2021} partially supports strong anonymity, as it achieves anonymity but does not support unlinkability. } 
On the contrary, $\UniHand$ supports all the crucial security properties such as mutual authentication, strong anonymity, perfect forward secrecy, key-escrow-free, KCI resilience, and secure revocation management while achieving secure handover. 

\begin{table}
    \caption{Features comparison.}
    \centering
    \scalebox{0.9}{
    \begin{tabular}{|p{0.19\linewidth}||p{0.06\linewidth}|p{0.09\linewidth}|p{0.09\linewidth}|p{0.08\linewidth}|p{0.08\linewidth}|p{0.08\linewidth}|p{0.08\linewidth}|}
    \hline
        \diagbox[width=6em]{Schemes}{Features} & \textbf{MA} &\textbf{SA} & \textbf{PFS} &\textbf{KEF} &\textbf{KCI} & \textbf{SRM} & \textbf{UHO} \\
        \hline \hline
        5G \cite{3rd_generation_partnership_project_3gpp_security_2020} & YES & NO & NO & NO & NO & NO & NO
        \\
        \hline
        CPPHA \cite{cao_cppha_2021} & YES & NO & NO & NO & NO & NO & YES
        \\
        \hline
        ReHand\cite{fan_rehand_2020} & YES & YES & NO & NO & NO & YES & NO
        \\
        \hline
        RUSH\cite{zhang_robust_2021} & YES & Partial & Partial & YES&  NO & NO & YES
        \\
        \hline
        Protocol of \cite{koutsos_5g-aka_2019} & YES & YES & NO & NO & NO & NO & NO
        \\
        \hline
        $\UniHand$ & YES & YES & YES & YES & YES & YES & YES
        \\
    \hline \hline
    \multicolumn {8}{|c|}{\textbf{MA}:Mutual Authentication, \textbf{SA}:Strong Anonymity, \textbf{PFS}: Perfect}\\
    \multicolumn {8}{|c|}{Forward Secrecy, \textbf{KEF}: Key-escrow Free, \textbf{KCI}: Key Compromise}\\
    \multicolumn {8}{|c|}{Impersonation, \textbf{SRM}: Secure Revocation Management,}\\
    \multicolumn {8}{|c|}{\textbf{UHO}: Universal HO}\\
    \hline
    
    \end{tabular}
    }
    \label{tab:sec.Features}

\end{table}

\ra{Now, to show the effectiveness of our scheme, we present and compare the computation cost of $\UniHand$ with the state-of-art handover protocols. To ensure fairness of comparison, we compare $\UniHand$ with  \cite{3rd_generation_partnership_project_3gpp_security_2020, fan_rehand_2020,zhang_robust_2021,cao_cppha_2021}, since like UniHand these schemes can also support handover}. Here we conduct simulations of the cryptographic operations used in $\UniHand$. In this context, we implemented the required cryptographic operations at the server level (as the aggregated network entities, i.e. $\AuC, \gNB$) on a Dell Inspiron machine with an i7 core, 2.30GHz CPU, and 16.0 GB RAM. As for measuring the computational cost at UE, we use Samsung Galaxy Note9, which runs the Android-10 mobile operating system and is equipped with octa-core 1.8GHz Quad-Core ARM Cortex-A55, 2.7GHz Quad-Core Mongoose M3 processors, and 6GB RAM. For implementation specifications, we use Java pairing-based cryptography (JPBC) \cite{de_caro_jpbc_2011} and Java Cryptography Extension (JCE) \cite{technology_network_java_2022}.

\begin{table*}[h]
\centering
\caption{\ra{Performance comparison based on computational cost}}
\label{Computational cost}
\scalebox{0.7}{
\begin{tabular}{|l|l|l|l||l|l|l|}

\hline
\textbf{Protocol}   &\textbf{Entity} & \textbf{Initial Authentication} & \makecell[l]{\textbf{Total} \\$(\boldsymbol{ms})$}& \textbf{Entity} & \textbf{Universal HO} & \makecell[l]{\textbf{Total} \\$(\boldsymbol{ms})$}\\ 
\hline

\multirow{2}{*}{\makecell[l]{\ra{\textbf{Conventional}}\\ \textbf{5G}\cite{3rd_generation_partnership_project_3gpp_security_2020}}}        
&$T_{UE}$ &$4{T_{PRG}} + 2{T_{MAC}} + {T_{AES}}$ & $\approx$ 2.817    
&$T_{UE}$ &$4{T_{AES}} + {T_{PRG}}$& $\approx$ 2.703 \\ 
\cline{2-7}
&$T_{Sys}$&$3{T_{PRG}} + 1{T_{MAC}} + 2{T_H} + {T_{ELG}}$&  $\approx$ 1.278     
&$T_{Sys}$&$4{T_{AES}} + {T_{PGR}}$&   $\approx$ 1.559   \\ \hline

\multirow{2}{*}{\makecell[l]{\ra{\textbf{CPPHA}\cite{cao_cppha_2021}}}}         
&$T_{UE}$ &$5{T_{PRG}} + 2{T_{MAC}} + {T_{AES}} + {T_{H}}$ & $\approx$ 3.687

&$T_{UE}$ &$2{T_{AES}} + {T_{PRG}} + 4T_H$& $\approx$ 3.19 \\ 
\cline{2-7}

&$T_{Sys}$&$4{T_{PRG}} + 1{T_{MAC}} + 3{T_H} + 2{T_{ELG}}+{T_{AES}}$& $\approx$ 2.536   
&$T_{Sys}$&${T_{AES}} + {T_{PGR}}+ 6{T_H} + 3{T_{ELG}}$&  $\approx$ 3.04   \\ \hline

\multirow{2}{*}{\ra{\textbf{ReHand}\cite{fan_rehand_2020}}}     
&$T_{UE}$&$2{T_{AES}} + 4{T_H} + {T_{PRG}}$&   $\approx$ 3.193  
&$T_{UE}$&${T_{PGR}} + 3{T_H} + {T_{AES}}$&  $\approx$ 2.232  \\ 
\cline{2-7}
&$T_{Sys}$&$3{T_{AES}} + 5{T_H} + {T_{PRG}}$&  $\approx$ 1.727   
&$T_{Sys}$&${T_{PRG}} + 5{T_H} + 2{T_{AES}}$&   $\approx$ 1.342    \\ \hline

\multirow{2}{*}{\ra{\textbf{RUSH}\cite{zhang_robust_2021}}   }
&$T_{UE}$& $7{T_{PRG}} + 2{T_{MAC}} + {T_{AES}} +3{T_H} + {T_E}+ 5{T_{Mod}} + 3{T_{SM}}$ &  $\approx$ 9.737  
&$T_{UE}$&$3{T_{PRG}} + {T_{SM}} + 5{T_H} + {T_E} + {T_{Mod}}$&   $\approx$ 5.42  \\ 
\cline{2-7}
 &$T_{Sys}$& $4{T_{PRG}} + 1{T_{MAC}} + 3{T_H} + {T_{ ELG}} + 2{T_{SM}} + 3{T_{Mod}}$&   $\approx$ 1.967    
&$T_{Sys}$&$2{T_{PRG}} + {T_{SM}} + 6{T_H} + {T_E} + {T_{Mod}}$&$\approx$ 1.365    \\ \hline

\multirow{2}{*}{\textbf{UniHand} } 
&$T_{UE}$& $3{T_{PRG}}+ 6{T_{AES}} + {T_H}+ 5{T_E}+ 5{T_{Mod}} $ &   $\approx$ 9.48 
&$T_{UE}$&$ 3{T_{PRG}} + 2{T_{AES}} + {T_H} + 3{T_E} + 3{T_{Mod}}$& $\approx$ 5.516   \\ 
\cline{2-7} 
&$T_{Sys}$& $8{T_{PRG}}+ 11{T_{AES}} + 4{T_H}+ 9{T_E}+ 9{T_{Mod}}$ &   $\approx$ 8.759 
&$T_{Sys}$&$ 2{T_{PGR}} + 2{T_{AES}} + {T_H}+ 3{T_E} + 3{T_{Mod}}$&  $\approx$  2.153  \\ 
\hline

\hline \hline

    \multicolumn {7}{|l|}{$\pmb{T_{PRG}}$: Random number generators[$T_{\UE}= 0.47$],[$T_{Sys}= 0.13$], $\pmb{T_{AES}}$:Symmetric encryption/decryption[$T_{\UE}= 0.56$],[$T_{Sys}= 0.39$], $\pmb{T_{MAC}}$: Message authentication}\\
    \multicolumn {7}{|l|}{(Hmac-SHA256) [$T_{\UE}= 0.195 $],[$T_{Sys}= 0.071$] $\pmb{T_{H}}$: Hash operations(SHA-256)[$T_{\UE}= 0.40$],[$T_{Sys}= 0.09$],  $\pmb{T_{E}}$: exponentiation operation[$T_{\UE}= 0.86$],[$T_{Sys}= 0.34$] }\\
    \multicolumn {7}{|l|}{$\pmb{T_{Mod}}$: Modular operations[$T_{\UE}= 0.002$],[$T_{Sys}= 0.001$], $\pmb{T_{ELG}}$: Elgamal Asymmetric encryption/decryption operations[$T_{\UE}= 1.176$],[$T_{Sys}= 0.648$]}\\
    \multicolumn {7}{|l|}{, $\pmb{T_{SM}}$: scalar multiplication [$T_{\UE}=1.148 $],[$T_{Sys}=0.235 $]}\\
        \hline

\end{tabular}
}

\end{table*}

\begin{table*}
\centering
    \caption{\ra{Performance comparison based on communication cost.}}
        \label{tab:Per. communication}
\begin{tabular}{|l|l|l|l|l||l|}
\hline
\textbf{Protocol}   & \textbf{Link} & \textbf{\makecell[l]{Total message size \\ (bits)}}&  \textbf{\makecell[l]{Transmission \\ time ($\mu s$)}} &  \textbf{\makecell[l]{Propagation \\ time ($\mu s$)}} &  \textbf{\makecell[l]{Total \\ time ($\mu s$)}}\\ \hline

\multirow{2}{*}{\makecell[l]{\ra{\textbf{Conventional}}\\ \textbf{5G}\cite{3rd_generation_partnership_project_3gpp_security_2020}}}         
& UP & 640 & 25.6 & 2.01& 27.64   \\ \cline{2-6} 
& Down & 256 & 5.12 &1.33 & 6.46  \\ \hline

\multirow{2}{*}{\makecell[l]{\ra{\textbf{CPPHA}\cite{cao_cppha_2021}}}}         
& UP & 1728 & 69.12 & 2.68&  71.8  \\ \cline{2-6} 
& Down & 1056 &21.12  &1.33 &  22.45 \\ \hline
                        
\multirow{2}{*}{\ra{\textbf{ReHand}\cite{fan_rehand_2020}} }
& UP & 832 & 33.28 & 1.34 &  34.62  \\ \cline{2-6} 
& Down & 768 & 15.36 &  0.67 & 39.8  \\ \hline

\multirow{2}{*}{\ra{\textbf{RUSH}\cite{zhang_robust_2021}}  }
& UP & 896  &35.84 & 1.33 & 37.17   \\ \cline{2-6} 
& Down & 896 & 17.92& 0.67 &  18.59   \\ \hline

\multirow{2}{*}{\textbf{UniHand}}  
& UP & 2109  & 84.36 & 0.67 & 85.03   \\ \cline{2-6} 
& Down & 3901  & 78 & 1.33 & 79.33     \\ \hline

\end{tabular}

\end{table*}

Table \ref{Computational cost} compares the performance based on the computations cost. It shows the used primitives in each protocol to measure the total required time to perform the protocol at the UE level ($T_{\UE}$) and system level ($T_{Sys}$).
\ra{$\UniHand$'s initial authentication protocol requires approximately 9.48${ms}$ and 8.759 ${ms}$ on $\UE$ and system level, respectively. On the other hand, the universal HO protocol requires approximately 5.561 ${ms}$ and 2.153 ${ms}$ on $\UE$ and system level, respectively.} 

 \begin{figure*}
 \centering
 \begin{floatrow}
\captionsetup{justification=centering}

    \ffigbox{\includegraphics[width=\columnwidth]{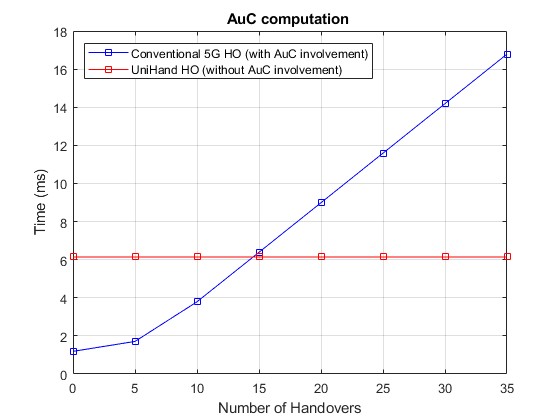}}
    {\caption{Computation overhead at AuC during HO for conventional 5G and UniHand}
    \label{fig:AuC com.}}

    \ffigbox{\includegraphics[width=\columnwidth]{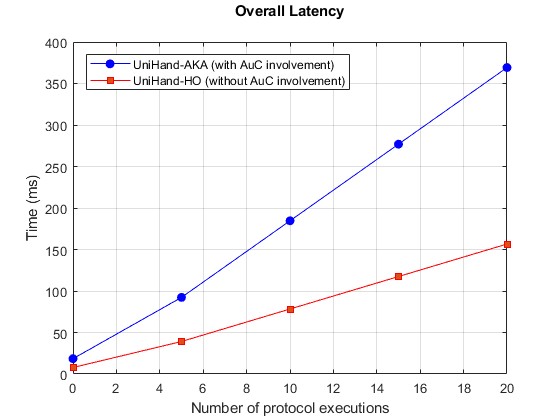}}
    {\caption{Overall latency at AuC during the executions of UniHand protocols}
    \label{fig:Overall latency}}

\end{floatrow}
\end{figure*}
\begin{comment}
    
\rabiah{Now, it should be noted that during the execution of the proposed universal HO protocol, $\gNB$ is the only system component to execute the HO with the $\UE$. Hence reduces the computation overhead on the $\AuC$.} 
\end{comment}

\rabiah{The AuC (Authentication Center) plays a crucial role in various important functions within the mobile network, including connectivity and mobility management, authentication and authorization, subscriber data management, and policy management.
As the number of connected and roaming users increases in 5G handovers, the communication to the AuC and its computation also grows.  Due to the higher number of small cells, handovers will occur more frequently compared to 4G. If the AuC is incorporated in each handover, it could potentially be overwhelmed \cite{fan_rehand_2020, parvez_survey_2017}. Our proposed $\UniHand$ scheme can resolve this issue by eliminating the need for the AuC's involvement during the execution of the handover protocol. This reduction in AuC's responsibilities effectively decreases or stabilizes the overall computation overhead on the AuC compared to the conventional 5G AKA and handover protocols (as shown in Figure \ref{fig:AuC com.}). 
Figure \ref{fig:AuC com.}  presents a comparative analysis between UniHand and the conventional 5G HO scheme \cite{3rd_generation_partnership_project_3gpp_security_2020} concerning the required computational cost at the AuC for authenticating a single user, including one execution of 5G-AKA and one execution of UniHand-AKA, while the user roams between several small cells (involving 35 HO protocol executions). In the conventional 5G scheme, where AuC assistance is obligatory for each handover, the computational cost at the AuC escalates proportionally with the number of handovers. Conversely, our proposed scheme (UniHand) maintains a consistent computational cost at the AuC, independent of the number of handovers, as it doesn't necessitate AuC intervention. As a result, our proposed scheme does not impose any additional computation cost on the AuC during the HO process. Additionally, when the AuC is not accessible or offline, our proposed HO protocol remains unaffected, representing a significant improvement from the current solution. }

\ra{Next, we analyse and compare the communication cost of the proposed $\UniHand$ universal HO protocol with respect to the existing handover protocols, including the conventional 5G \cite{3rd_generation_partnership_project_3gpp_security_2020}, ReHand \cite{fan_rehand_2020}, RUSH \cite{zhang_robust_2021} and CPPHA \cite{cao_cppha_2021}. In this context, we measure the transmitted message's propagation and transmission time.  We consider the size and the network's data rate for accurate computation of transmission delay. Based on the 3GPP specification \cite{3rd_generation_partnership_project_3gpp_security_2020}, in a wide-area scenario, the data rate of uplink data rate is 25 Mbps, and the downlink is 50 Mbps. Accordingly, here we use the above-specified measurements to compute the transmission delay of transmitted messages in all protocols. For the message size,  we use the recommended message size in each state-of-the-art protocol for a head-to-head comparison. In $\UniHand$ the certificate $\cert$ is of size 192 bits, the signature is 1533 bits, ECDH key size is 256 bits, and finally, the non-membership witness is 2048 bits.}
\ra{On the other hand, we consider the spectrum wave speed and the distance between the user and the nearest connected gNB to measure the propagation delay. In this context, we follow the specified propagation speed and distance in 3GPP specification \cite{3rd_generation_partnership_project_3gpp_security_2020}, which is  $3 \times 108 m/s$ and 200 m for the propagation speed and the approximate maximum cell size in 5G, respectively. Hence the propagation delay in sending one message in 5G is approximately 0.67 $\mu s$. Table \ref{tab:Per. communication} shows the communication cost of UniHand and the related schemes. From the  Table \ref{tab:Per. communication} we can see that $\UniHand$ has a little more communication overhead compared to the existing state-of-the-art protocols \cite{3rd_generation_partnership_project_3gpp_security_2020},  \cite{fan_rehand_2020},  \cite{zhang_robust_2021}, \cite{cao_cppha_2021}.}

\begin{remark}
    \ra{The rise in the communication cost is directly associated with the size of the used certificate alongside its signature
    and the non-membership witness for user revocation. Nevertheless, the combination of signatures and certificates enables our scheme to achieve a secure universal handover without the assistance of the AuC. While the addition of a non-membership witness enables $\UniHand$ to manage revoked users in the network. It is also worth noting that our comparison is not quite fair, as only $\UniHand$ counts the communication cost of the revocation scheme, which adds more cost on $\UniHand$.}
\end{remark}

\rabiah{Although both the proposed protocols (Initial Authentication and Handover) ensure all the specified security properties, there is a notable distinction between them. Unlike the Initial Authentication protocol, the execution of our proposed Handover protocol does not involve the AuC. Figure \ref{fig:Overall latency} provides an overview of the overall latency cost for each protocol in a handover-based scenario, including both communication and computation costs. To simulate this scenario, where a user moves between cells (0-20 cells), we execute both UniHand protocols (Initial Authentication and Handover) and measure the overall latency required for 0-20 executions of each protocol. From the observations in Figure \ref{fig:Overall latency}, it is evident that the execution time of our proposed Handover protocol (which does not require the AuC) is considerably less than our Initial Authentication protocol (which does). This reduction in execution time demonstrates the efficiency and benefits of our Handover protocol, which eliminates the need for AuC involvement. Despite this advantage, our proposed scheme ensures several key security properties (such as PFS, KEF, and KCI) that the conventional 5G scheme fails to achieve. By striking this balance between efficient performance and enhanced security features, UniHand represents a promising approach to address the limitations of state-of-the-art protocols.}

\begin{remark}
\rabiah{In general, the 5G networks have an acceptable latency range defined by ETSI \cite{dahmen-lhuissier_5g_nodate} of 1 ms to 100 ms, with further discussion of the ETSI standards in \cite{parvez_survey_2017,thales_group_5g_2023}. This latency is particularly relevant for scenarios that require low latency, such as VR-assisted tele-surgery, Intelligent Transportation Systems, and factory automation. The latency rates are influenced by both computational and communication costs. In the case of the proposed UniHand handover, the total computational cost including UE and Sys is 7.66 $ms$, while the total communication cost for upload and download is 164.36 $\mu s$. Consequently, the overall latency cost incurred by the proposed scheme is only 7.83 $ms$,  which perfectly falls within the acceptable range of latency requirement for 5G networks and remarkably close to the absolute minimum range of 1 ms. Therefore, our handover achieves accepted latency rates as defined by the ETSI standards. }
\end{remark}

\begin{comment}

\begin{figure}
    \includegraphics [scale=0.5]{AuC computation 3.png}
    \caption{AuC computation Overhead during HO}
    \label{fig:AuC com.}

\end{figure}
\end{comment}

\iffullversion
\subsection{Informal Security Analysis}
Due to the importance of security and privacy in all cellular networks, including 5G, we highlight that non of the previous related works support \textit{all} desired security requirements in 5G. Table \ref{tab:sec.Features}, provides a comparison on security requirements (discussed in \ref{Design Goals}) between $\UniHand$ and the state of the art protocols \cite{3rd_generation_partnership_project_3gpp_security_2020},\cite{fan_rehand_2020},\cite{zhang_robust_2021},\cite{yan_lightweight_2021}.

In this part, based on the discussion in Section \ref{Design Goals}, we verify various security properties of $\UniHand$. Here we demonstrate that our proposed scheme can achieve the required security features such as MA, strong anonymity, PFS, key escrow and KCI, which are formally proved in section \ref{Formal Security Analysis}. 
\paragraph{\textbf{Accomplishment of Mutual Authentication}:}
    All participants in this protocol ($\UE,\gNB \&\AuC$) authenticate each other using certificates that are signed using $\SanSig$ algorithm. In the initial authentication protocol, the   $\AuC$ issues and signs all certificates in the network and delegates sanitising capabilities to the certificates' owners. In $\UniHand$,  including Initial authentication and Universal HO, certificate verification ensures the authenticity of communicating parties in both protocols. If any certificate verification is invalid, then the protocol will be terminated. Formally, according to theorem \ref{theorem: MA sec- initial auth} and \ref{theorem: MA sec- HO}, no $\adv$ can break $\MA$ security of both protocols assuming the security of EUFCMA, AE, KDF and DDH cryptographic primitives. The formal security analysis of the achievement of MA can be found in [\ref{MA of Initial Authentication}$\&$ \ref{MA of Universal Handover}].

\paragraph{ \textbf{Accomplishment of Strong anonymity:}}
    The Initial authentication protocol maintains a pseudo ($\pid$) and a temporary identity ($\tid$)to preserve user privacy, where there is no direct relationship between the aliases. The $\pid$ is always encrypted twice using two private keys, the long-term and session keys, and signed using a third key for integrity purposes. The $\tid$ is only used once per run of the Initial authentication protocol. Also, it is always encrypted using per session encryption key. Additionally, the $\tid $ is always updated after a successful run of the initial authentication. Hence, achieving unlinkability of users' activities. Similarly, users' certificates are constantly updated using the sanitising keys in the universal Ho protocol. Formally, based on theorem \ref{theorem: UNlink sec- HO} and \ref{theorem: UNlink sec- initial auth.} no $\adv$ can break $\Unlink$ security of the $\UniHand$ protocols assuming the security of EUFCMA, AE, KDF and DDH cryptographic primitives. The formal security analysis of the achievement of $\Unlink$ can be found in [\ref{UNlink of Initial Authentication}].
    
\paragraph{\textbf{Accomplishment of Perfect Forward Secrecy}:}
    To achieve PFS, we leverage the notion of DH key agreement and generate session keys by using two ephemeral keys from each participant.
    
\paragraph{\textbf{Accomplishment of Key escrow free}:}
    In our proposed protocol, only the long-term key (symmetric) is generated by the $\AuC$; the entity itself generates all other sanitising keys. Hence, no escrow/ third party manages all keys in our protocol, which reduces dependence on one single entity and provides resistance against a compromised third party.

\paragraph{ \textbf{Accomplishment of security against Key Compromise Impersonation}:}
    The proposed protocol provides resistance against the KCI attack; compromising a single secret key will not enable the adversary to make an impersonation attack. In order to impersonate any participant to the compromised device, the adversary needs to compromise both the long-term key with the signing/sanitising private key. In other words, the security of our scheme is based on two private keys, a long-term key shared between $\UE \& \AuC$ ($k_i$) and private signing/sanitising keys ($\sk_{sig}^{\AuC}, \sk_{san}^{\gNB/\UE}$). We are not allowing the $\adv$ to compromise \textit{both} keys before the protocol accepts. Hence the $\adv$ cannot lunch a KCI attack on our scheme, where the protocol will terminate if one of the private keys is corrupted.

    The formal security analysis of the KCI resistance can be found in [\ref{MA of Initial Authentication}].
\fi

\section{Conclusion}\label{sec:conclusion}
This article proposed a new AKA handover scheme ($\UniHand$) to achieve secure, privacy-preserving universal authentication for small cell networks in 5G mobile networks. Our proposal $\UniHand$ can guarantee all the essential security properties (as mentioned in Section \ref{Design Goals}). It can tackle all the security vulnerabilities of the existing schemes and weaknesses in the conventional 5G-AKA. Our proposed scheme has been designed based on sanitisable signatures, ephemeral Diffie-Hellman key exchange, key derivation functions, authenticated encryption, and dynamic accumulator. We conducted a formal security and privacy analysis of the proposed scheme and compared the security features of the proposed $\UniHand$ scheme with the existing state-of-the-art 5G-AKA protocols. Finally, We evaluated the performance of the $\UniHand$ scheme, which shows a reasonable computation and communication cost while achieving all required security and privacy properties.

\bibliographystyle{splncs04}
\bibliography{mybibliography}

\begin{thebibliography}{10}
\providecommand{\url}[1]{\texttt{#1}}
\providecommand{\urlprefix}{URL }
\providecommand{\doi}[1]{https://doi.org/#1}

\bibitem{ateniese_sanitizable_2005}
Ateniese, G., Chou, D.H., de~Medeiros, B., Tsudik, G.: Sanitizable {Signatures}. In: di~Vimercati, S.d.C., Syverson, P., Gollmann, D. (eds.) Computer {Security} – {ESORICS} 2005. pp. 159--177. Lecture {Notes} in {Computer} {Science}, Springer, Berlin, Heidelberg (2005). \doi{10.1007/11555827-10}

\bibitem{authors_unihand_2023}
Authors, A.: {UniHand} supplementary material (2023), \url{https://drive.google.com/file/d/1UZtLpNMOW21ptTNMBAK34818gtRaasFT/view?usp=share_link}

\bibitem{baldimtsi_accumulators_2017}
Baldimtsi, F., Camenisch, J., Dubovitskaya, M., Lysyanskaya, A., Reyzin, L., Samelin, K., Yakoubov, S.: Accumulators with {Applications} to {Anonymity}-{Preserving} {Revocation}. In: 2017 {IEEE} {European} {Symposium} on {Security} and {Privacy} ({EuroS}\&{P}). pp. 301--315 (Apr 2017). \doi{10.1109/EuroSP.2017.13}

\bibitem{basin_formal_2018}
Basin, D., Dreier, J., Hirschi, L., Radomirovic, S., Sasse, R., Stettler, V.: A {Formal} {Analysis} of {5G} {Authentication}. In: Proceedings of the 2018 {ACM} {SIGSAC} {Conference} on {Computer} and {Communications} {Security}. pp. 1383--1396. {CCS} '18, Association for Computing Machinery, New York, NY, USA (Oct 2018). \doi{10.1145/3243734.3243846}, \url{http://doi.org/10.1145/3243734.3243846}

\bibitem{bellare_authenticated_2008}
Bellare, M., Namprempre, C.: Authenticated {Encryption}: {Relations} among {Notions} and {Analysis} of the {Generic} {Composition} {Paradigm}. Journal of Cryptology  \textbf{21}(4),  469--491 (Oct 2008). \doi{10.1007/s00145-008-9026-x}, \url{https://doi.org/10.1007/s00145-008-9026-x}

\bibitem{bellare_security_2006}
Bellare, M., Rogaway, P.: The {Security} of {Triple} {Encryption} and a {Framework} for {Code}-{Based} {Game}-{Playing} {Proofs}. In: Vaudenay, S. (ed.) Advances in {Cryptology} - {EUROCRYPT} 2006. pp. 409--426. Lecture {Notes} in {Computer} {Science}, Springer, Berlin, Heidelberg (2006). \doi{10.1007/11761679-25}

\bibitem{benaloh_one-way_1994}
Benaloh, J., de~Mare, M.: One-{Way} {Accumulators}: {A} {Decentralized} {Alternative} to {Digital} {Signatures}. In: Helleseth, T. (ed.) Advances in {Cryptology} — {EUROCRYPT} ’93. pp. 274--285. Lecture {Notes} in {Computer} {Science}, Springer, Berlin, Heidelberg (1994). \doi{10.1007/3-540-48285-7-24}

\bibitem{borgaonkar_new_2018}
Borgaonkar, R., Hirschi, L., Park, S., Shaik, A.: New {Privacy} {Threat} on {3G}, {4G}, and {Upcoming} {5G} {AKA} {Protocols}. Tech. Rep.~1175 (2018), \url{http://eprint.iacr.org/2018/1175}

\bibitem{braeken_symmetric_2020}
Braeken, A.: Symmetric key based {5G} {AKA} authentication protocol satisfying anonymity and unlinkability. Computer Networks  \textbf{181},  107424 (Nov 2020). \doi{10.1016/j.comnet.2020.107424}, \url{https://www.sciencedirect.com/science/article/pii/S1389128620311130}

\bibitem{brzuska_security_2009}
Brzuska, C., Fischlin, M., Freudenreich, T., Lehmann, A., Page, M., Schelbert, J., Schröder, D., Volk, F.: Security of {Sanitizable} {Signatures} {Revisited}. In: Jarecki, S., Tsudik, G. (eds.) Public {Key} {Cryptography} – {PKC} 2009. pp. 317--336. Lecture {Notes} in {Computer} {Science}, Springer, Berlin, Heidelberg (2009). \doi{10.1007/978-3-642-00468-1-18}

\bibitem{brzuska_santizable_2009}
Brzuska, C., Fischlin, M., Lehmann, A., Schröder, D.: Santizable {Signatures}: {How} to {Partially} {Delegate} {Control} for {Authenticated} {Data}. pp. 117--128 (Jan 2009)

\bibitem{brzuska_unlinkability_2010}
Brzuska, C., Fischlin, M., Lehmann, A., Schröder, D.: Unlinkability of {Sanitizable} {Signatures}. pp. 444--461 (May 2010). \doi{10.1007/978-3-642-13013-7-26}

\bibitem{brzuska_non-interactive_2013}
Brzuska, C., Pöhls, H.C., Samelin, K.: Non-interactive {Public} {Accountability} for {Sanitizable} {Signatures}. In: De~Capitani~di Vimercati, S., Mitchell, C. (eds.) Public {Key} {Infrastructures}, {Services} and {Applications}. pp. 178--193. Lecture {Notes} in {Computer} {Science}, Springer, Berlin, Heidelberg (2013). \doi{10.1007/978-3-642-40012-4-12}

\bibitem{brzuska_efficient_2014}
Brzuska, C., Pöhls, H.C., Samelin, K.: Efficient and {Perfectly} {Unlinkable} {Sanitizable} {Signatures} without {Group} {Signatures}. In: Katsikas, S., Agudo, I. (eds.) Public {Key} {Infrastructures}, {Services} and {Applications}. pp. 12--30. Lecture {Notes} in {Computer} {Science}, Springer, Berlin, Heidelberg (2014). \doi{10.1007/978-3-642-53997-8-2}

\bibitem{cao_cppha_2021}
Cao, J., Ma, M., Fu, Y., Li, H., Zhang, Y.: {CPPHA}: {Capability}-{Based} {Privacy}-{Protection} {Handover} {Authentication} {Mechanism} for {SDN}-{Based} {5G} {HetNets}. IEEE Transactions on Dependable and Secure Computing  \textbf{18}(3),  1182--1195 (May 2021). \doi{10.1109/TDSC.2019.2916593}, conference Name: IEEE Transactions on Dependable and Secure Computing

\bibitem{chalkias_two_2009}
Chalkias, K., Baldimtsi, F., Hristu-Varsakelis, D., Stephanides, G.: Two {Types} of {Key}-{Compromise} {Impersonation} {Attacks} against {One}-{Pass} {Key} {Establishment} {Protocols}. In: Filipe, J., Obaidat, M.S. (eds.) E-business and {Telecommunications}. pp. 227--238. Communications in {Computer} and {Information} {Science}, Springer, Berlin, Heidelberg (2009). \doi{10.1007/978-3-540-88653-2-17}

\bibitem{cremers_component-based_2019}
Cremers, C., Dehnel-Wild, M.: Component-{Based} {Formal} {Analysis} of {5G}-{AKA}: {Channel} {Assumptions} and {Session} {Confusion}. San Diego, CA, USA (Feb 2019), \url{https://publications.cispa.saarland/2758/}

\bibitem{dahmen-lhuissier_5g_nodate}
Dahmen-Lhuissier, S.: {5G}, \url{https://www.etsi.org/technologies/5G}

\bibitem{de_caro_jpbc_2011}
De~Caro, A., Iovino, V.: {jPBC}: {Java} pairing based cryptography. Proceedings of the 16th {IEEE} {Symposium} on {Computers} and {Communications}, {ISCC} 2011, IEEE (2011), \url{http://gas.dia.unisa.it/projects/jpbc/}

\bibitem{fan_rehand_2020}
Fan, C.I., Huang, J.J., Zhong, M.Z., Hsu, R.H., Chen, W.T., Lee, J.: {ReHand}: {Secure} {Region}-{Based} {Fast} {Handover} {With} {User} {Anonymity} for {Small} {Cell} {Networks} in {Mobile} {Communications}. IEEE Transactions on Information Forensics and Security  \textbf{15},  927--942 (2020). \doi{10.1109/TIFS.2019.2931076}

\bibitem{3rd_generation_partnership_project_3gpp_security_2020}
3rd Generation Partnership Project~(3GPP), E.: Security architecture and procedures for {5G} {System}. Technical {Specification} version 16.3.0 Release 16, 3GPP (Aug 2020), \url{https://www.etsi.org/deliver/etsi_ts/133500_133599/133501/16.03.00_60/ts_133501v160300p.pdf}

\bibitem{thales_group_5g_2023}
thales group: {5G} technology and networks (speed, use cases, rollout) (Feb 2023), \url{https://www.thalesgroup.com/en/markets/digital-identity-and-security/mobile/inspired/5G}

\bibitem{khan_identity_2018}
Khan, H., Dowling, B., Martin, K.M.: Identity {Confidentiality} in {5G} {Mobile} {Telephony} {Systems}. Tech. Rep.~876 (2018), \url{http://eprint.iacr.org/2018/876}

\bibitem{koutsos_5g-aka_2019}
Koutsos, A.: The {5G}-{AKA} {Authentication} {Protocol} {Privacy}. In: 2019 {IEEE} {European} {Symposium} on {Security} and {Privacy} ({EuroS}\&{P}). pp. 464--479 (Jun 2019). \doi{10.1109/EuroSP.2019.00041}

\bibitem{lamacchia_stronger_2007}
LaMacchia, B., Lauter, K., Mityagin, A.: Stronger {Security} of {Authenticated} {Key} {Exchange}. In: Susilo, W., Liu, J.K., Mu, Y. (eds.) Provable {Security}. pp. 1--16. Lecture {Notes} in {Computer} {Science}, Springer, Berlin, Heidelberg (2007). \doi{10.1007/978-3-540-75670-5}

\bibitem{li_universal_2007}
Li, J., Li, N., Xue, R.: Universal {Accumulators} with {Efficient} {Nonmembership} {Proofs}. In: Katz, J., Yung, M. (eds.) Applied {Cryptography} and {Network} {Security}. pp. 253--269. Lecture {Notes} in {Computer} {Science}, Springer, Berlin, Heidelberg (2007). \doi{10.1007/978-3-540-72738-5-17}

\bibitem{nyberg_fast_1996}
Nyberg, K.: Fast accumulated hashing. In: Gollmann, D. (ed.) Fast {Software} {Encryption}. pp. 83--87. Lecture {Notes} in {Computer} {Science}, Springer, Berlin, Heidelberg (1996). \doi{10.1007/3-540-60865-6-45}

\bibitem{parvez_survey_2017}
Parvez, I., Rahmati, A., Guvenc, I., Sarwat, A., Dai, H.: A {Survey} on {Low} {Latency} {Towards} {5G}: {RAN}, {Core} {Network} and {Caching} {Solutions}. IEEE Communications Surveys \& Tutorials  \textbf{PP} (Aug 2017). \doi{10.1109/COMST.2018.2841349}

\bibitem{peltonen_comprehensive_2021}
Peltonen, A., Sasse, R., Basin, D.: A comprehensive formal analysis of {5G} handover. In: Proceedings of the 14th {ACM} {Conference} on {Security} and {Privacy} in {Wireless} and {Mobile} {Networks}. pp. 1--12. {WiSec} '21, Association for Computing Machinery, New York, NY, USA (Jun 2021). \doi{10.1145/3448300.3467823}, \url{http://doi.org/10.1145/3448300.3467823}

\bibitem{technology_network_java_2022}
Technology~Network., O.: Java {Cryptography} {Architecture} ({JCA}) {Reference} {Guide} (Sep 2022), \url{https://docs.oracle.com/javase/8/docs/technotes/guides/security/crypto/CryptoSpec.html}

\bibitem{yan_lightweight_2021}
Yan, X., Ma, M.: A lightweight and secure handover authentication scheme for {5G} network using neighbour base stations. Journal of Network and Computer Applications  \textbf{193},  103204 (Nov 2021). \doi{10.1016/j.jnca.2021.103204}, \url{https://www.sciencedirect.com/science/article/pii/S1084804521002095}

\bibitem{zhang_robust_2021}
Zhang, Y., Deng, R.H., Bertino, E., Zheng, D.: Robust and {Universal} {Seamless} {Handover} {Authentication} in {5G} {HetNets}. IEEE Transactions on Dependable and Secure Computing  \textbf{18}(2),  858--874 (Mar 2021). \doi{10.1109/TDSC.2019.2927664}

\end{thebibliography}

\end{document}